\documentclass[aps,pra,amsmath,amssymb,floatfix,twocolumn,groupedaddress,nofootinbib,longbibliography,10pt]{revtex4-1}

\usepackage{times}
\usepackage{bbold}
\usepackage{bm}
\usepackage{graphicx} 
\usepackage{hyperref}
\usepackage{textcomp}
\usepackage{amssymb} 
\usepackage{wasysym}
\usepackage{epstopdf}

\newcommand{\abulk}{\alpha_{\rm bulk}}
\newcommand{\aend}{\alpha_{\rm end}}          
\newcommand{\aimp}{\alpha_{\rm imp}}

\newcommand{\aendj}{\alpha_{j}}      
\newcommand{\aendone}{\alpha_{1}}   
\newcommand{\aendtwo}{\alpha_{2}}   
\newcommand{\deltaj}{\delta\alpha_{j}}
\newcommand{\deltaone}{\delta\alpha_{1}} 
\newcommand{\deltatwo}{\delta\alpha_{2}}  

\newcommand{\dosbulk}{\beta^{\rm bulk}_{r\sigma}}
\newcommand{\dosend}{\beta_{r\sigma}}
\newcommand{\dosendj}{\beta_{j r \sigma}}

\newcommand{\Gbulk}{G_{\rm bulk}}
\newcommand{\Gend}{G_{\rm end}}            
\newcommand{\Gimp}{G_{\rm imp}}
\newcommand{\Gsingleimp}{ \delta G_{\rm 1}}
\newcommand{\Rsingleimp}{ \delta R_{\rm 1}}

\newcommand{\TLL}{Luttinger liquid}
\newcommand{\TLLadj}{Luttinger-liquid}
\newcommand{\TLLabr}{LL}
\newcommand{\TL}{Tomonaga-Luttinger}

\begin{document}

\title{Charge transport of a spin-orbit-coupled Luttinger liquid}

\author{Chen-Hsuan Hsu$^{1}$}
\author{Peter Stano$^{1,2,3}$}
\author{Yosuke Sato$^{2}$}
\author{Sadashige Matsuo$^{1,2,4}$}
\author{Seigo Tarucha$^{1,2}$}
\author{Daniel Loss$^{1,5}$}
 
\affiliation{$^{1}$RIKEN Center for Emergent Matter Science (CEMS), Wako, Saitama 351-0198, Japan}
\affiliation{$^{2}$Department of Applied Physics, School of Engineering, University of Tokyo, 7-3-1 Hongo, Bunkyo-ku, Tokyo 113-8656, Japan}
\affiliation{$^{3}$Institute of Physics, Slovak Academy of Sciences, 845 11 Bratislava, Slovakia}
\affiliation{$^{4}$JST, PRESTO, 4-1-8 Honcho, Kawaguchi, Saitama 332-0012, Japan}
\affiliation{$^{5}$Department of Physics, University of Basel, Klingelbergstrasse 82, CH-4056 Basel, Switzerland}
 
\date{\today}

\begin{abstract}
The charge transport of a (Tomonaga-)Luttinger liquid with tunnel barriers exhibits universal scaling: the current-voltage curves measured at various temperatures collapse into a single curve upon rescaling. The exponent characterizing this single curve can be used to extract the strength of electron-electron interaction.
Motivated by a recent experiment on InAs nanowires [Sato~{\it et~al.}, Phys. Rev. B {\bf 99}, 155304 (2019)], we theoretically investigate the analogous behavior of a {\it spin-orbit-coupled} {\TLL}.
We find that the scaling exponent differs for different impurity strengths, being weak (disorder potential) or strong (tunnel barriers), and their positions, either in the bulk or near the edge of the wire. For each case we quantify the exponent of the universal scaling and its modification due to the spin-orbit coupling. 
Our findings serve as a guide in the determination of the interaction strength of quasi-one-dimensional spin-orbit-coupled quantum wires from transport measurements. 
\end{abstract}

\maketitle

\section{Introduction}
One-dimensional interacting electron systems, in contrast to their higher-dimensional counterparts, invalidate the Fermi-liquid description.
Instead, they can be described as (Tomonaga-)Luttinger liquids~\cite{Tomonaga:1950,Luttinger:1963,Haldane:1981,Voit:1995,vonDelft:1998,Giamarchi:2003}. Among other interesting features, the theory predicts unusual transport properties of the {\TLL}.
Namely, a {\it clean} {\TLL} connected to Fermi-liquid leads has an interaction-independent conductance~\cite{Maslov:1995,Ponomarenko:1995,Safi:1995}. However, defects and impurities (in the form of either tunnel barriers or weak potential disorder) alter the conductance, which becomes dependent on the interaction strength~\cite{Kane:1992a,Kane:1992,Furusaki:1993,Matveev:1993,Maslov:1995b,Sandler:1997,Balents:1999}.
The predicted conductance therefore allows one to confirm the {\TLLadj} nature of the system and to deduce its interaction strength through transport measurements.
 
A typical realization of {\TLL}s is provided by nanowires~\cite{Tarucha:1995,Giamarchi:2003}, in which electrons are confined in two spatial dimensions and are free to move along the third dimension.
Here, the spin-orbit coupling\footnote{Throughout this article, we use ``interaction'' for electron-electron interaction and ``coupling'' for spin-orbit coupling (with very few exceptions), to discriminate the two terms more easily for the readers.} is an important ingredient for exploiting nanowires as elements in spintronics devices~\cite{Datta:1990,Nitta:1997,Fasth:2007,Fabian:2007,Nadj-Perge:2010},
including, more recently, topological states of matter~\cite{Lutchyn:2010,Oreg:2010,Mourik:2012,Das:2012,Deng:2012,Rokhinson:2012,Churchill:2013}.
In the latter examples, an external magnetic field induces Majorana bound states at the ends of a spin-orbit-coupled nanowire in proximity of a superconductor. 
Even though most theoretical works on these Majorana nanowires use the single-particle picture in which electron-electron interaction is ignored, nanowires with both strong electron-electron interaction and strong spin-orbit coupling might have substantial advantages: it has been suggested that they are capable of hosting Majorana Kramers pairs~\cite{Gaidamauskas:2014,Ebisu:2016,Reeg:2017,Schrade:2017,Thakurathi:2018} and computationally more powerful parafermions~\cite{Klinovaja:2014a,Klinovaja:2014,Oreg:2014,Alicea:2016} without applying magnetic fields.
In these proposals, sufficiently strong electron-electron interaction is required to establish topological states. Namely, nonlocal pairing should dominate over local pairing, or, equivalently, the Cooper pair splitting efficiency should exceed unity. 
While such high splitting efficiency has been observed in a Josephson junction made of InAs double nanowires~\cite{Ueda:2018}, the interaction strength of that device remains undetermined.
Since most of the proposals employ the {\TL} model, it is crucial to establish a reliable experimental approach to characterize nanowires with strong spin-orbit coupling within this formalism.

Recently, Ref.~\cite{Sato:2019} attempted to determine the interaction strength in nanowires made of InAs.
There, it was found that the current-bias curves fit well to the universal scaling formula of a {\TLL}~\cite{Balents:1999,Bockrath:1999} for charge density spanning an appreciable range. Accordingly, the universal scaling formula was used to deduce the interaction strength in these nanowires. Since it is well established that the spin-orbit coupling of InAs nanowires is considerably strong~\cite{Luo:1988,Luo:1990,Grundler:2000,Fasth:2007,Heedt:2017}, it is necessary to clarify whether and how the universal scaling formula is affected by it. 

In Ref.~\cite{Sato:2019}, we listed some of the results which we derive here, namely those that were needed to interpret the data of, and in the parameter limits appropriate for, that experiment. In a nutshell, in Ref.~\cite{Sato:2019}, we used the formulas for wires with zero spin-orbit coupling. Here, we provide the theory for wires with arbitrary spin-orbit coupling strength. Specifically, we give the derivations of the formulas, 
analyze their general trends, and focus on the changes induced by the finite value of the spin-orbit coupling.

With this motivation, we investigate theoretically the transport properties of a spin-orbit-coupled {\TLL}.
Despite the fact that, in a purely one-dimensional system, the spin-orbit coupling can be removed by a gauge transformation~\cite{Braunecker:2010,Meng:2014,Kainaris:2015}, in realistic nanowires the presence of transverse degrees of freedom makes the removal argument invalid.\footnote{In addition to a strictly one-dimensional system, the removal requires both a zero external magnetic field and a linear-in-momentum spin-orbit coupling.} As a result, the spin-orbit coupling can cause band distortion~\cite{Moroz:1999,Governale:2002}.
In the bosonization formalism, it translates into an additional term breaking the charge-spin separation in the Hamiltonian~\cite{Moroz:2000,Moroz:2000b}.
Our goal is to analyze how this charge-spin mixing affects the transport properties.

Here it is in order to comment on theoretical works taking into account the backscattering process of the electron-electron interaction (referred to as $g_{1}$ term). It was shown to be irrelevant in the renormalization-group (RG) sense in Ref.~\cite{Moroz:2000b}. 
Among the subsequent works, Refs.~\cite{Gritsev:2005,Kainaris:2015} concluded that the spin-orbit coupling combined with the $g_{1}$ term can induce a gap in the energy spectrum and thus destabilize the {\TLL} phase, whereas Ref.~\cite{Schulz:2009} arrived at the opposite conclusion.
Including external magnetic fields~\cite{Sun:2007,Gangadharaiah:2008,Schmidt:2013,Tretiakov:2013}, intersubband spin-orbit coupling with the chemical potential close to the subband crossing point~\cite{Meng:2014}, 
or spin-umklapp scattering~\cite{Pedder:2016,Pedder:2017} can also lead to anti-crossings or partial gaps in the spectrum.\footnote{Recently a partial gap in the lowest subband of an InAs nanowire was observed in the absence of magnetic fields~\cite{Heedt:2017}.}
Instead of entering this debate, we take a pragmatic approach. Our theory is supposed to provide interpretation for experiments which show no sign of a gap, such as Ref.~\cite{Sato:2019}. Therefore, we adopt the model in Refs.~\cite{Moroz:2000,Moroz:2000b}, incorporating the spin-orbit coupling as a band distortion.

\begin{figure}[t]
 \includegraphics[width=\linewidth]{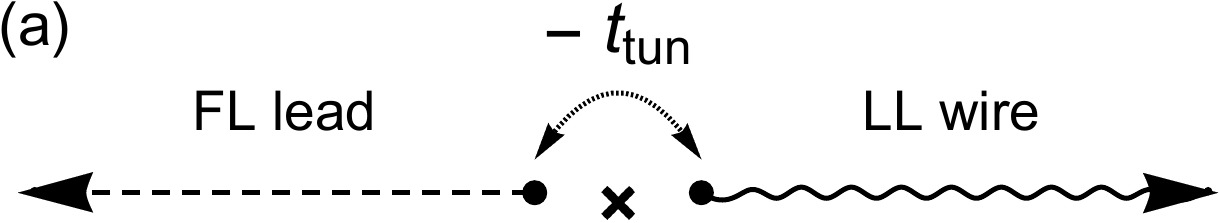} \\ ~\\
 \includegraphics[width=\linewidth]{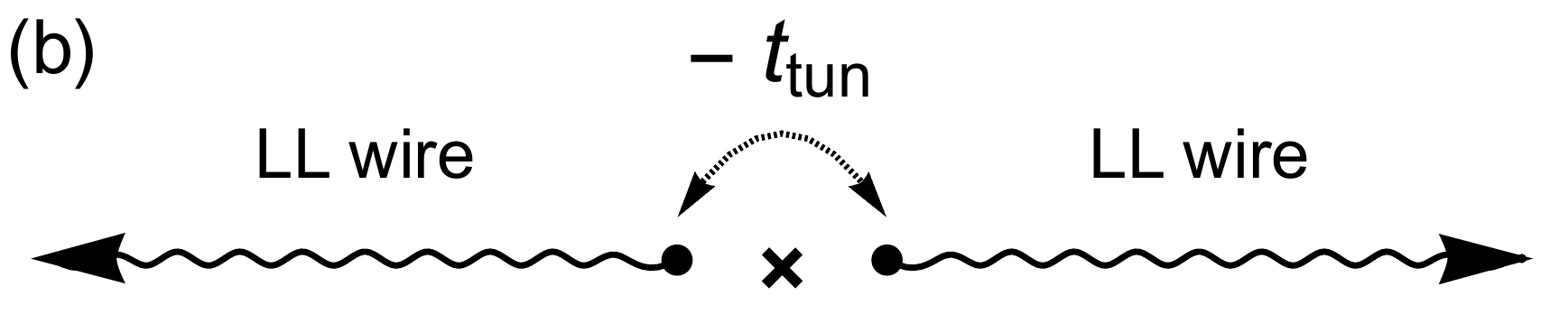} \\ ~ \\
 \includegraphics[width=\linewidth]{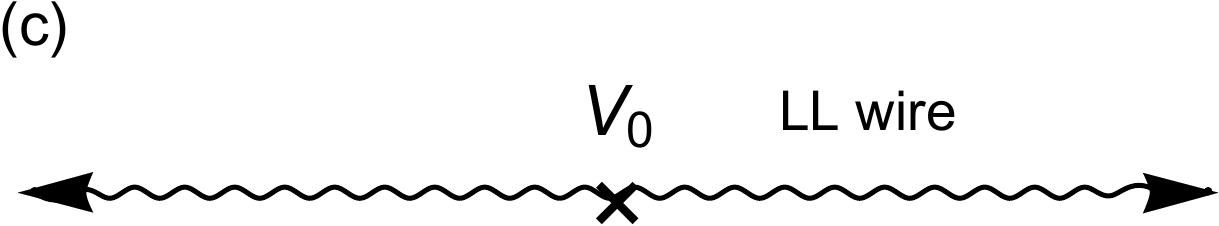}
 \caption{Illustrations of investigated impurity types: (a) a boundary barrier, (b) a bulk barrier, and (c) a weak backscattering center. 
In all the panels, we mark a single impurity as $\times$, and assume that it is located at the origin $x=0$.
In Panel (a), a strong impurity is located between a Fermi-liquid lead (FL lead, plotted as a dashed line) and a {\TLLadj} wire ({\TLLabr} wire, plotted as a solid wavy line), which are tunnel coupled with the tunnel amplitude $t_{\rm tun}$.
In Panel (b), a strong impurity breaks the {\TLLadj} wire into two segments. 
In Panel (c), a weak impurity acts as a backscattering center with the potential strength $V_{0}$.
}
 \label{Fig:illustration}
 \end{figure}

We further note works on signatures of the spin-orbit coupling in the correlation functions of the {\TLL}~\cite{Moroz:2000,Moroz:2000b,Iucci:2003}. In contrast to those, we investigate the influence of spin-orbit coupling on {\it charge transport properties}. 
Specifically, we calculate the temperature and bias-voltage dependence of the tunnel current and/or the (differential) conductance of a {\TLL} containing impurities in the following scenarios: (i) when the impurities are strong and treated as tunnel barriers located either (ia) near the boundary (wire end) or (ib) in the bulk of the wire; 
(ii) when they are weak, treated as potential disorder,
(iii) when both types (i) and (ii) are present.
These impurity types are illustrated in Fig.~\ref{Fig:illustration}.
In addition to the universal scaling behavior of the tunnel current in scenario (i), we find the differential conductance as a power law of the temperature and bias voltage in the high-temperature and high-bias limits, respectively, in the above scenarios. 
One of our main conclusions is that, for realistic strengths of the spin-orbit coupling, the current-voltage curve follows the universal scaling relation of a standard\footnote{That is, the one without any spin-orbit coupling effects.} {\TLL} with modified parameters. The interaction strength can be therefore obtained reliably from the universal scaling behavior, upon fitting the power law.
Furthermore, in the strong-interaction regime the modifications due to the spin-orbit coupling are negligible.
Nevertheless, in general the effects of spin-orbit coupling enter, and the charge transport is additionally complicated by the character of impurities in the wires. 
Our analysis incorporating various impurity types and locations resolves these complications.

The paper is organized as follows. In Sec.~\ref{Sec:Model} we introduce our model. We review the properties of a quasi-one-dimensional spin-orbit-coupled wire in Sec.~\ref{Sec:Spectrum}, and present our bosonized model, incorporating the effects of the spin-orbit coupling and the quasi-one dimensionality in Sec.~\ref{Sec:Hamiltonian}. 
In Sec.~\ref{Sec:Conductance} we consider various types of impurities. 
For the strong-impurity scenario considered in Sec.~\ref{SubSec:Strong}, we calculate the universal scaling formula for the tunnel current in the case of (a) boundary barriers and (b) bulk barriers. 
Through the RG analysis, we compute also the conductance in the high-temperature and high-bias regimes, and show consistency with the tunnel current calculation.
In Sec.~\ref{SubSec:Weak} we compute the conductance in the high-temperature and high-bias regimes for the weak-impurity scenario and propose an interpolation formula for arbitrary temperature and bias.
Finally, in Sec.~\ref{SubSec:Coexist}, we consider the scenario in which both strong and weak impurities are present, and reveal a transition between different power-laws upon varying the interaction strength. 
We discuss generalization of our calculation and robustness of the transport signatures for a {\TLL} in Sec.~\ref{Sec:Discussion}.
In Sec.~\ref{Sec:Summary}, we summarize our main results in Table~\ref{Table:summary}.
Appendix~\ref{App:Correlator} gives the details on the derivation of the single-particle correlation function.
In Appendix~\ref{App:FT} we present the derivation of the universal scaling relation of the current-voltage curve, and its asymptotic behavior in the high-temperature and high-bias limits.
In Appendix~\ref{App:GZ}, we discuss the density of states of a bosonized model proposed in Ref.~\cite{Governale:2004}. 
In Appendix~\ref{App:weak} we discuss an alternative approach for the analysis in the weak-impurity regime.

\begin{figure}[t]
 \includegraphics[width=\linewidth]{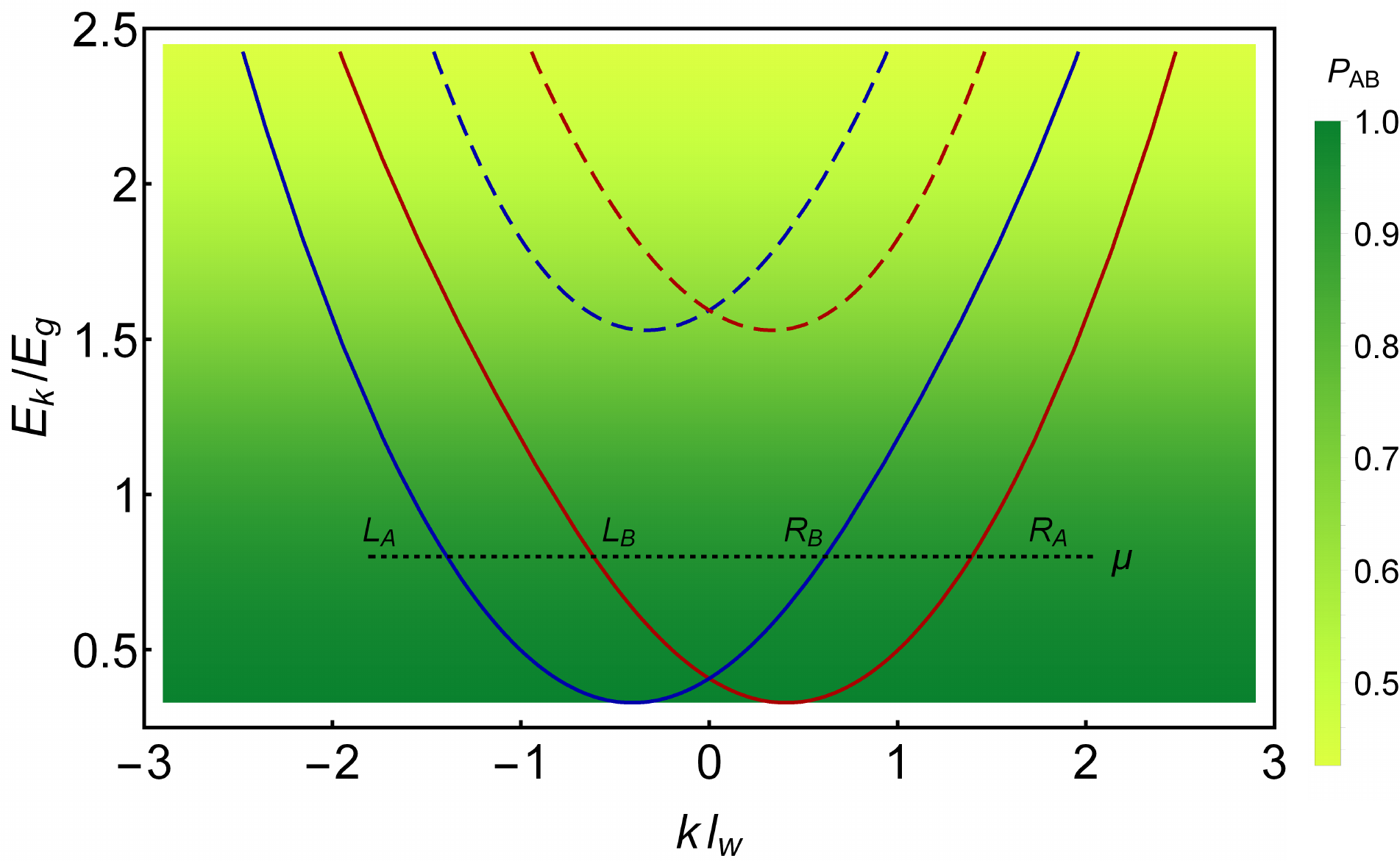}
 \caption{Energy spectrum ($E_k$) of a quasi-one-dimensional spin-orbit-coupled wire for $k_{\rm so} l_w = 0.45$. The horizontal axis labels the momentum $k$ along the wire (multiplied by the wire width $l_w$), and the vertical axis is scaled with $E_{g} \equiv \hbar^2/(m l_w^2)$. When the chemical potential $\mu$ (dotted line) intersects with the lowest transverse subband (solid curves) but not the upper subband (dashed curves), there exist four distinct branches, which we label as $L_A$, $L_B$, $R_B$, and $R_A$, according to their Fermi velocities. The color in the background shows the value of the spinor overlap $P_{AB}$ [see Eq.~\eqref{Eq:PAB}]. 
}
 \label{Fig:spectrum}
 \end{figure}

\section{Hamiltonian of a clean system~\label{Sec:Model}}

\subsection{Energy spectrum and spin orientation~\label{Sec:Spectrum}}
Owing to their potential application in spintronics and topological matter, the spectral properties of semiconductor nanowires with spin-orbit coupling have been widely studied in the literature~\cite{Moroz:1999,Moroz:2000,Moroz:2000b,Governale:2002,Governale:2004,Braunecker:2010,Meng:2014,Rainis:2014,Kainaris:2015}. Here we review the basic properties of a quasi-one-dimensional spin-orbit-coupled wire~\cite{Moroz:1999,Moroz:2000,Moroz:2000b,Governale:2002,Governale:2004} that are essential for our analysis.  
We assume that the wire lies along $x$ direction, and its transverse directions ($y$ and $z$) are subject to confinement potentials, taken as anisotropic harmonic for specificity. The $y$-axis confinement is assumed to be much softer than $z$ direction, so the $z$ transverse degrees of freedom can be neglected. The confinement energy scale $E_g$ is thus determined by the wire width $l_w$ along $y$ direction. 
It is known that in the presence of a relatively weak transverse confinement (meaning that $E_g$ is not very large), the spin-orbit coupling can cause appreciable band distortion~\cite{Moroz:1999}. 
More precisely, the Rashba spin-orbit coupling mixes the opposite spin states of the neighboring transverse subbands, making the energy spectrum spin-dependent and distorting its originally quadratic dispersion.

\begin{figure}[t]
 \includegraphics[width=\linewidth]{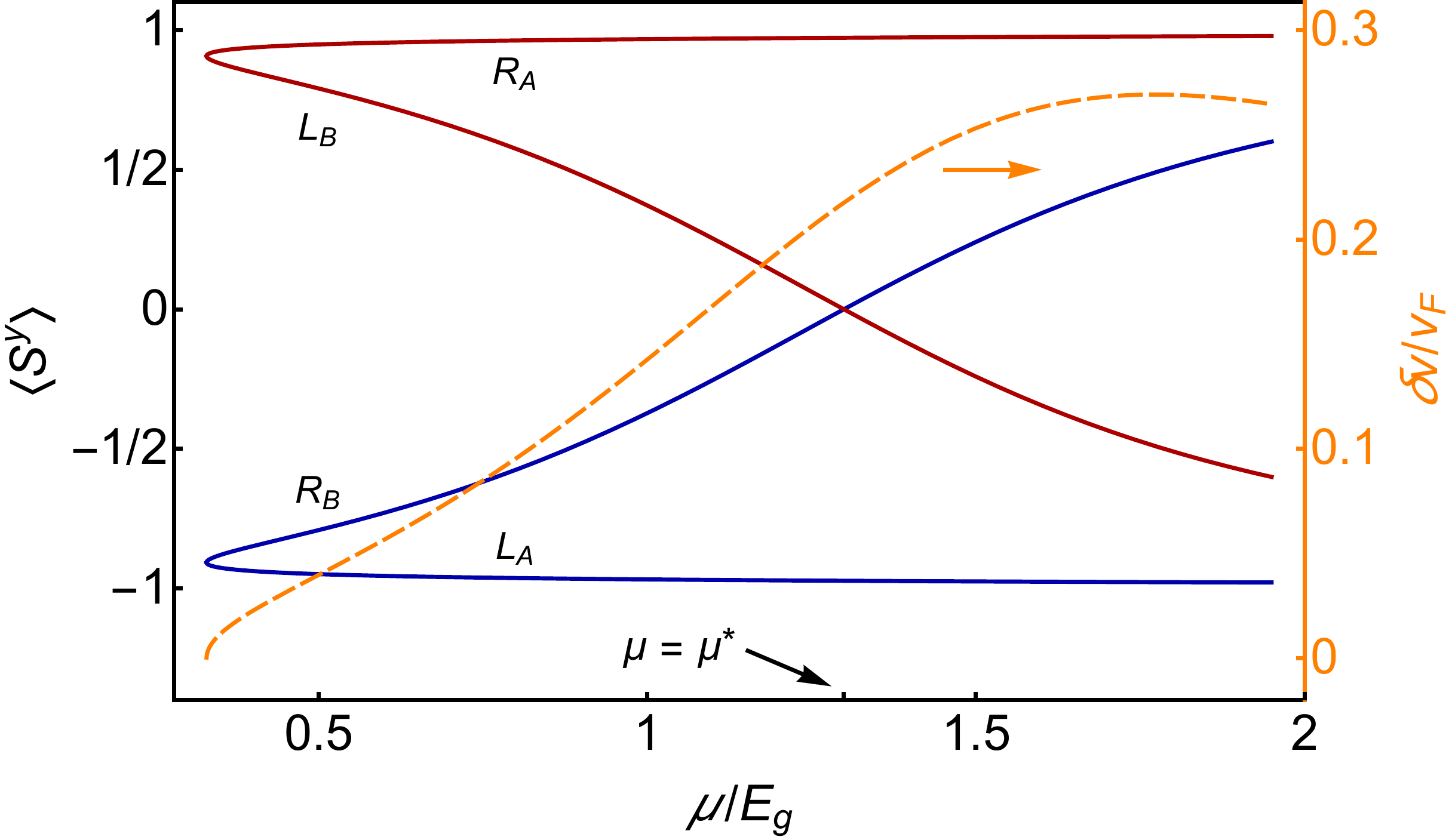}
 \caption{Chemical potential ($\mu$) dependence of the spin expectation value $\langle S^{y} \rangle$ (left axis, in unit of $\hbar/2$) of the eigenstates in the lowest subband, as well as the band distortion parameter $\delta v / v_F$ (right axis).  
For $\mu<\mu^{*}$, the spins of $R_A$ and $R_B$ are of opposite signs, so are the spins of $L_A$ and $L_B$.
The adopted parameters are the same as those used in Fig.~\ref{Fig:spectrum}. 
}
 \label{Fig:orientation}
 \end{figure}

Aiming at a quantitative description, we take the two-subband model introduced in Refs.~\cite{Governale:2002,Governale:2004}.
The energy spectrum depends on the dimensionless parameter $k_{\rm so} l_w$, with $k_{\rm so} \equiv m|\alpha_{\rm R}|/\hbar^2$ being determined by the Rashba coefficient $\alpha_{\rm R}$ and effective mass $m$ of the material. 
For the parameter values $|\alpha_{\rm R}| =$ 100--200~meV\AA, $m=0.023~m_{e}$ with the electron mass $m_{e}$, and $l_w = 100~$nm, we obtain $k_{\rm so} l_w =$0.3--0.6. 
For illustration, Fig.~\ref{Fig:spectrum} shows the spectrum for $k_{\rm so} l_w = 0.45$.
We label the outer (inner) branch of the energy dispersion as $A$ $(B)$ and the right- (left-)moving electron as $R$ ($L$) for the chemical potential $\mu$ located within the lowest-subband regime, as indicated in Fig.~\ref{Fig:spectrum}.
Due to the band distortion induced by the spin-orbit coupling, the Fermi velocities of the branches $A$ and $B$ are different.
We define the band distortion parameter as the ratio $\delta v / v_F$ with
\begin{subequations}
\label{Eq:deltaV}
\begin{eqnarray}
\delta v &\equiv&  v_{A} - v_{B} , \\
 v_F  &\equiv&   (v_{A} + v_{B})/2,
\end{eqnarray}
\end{subequations}
where $v_{A}$ ($v_{B}$) is the Fermi velocity of the branch $A$ ($B$). 
The band distortion parameter is plotted as a function of $\mu$ in Fig.~\ref{Fig:orientation}. For parameters relevant to Ref.~\cite{Sato:2019}, we obtain $\delta v / v_F \apprle 0.1$ in the strong-interaction regime (that is, close to the bottom of the lowest subband).

In addition to the band distortion, the spin orientation of the electrons is also affected by the spin-orbit coupling. 
Whereas the spins of the right- and left-moving electrons of the same branch (either $A$ or $B$) must be opposite due to the time-reversal symmetry, there is in general no relation between the spins of electrons moving in the same direction (namely, between $R_A$ and $R_B$, or, equivalently, between $L_A$ and $L_B$). In the literature, Moroz~{\it et~al.} assigned the electron spins of the same direction of motion as antiparallel~\cite{Moroz:2000,Moroz:2000b}, while Governale and Z{\"u}licke assigned them as parallel~\cite{Governale:2002,Governale:2004}. In fact, due to the spin-orbit mixing, the spins of $R_A$ and $R_B$ are in general neither exactly parallel nor exactly antiparallel. The relative orientation between these two spins depends on the strengths of the spin-orbit coupling and the transverse confinement, as well as the chemical potential.

To demonstrate it, in Fig.~\ref{Fig:orientation} we plot the $\mu$ dependence of the spin expectation values $\langle S^{y} \rangle$ with respect to the transverse wave functions of the four labeled branches.
Similar to the band distortion, the spin orientation also shows strong $\mu$ dependence. 
Upon increasing $\mu$, the spins of $R_A$ and $R_B$ evolve from nearly antiparallel close to the bottom of the lowest subband, to not very well aligned when approaching the bottom of the upper subband. 
The crossover between these two regimes occurs at $\mu^{*}$, where the expectation values $\langle R_B | S^{y} | R_B \rangle$ and $\langle L_B | S^{y} | L_B \rangle$ vanish.
As a result, the two sets of seemingly contradicting references (Refs.~\cite{Moroz:2000,Moroz:2000b} and Refs.~\cite{Governale:2002,Governale:2004}) can be reconciled--whether the former or the latter gives a better description depends on which parameter regime we are looking at. 
Therefore, to investigate the strong-interaction regime in Ref.~\cite{Sato:2019}, which corresponds to the low-$\mu$ regime in Fig.~\ref{Fig:spectrum}, it is more appropriate to consider the scenario described in Refs.~\cite{Moroz:2000,Moroz:2000b}. 

After examining the parameter dependence of the spin orientation, we discuss the implication of the distinct spin-state assignments in Refs.~\cite{Moroz:2000,Moroz:2000b,Governale:2002,Governale:2004} on our analysis. 
To proceed, we denote the spin states of the time-reversal pairs $R_{A}$ and $L_{A}$ as $\sigma = \uparrow$ and $\downarrow$, respectively. 
Similarly, the other pairs $R_{B}$ and $L_{B}$ can be labeled as $\sigma^{\prime} = \uparrow^{\prime}$ and $\downarrow^{\prime}$, respectively.
As mentioned above, there is in general no relation between $\sigma$ and $\sigma^{\prime}$. 
Concerning transport properties, backscattering on charge impurities is feasible only between counter-propagating electrons with nonzero spinor overlap.
Since the spin assignment $\sigma = \sigma^\prime$ in Refs.~\cite{Governale:2002,Governale:2004} corresponds to a {\it helical channel}, it is immune against elastic single-particle backscattering on charge impurities.
To facilitate backscattering of electrons in a helical channel, an alternative mechanism has to be involved, which may arise from broken time-reversal symmetry, higher-order scattering, or inelastic process (for example, see Refs.~\cite{Hsu:2017,Hsu:2018} and references therein). In the present work, however, since we consider charge impurities as the dominating mechanism responsible for the transport, 
 only backscattering processes between two non-orthogonal states (that is, $R_{A} \leftrightarrow L_{B}$ and $R_{B} \leftrightarrow L_{A}$) can cause finite resistance. In consequence, it is more appropriate to consider the spin-state assignment $\sigma = -\sigma^\prime$ suggested in Refs.~\cite{Moroz:2000,Moroz:2000b} for our analysis.\footnote{Nonetheless, one may be interested in the spin-orbit effect on the bosonized model of the helical channel presented in Ref.~\cite{Governale:2004}. In Appendix~\ref{App:GZ}, we analyze the power-law density of states of that model, which turns out to be weakly dependent on the band distortion.}
To further quantify the backscattering strength, we compute the scalar product of two non-orthogonal states in different branches for a fixed $\mu$,
\begin{eqnarray}
P_{AB} &\equiv& |\langle R_{A} | L_{B} \rangle| = |\langle R_{B} | L_{A} \rangle|,
\label{Eq:PAB}
\end{eqnarray}
which is shown in the background color of Fig.~\ref{Fig:spectrum}. In spite of the misaligned spins, the strength of backscattering between branches decreases only modestly when the chemical potential is increased. 

Based on the above consideration, we are motivated to introduce the following fermion operators for the four branches,
\begin{eqnarray}
L_A \to \psi_{L \downarrow}, \; L_B \to \psi_{L \uparrow}, \; 
R_B \to \psi_{R \downarrow}, \; R_A \to \psi_{R \uparrow} ,
\end{eqnarray}
where we have removed the redundant prime for branch $B$. 
With this electron spin assignment, the energy spectrum can be linearized and the fields $\psi_{r\sigma}$ can be bosonized in the standard way [see Eq.~\eqref{Eq:bosonization} below].
The strength of the impurity-induced backscattering is affected by the spin misalignment, which influences the prefactors of the conductance and current. In our calculation, it can be incorporated through renormalized coupling constants by  replacing the tunnel amplitude $t_{\rm tun}$ in Eq.~\eqref{Eq:I_tun2} with (assuming that $P_{AB}$ is well above zero)
\begin{eqnarray}
t_{\rm tun}^2 \to t_{\rm tun}^2 \Big( \frac{1+ P_{AB}^2}{2} \Big),
\label{Eq:tun}
\end{eqnarray}
and the backscattering strength $V_0$ in Eq.~\eqref{Eq:Gsingle} with
\begin{eqnarray}
V_{0}^2 \to V_{0}^2 P_{AB}^2.
\label{Eq:V0}
\end{eqnarray}
Nevertheless, the spin misalignment does not affect the universal scaling exponents that we aim to determine.

In consequence, the distinct Fermi velocities of the two spin branches are symptomatic for the spin-orbit-induced band distortion discussed here, with the ratio $\delta v / v_F $ depending on the specific model used to compute the spectrum. 
In order to keep our analysis general, from now on we take $\delta v / v_F $ as a phenomenological parameter quantifying the band distortion effect.  
In the bosonized Hamiltonian, this band distortion leads to a charge-spin mixing term~\cite{Moroz:2000,Moroz:2000b}, which we present in the next subsection.

\subsection{Bosonized Hamiltonian~\label{Sec:Hamiltonian}}
We now present our model based on the bosonization formalism~\cite{Giamarchi:2003}. 
We introduce the Hamiltonian $H=H_{\textrm{0}}+H_{\textrm{so}}$ as a model of a clean spin-orbit-coupled wire. We postpone the discussion of additional terms induced by impurities to Sec.~\ref{Sec:Conductance}. 
The first term, $H_{\textrm{0}}$, describes a standard, spinful {\TLL}\footnote{Here we assume the wire length to be much longer than any other length scale, such as Fermi wavelength, thermal length, and average impurity separation, so that the wire can be regarded as a {\TLL} extending over the entire space. Moreover, in such a long wire the $g_1$ backscattering term is renormalized to a vanishing contribution to the effective action for any repulsive interaction~\cite{Moroz:2000b}, so we neglect the $g_1$ term here.} 
\begin{align}
H_{\textrm{0}} = \sum_{\nu} \int \frac{\hbar dx}{2\pi} \left\{
u_{\nu} g_{\nu} \left[ \partial_x \theta_{\nu}(x) \right]^2 + \frac{u_{\nu}}{g_{\nu}} \left[ \partial_x \phi_{\nu}(x) \right]^2
\right\}.
\label{Eq:H_TLL} 
\end{align}
Here, $g_{\nu}$ is the interaction parameter in the $\nu$ sector (with the index $\nu \in \{c,s\}$ referring to the charge and spin sectors, respectively) and $u_{\nu} = v_F/g_{\nu}$ is the corresponding renormalized velocity with $v_F$ defined in Eq.~\eqref{Eq:deltaV}.
The boson fields $(\phi_{\nu},\theta_{\nu})$ are connected to the fermion fields through the standard bosonization formula,
\begin{eqnarray}
\psi_{r\sigma}(x) &=& \frac{1}{\sqrt{2 \pi a}} e^{i r k_F x} e^{-\frac{i}{\sqrt{2}} [ r\phi_{c}(x) - \theta_{c} (x) + r\sigma\phi_{s}(x) -\sigma \theta_{s} (x) ] }. \nonumber \\
\label{Eq:bosonization}
\end{eqnarray}
Here, the Klein factor is omitted, $a$ is the short-distance cutoff, $k_{F}$ is the Fermi wave vector,\footnote{Similar to the definition of $v_F$, the parameter $k_F$ becomes the average of the Fermi wave vectors in the two spin branches when the spin-orbit coupling is included.}
 and the index $r\in \{ R \equiv +1,~ L \equiv -1 \}$ refers to the fermion operator describing the right- and left-moving particle, respectively. 
The boson fields satisfy the following commutation relation~\cite{Giamarchi:2003},
\begin{eqnarray}
\left[ \phi_{\nu}(x), \theta_{\nu'}(x')\right] &=& i \frac{\pi}{2} \textrm{sign}(x'-x) \delta_{\nu\nu'}. 
\label{Eq:commutator}
\end{eqnarray}
Equation~\eqref{Eq:H_TLL} itself describes a system in which the spin-orbit-induced band distortion is absent, and reveals the separation of the charge and spin sectors, the hallmark of the standard {\TLL}. 
Throughout this article, we will constantly compare the known formula derived from Eq.~\eqref{Eq:H_TLL} with our results including the influence of the band distortion. 

The term $H_{\textrm{so}}$ incorporates the band distortion induced by the spin-orbit coupling~\cite{Moroz:1999,Moroz:2000,Moroz:2000b} 
\begin{align}
H_{\textrm{so}} =&
 \delta v \int \frac{\hbar dx}{2\pi} \Big\{ \left[ \partial_x \phi_{c}(x)\right] \left[\partial_x \theta_{s}(x) \right] 
 \nonumber \\
& \hspace{42pt} 
 + \left[ \partial_x \phi_{s}(x) \right] \left[\partial_x \theta_{c}(x) \right] \Big\},
\label{Eq:H_SOI}
\end{align}
as a mixing between the charge and spin sectors. 
As demonstrated in Fig.~\ref{Fig:orientation}, the typical values of $\delta v$ are small compared to the averaged Fermi velocity $v_{F}$. 

Since the Hamiltonian $H_{\textrm{0}}+H_{\textrm{so}}$ is quadratic in the boson fields, it can be diagonalized by using new boson fields,
\begin{subequations}
\label{Eq:NewFields}
\begin{eqnarray}
\left( \begin{array}{c} \phi_s^{\prime} (x) \\ \theta_c^{\prime} (x) \end{array} \right) &=& 
\left( \begin{array}{cc} \cos \theta & -g_0 \sin \theta \\ \frac{1}{g_0}\sin \theta & \cos \theta  \end{array} \right) 
\left( \begin{array}{c} \phi_s (x) \\ \theta_c (x) \end{array} \right), \\
\left( \begin{array}{c} \phi_c^{\prime} (x) \\ \theta_s^{\prime} (x) \end{array} \right) &=& 
\left( \begin{array}{cc} \cos \theta & g_0 \sin \theta \\ -\frac{1}{g_0}\sin \theta & \cos \theta  \end{array} \right) 
\left( \begin{array}{c} \phi_c (x) \\ \theta_s (x) \end{array} \right),
\end{eqnarray}
\end{subequations}
with the parameters
\begin{subequations}
\label{Eq:parameters}
\begin{eqnarray}
g_0 &=& \frac{\sqrt{2} g_c g_s }{ \sqrt{g_c^2 + g_s ^2}}, \\
\theta &=& \frac{1}{2} \arctan \left( \frac{\delta v }{v_F}  \frac{ \sqrt{2} g_c g_s \sqrt{g_c^2 + g_s^2} }{ g_s^2 - g_c^2} \right).
\end{eqnarray}
\end{subequations}
It can be checked that these new boson fields satisfy the commutation relation \eqref{Eq:commutator} upon replacing the fields $(\phi_{\nu},\theta_{\nu}) \rightarrow (\phi_{\nu}^{\prime},\theta_{\nu}^{\prime})$.
In terms of the new fields, the Hamiltonian reads
\begin{eqnarray}
H &=& \sum_{\nu} \int \frac{\hbar dx}{2\pi} \left\{
u_{\nu}^{\prime} g_{\nu}^{\prime} \left[ \partial_x \theta_{\nu}^{\prime}(x) \right]^2 + \frac{u_{\nu}^{\prime}}{g_{\nu}^{\prime}} \left[ \partial_x \phi_{\nu}^{\prime}(x) \right]^2
\right\} , \nonumber \\
\label{Eq:diag}
\end{eqnarray}
where the modified interaction parameters and velocities by the spin-orbit coupling are given by 
\begin{subequations}
\label{Eq:g_nu_prime}
\begin{eqnarray}
g_{c}^{\prime} &=& \frac{g_c g_0}{g_s} \left[ \frac{g_s^2 - (g_0^2 +g_s^2) \sin^2 \theta } {g_0^2 - (g_0^2 +g_c^2) \sin^2 \theta}\right]^{1/2}, \\
g_{s}^{\prime} &=& \frac{g_s g_0}{g_c} \left[ \frac{g_c^2 - (g_0^2 +g_c^2) \sin^2 \theta } {g_0^2 - (g_0^2 +g_s^2) \sin^2 \theta}\right]^{1/2},
\end{eqnarray}
and
\begin{eqnarray}
u_{c}^{\prime} &=& \frac{u_c}{g_0 g_s \cos (2\theta)} \left[g_0^2 - (g_0^2 +g_c^2) \sin^2 \theta \right]^{1/2} \nonumber\\
&& \hspace{48pt} \times \left[ g_s^2 - (g_0^2 +g_s^2) \sin^2\theta \right]^{1/2},\\
u_{s}^{\prime} &=& \frac{u_s}{g_0 g_c \cos (2\theta)} \left[g_0^2 - (g_0^2 +g_s^2) \sin^2 \theta \right]^{1/2} \nonumber\\
&& \hspace{48pt} \times  \left[ g_c^2 - (g_0^2 +g_c^2) \sin^2\theta \right]^{1/2},
\end{eqnarray}
\end{subequations}
respectively.
In the absence of spin-orbit coupling (that is, when $\delta v,~ \theta \rightarrow 0$), we recover the limit $(g_{c}^{\prime}, g_{s}^{\prime}, u_{c}^{\prime}, u_{s}^{\prime}) \rightarrow (g_{c}, g_{s}, u_{c}, u_{s})$, as expected. 
The above formulas quantify the effects of the band distortion, Eq.~\eqref{Eq:H_SOI}, on the interaction parameters and velocities. 
For fixed $\delta v$, the parameters $\theta$ and $g_{0}$ decrease with a decreasing $g_c$. 
Therefore, $g_{\nu}^\prime$ approaches its value at zero spin-orbit coupling ($g_{\nu}$) when $g_c$ approaches zero. 
As a result, the modification of the interaction parameters by the band distortion is smaller for more strongly interacting wires.

For a moderate strength of the interaction, since the modified interaction parameters enter the exponents of the correlation functions, we expect to see influence on observable quantities. In particular, we are interested in the charge transport of a spin-orbit-coupled wire for arbitrary strength of interaction. 
Since the ballistic conductance of a clean spin-orbit-coupled system displays no signature for a {\TLL}~\cite{Moroz:2000b}, as originally found for systems without spin-orbit coupling~\cite{Maslov:1995,Ponomarenko:1995,Safi:1995}, in the following we seek for signatures in the presence of impurities.

\section{Transport properties in the presence of various impurities~\label{Sec:Conductance}}
We aim at finding out how the charge-spin mixing term in Eq.~\eqref{Eq:H_SOI} influences the transport properties of the system. To this end, we compute the current and/or the differential conductance of the system described by Eq.~\eqref{Eq:diag} in the presence of various types of impurities illustrated in Fig.~\ref{Fig:illustration}.
We consider a wire adiabatically connected to the leads, a common assumption as in, e.g., Refs.~\cite{Ponomarenko:1995,Maslov:1995b,Sandler:1997,Egger:1998b,Aseev:2018}; for the effect of an abrupt contact in a microscopic model, see Refs.~\cite{Meden:2003,Janzen:2006}.

\subsection{Strong impurities--tunnel barriers~\label{SubSec:Strong}}
We begin with an isolated strong impurity and model it as a weak tunnel barrier. Assuming that the barrier is located at the origin, the two sides of the barrier are described by
\begin{subequations}
\label{Eq:H_12}
\begin{align}
H_{1} = &\sum_{\nu} \int_{-\infty}^{0}   \frac{\hbar u_{1\nu}^{\prime} dx}{2\pi}  \; \left\{
g_{1\nu}^{\prime} \left[ \partial_x \theta_{1\nu}^{\prime}(x) \right]^2 + \frac{ \left[ \partial_x \phi_{1\nu}^{\prime}(x) \right]^2
 }{g_{1\nu}^{\prime}} \right\} , \\
H_{2} = &\sum_{\nu} \int_{0}^{\infty}   \frac{\hbar u_{2\nu}^{\prime} dx}{2\pi}  \; \left\{
g_{2\nu}^{\prime} \left[ \partial_x \theta_{2\nu}^{\prime}(x) \right]^2 + \frac{ \left[ \partial_x \phi_{2\nu}^{\prime}(x) \right]^2
 }{g_{2\nu}^{\prime}} \right\} , 
\end{align}
\end{subequations}
obtained by generalizing Eq.~\eqref{Eq:diag} to possibly different parameters on the two sides. 
We include an additional index $j \in \{1,2\}$ to label the semi-infinite subsystem on the left and right side of the barrier, respectively. 
The two subsystems are connected through a tunneling process described by
\begin{align}
H_{\rm tun} = -t_{\rm tun} \sum_{\sigma} c_{1\sigma}^\dagger c_{2\sigma} + {\rm H.c.},
\label{Eq:H_tun}
\end{align}
where $t_{\rm tun}$ is the tunnel amplitude and $c_{j\sigma}$ is the fermion operator with spin $\sigma \in \{\uparrow,\downarrow\}$ at $x=0$ in the left ($j=1$) or right ($j=2$) side. It is related to the right- and left-movers in Eq.~\eqref{Eq:bosonization} by 
\begin{align}
c_{j\sigma} = \psi_{jR\sigma}(0)+\psi_{jL\sigma}(0),
\end{align}
where we generalize the field in Eq.~\eqref{Eq:bosonization} to $\psi_{jr\sigma}$ by including the subsystem index $j$.

The tunnel current through the barrier depends on whether the barrier is located near the boundary\footnote{We stress that the ``boundary (end) barrier'' may be located close to, but not necessarily precisely at, the wire end, as discussed in Refs.~\cite{Balents:1999,Sato:2019}.}
 or in the bulk of the wire.
When it is near the boundary, the barrier corresponds to a junction between a {\TLL} wire and a Fermi-liquid lead. In this case, one side of the barrier (say, for $x<0$) is described by the parameters of a lead with negligible interaction\footnote{For $(g_{1c},~ g_{1s}) = (1,~ 1)$, we get $(g_{1c}^{\prime},~ g_{1s}^{\prime}) = (1,~ 1)$ for arbitrary value of spin-orbit coupling. Our results therefore include the possibility that the strength of spin-orbit coupling is different in the wire and in the Fermi-liquid lead.} 
such that $(g_{1c}^{\prime},~ g_{1s}^{\prime}) = (1,~ 1)$, whereas on the other side (for $x>0$) the wire parameters are $(g_{2c}^{\prime},~ g_{2s}^{\prime}) = (g_{c}^{\prime},~ g_{s}^{\prime})$.
When the barrier is in the bulk of the wire, on the other hand, the barrier corresponds to a junction between two {\TLL}s. Therefore, both $H_1$ and $H_2$ have parameters $(g_{1c}^{\prime},~ g_{1s}^{\prime})= (g_{2c}^{\prime},~ g_{2s}^{\prime}) =(g_{c}^{\prime},~ g_{s}^{\prime})$.
In the following, we first keep general parameters for the two subsystems $j$, and specify them later. 

To the leading order, the tunnel current through the barrier is given by~\cite{Schrieffer:1963,Wen:1991}
\begin{widetext}
\begin{align}
I = \frac{e t_{\rm tun}^2}{\hbar^2} \sum_{\sigma} \int_{0}^{\infty} dt \; \Big\{ e^{-ieVt/\hbar} 
\Big\langle \Big[ c_{1\sigma}^\dagger(t) c_{2\sigma} (t), c_{2\sigma}^{\dagger}(0) c_{1\sigma}(0) \Big] \Big\rangle
- e^{ieVt/\hbar} \Big\langle \Big[ c_{2\sigma}^{\dagger}(t) c_{1\sigma}(t), c_{1\sigma}^\dagger(0) c_{2\sigma}(0) \Big] \Big\rangle \Big\},
\label{Eq:I_tun}
\end{align}
\end{widetext}
with the elementary charge $e$, the reduced Planck constant $\hbar$, and the voltage difference $V$ between the two sides of the barrier. 
Here, the notation $[\cdots,\cdots]$ is the commutator and $\langle \cdots \rangle$ is the average with respect to the unperturbed action [before introducing Eq.~\eqref{Eq:H_tun}].
It is convenient to write Eq.~\eqref{Eq:I_tun} as
\begin{align}
I =& -\frac{2e t_{\rm tun}^2}{\hbar^2} \sum_{\sigma} {\rm Im} \Big[ \chi_{\sigma}^{\rm ret} (- eV/\hbar )\Big],
\label{Eq:I_tun2}
\end{align}
with ${\rm Im} [\cdots]$ being the imaginary part and the following correlation functions,
\begin{subequations}
\label{Eq:chi_all} 
\begin{eqnarray}
\chi_{\sigma}^{\rm ret} (\omega ) &\equiv& -i \int_{0}^{\infty} dt \; e^{i \omega t} \Big[ \chi_{\sigma}(t) - \chi_{\sigma}(-t)\Big], 
\label{Eq:chi_ret} \\
\chi_{\sigma}(t) &\equiv& \Big\langle c_{1\sigma}^\dagger (t) c_{1\sigma} (0) \Big\rangle_{1} \Big\langle c_{2\sigma} (t) c_{2\sigma}^\dagger (0) \Big\rangle_{2}, 
\label{Eq:chi} 
\end{eqnarray}
\end{subequations}
where $\langle \cdots \rangle_{j}$ is the average corresponding to $H_{j}$ in Eq.~\eqref{Eq:H_12}. 

The single-particle equal-space correlation function in the above formula is defined at the boundary of the {\TLL}. To take the boundary into account properly, we treated it along the lines of Ref.~\cite{Giamarchi:2003}, as presented in Appendix~\ref{App:Correlator}. At finite temperature $T$, the single-particle correlation function is
\begin{eqnarray}
\Big\langle c_{j\sigma}^\dagger (t) c_{j\sigma} (0) \Big\rangle_{j} &=&  \frac{1}{2\pi a} \sum_{r=\pm} \Big[ \frac{\pi k_B T/\Delta_{a}}{ i \sinh(\pi k_B T t/\hbar)} \Big]^{\dosendj+1 }, \nonumber \\
\end{eqnarray}
with the Boltzmann constant $k_{B}$ and the bandwidth $\Delta_{a} \equiv \hbar v_F /a$. The exponent $\dosendj$, given in Eq.~\eqref{Eq:alpha_DOS_end_j}, corresponds to the density of states at the boundary of the subsystem $j$. 
Plugging the above formula into Eq.~\eqref{Eq:chi_all}, with some algebra and approximations presented in Appendix~\ref{App:FT}, we get the current-voltage curve at finite temperature as
\begin{eqnarray}
I &\propto& T^{\alpha+1} \sinh\left( \frac{ eV}{2k_B T} \right) \left| \Gamma \left( 1 + \frac{ \alpha}{2} + i \frac{eV}{2 \pi k_B T} \right)\right|^2, \nonumber\\
\label{Eq:Balents} 
\end{eqnarray}
with the gamma function $\Gamma(x)$. The current-voltage relation has the same form as in Ref.~\cite{Balents:1999} except for a parameter modified by the spin-orbit coupling,
\begin{align}
\alpha = \aendone + \aendtwo.
\label{Eq:alpha_general}
\end{align}
The explicit form of $\aendj$ depends on the density of states on the two sides of the barrier and thus the location of the barrier, which will be specified later. 

So far we have considered a wire with a single barrier which causes a voltage drop $V$. 
As discussed in Ref.~\cite{Bockrath:1999}, assuming that there are $N_{\rm b}$ independent barriers\footnote{See, however, Ref.~\cite{Jakobs:2007} for calculations beyond this assumption.} in the wire, each of which causes a similar voltage drop with $V$ being the bias voltage across the entire wire, Eq.~\eqref{Eq:Balents} is valid upon replacing $V \rightarrow V/N_{\rm b}$. Here, the barrier number $N_{\rm b}$ corresponds to the parameter $\gamma = 1/N_{\rm b}$ in Refs.~\cite{Bockrath:1999,Sato:2019}.
As a result, it is possible to experimentally determine the number of the barriers in the wire through the current-voltage characteristics.

Before specifying the barrier type to obtain the exponent in Eq.~\eqref{Eq:alpha_general} in terms of the parameters in Eqs.~\eqref{Eq:parameters} and \eqref{Eq:g_nu_prime}, we make four remarks on 
Eqs.~\eqref{Eq:Balents}--\eqref{Eq:alpha_general}.
First, the current-voltage curve exhibits a universal scaling behavior, analogous to the zero spin-orbit coupling case~\cite{Balents:1999}.
Namely, the rescaled current, $I/T^{\alpha+1}$, is a function of the ratio $V/T$. Therefore, the curves $I/T^{\alpha+1}$ plotted versus $V/T$ for various bias voltages and temperatures collapse onto a single curve. 

Second, in addition to the full dependence on the temperature and bias voltage, the asymptotic behavior of the (differential) conductance\footnote{In the presence of the strong impurities (tunnel barriers), the resistance contribution from the barriers dominates over the contact resistance between the lead and the wire. We therefore neglect the latter and evaluate the wire conductance as the derivative of Eq.~\eqref{Eq:Balents}.
}
 $G \equiv dI/dV$ may be of interest. In the $eV \ll k_BT$ regime, Eq.~\eqref{Eq:Balents} gives the linear-response conductance with a power-law temperature dependence, whereas in the opposite limit
we find a nonlinear current-voltage curve; the detailed discussions on the asymptotic behavior of Eq.~\eqref{Eq:Balents} are presented in Appendix~\ref{App:FT}. The behavior in these limits can be summarized as
\begin{eqnarray}
\label{Eq:dI/dV}
G \equiv \frac{d I} {d V} &\propto&  \left\{
\begin{array}{l} 
T^{\alpha}, \;\; {\rm for} \;\; eV\ll k_BT, \\
V^{\alpha}, \;\; {\rm for} \;\; eV\gg k_BT,
 \end{array} \right. 
\end{eqnarray}
that is, a power-law conductance with the identical exponent $\alpha$ in the high-temperature and high-bias regimes.

Third, the value of $\alpha$, which parametrizes the universal scaling relation Eq.~\eqref{Eq:Balents}, depends on the interaction parameters, so that the current-voltage curve can be used to extract the strength of the electron-electron interaction in the wire. Such characterization, however, strongly depends on the location of the barrier, as discussed below. 

Finally, it may be tempting to guess the parameter $\alpha$ in Eq.~\eqref{Eq:alpha_general} using the corresponding form in the absence of the spin-orbit coupling. Namely, one may naively replace $(g_c,g_s) \rightarrow (g_{c}^{\prime} , g_{s}^{\prime})$ in the following expressions~\cite{Kane:1992,Furusaki:1993,Matveev:1993} 
\begin{subequations}
\label{Eq:alpha_zero-so}
\begin{eqnarray}
\aend (\delta v = 0) &=&  \frac{1}{2g_{c}} + \frac{1}{2g_{s}} -1, \\
\abulk (\delta v = 0) &=&  \frac{1}{g_{c}} + \frac{1}{g_{s}} -2,
 \end{eqnarray}
\end{subequations}
for boundary and bulk barriers, respectively.
Such a replacement would, however, have given an incorrect result. The reason can be traced back to the fact that the tunnel Hamiltonian is written with the original fermions. When expressing these fermions in terms of the new fields $\phi_{\nu}^\prime$ and $\theta_{\nu}^\prime$, additional coefficients arise from the transformation from the fields $\phi_{\nu}$ and $\theta_{\nu}$ into the new ones using Eq.~\eqref{Eq:NewFields}. As a result, $\alpha$ as a function of $g_c^\prime$ and $g_s^\prime$ has a different functional than Eqs.~\eqref{Eq:alpha_zero-so}, as we now demonstrate.

Let us consider the boundary barrier, so that there is a noninteracting lead to the left ($j=1$) and a spin-orbit-coupled {\TLL} to the right ($j=2$) of the barrier, as illustrated in Fig.~\ref{Fig:illustration}(a). 
With the exponent of the single-particle correlation function $\aendj$ derived in Appendix~\ref{App:FT}, we obtain the current-voltage curve Eq.~\eqref{Eq:Balents}, with $\alpha$ given by
\begin{eqnarray}
\aend &=& \frac{1}{2}\left( \frac{1}{g_{c}^{\prime}} + \frac{1}{g_{s}^{\prime}} \right) \left( \cos^2 \theta + g_0^2 \sin^2 \theta \right) -1.
\label{Eq:alpha_boundary}
\end{eqnarray}
It goes over to Eq.~\eqref{Eq:alpha_zero-so} in the limit of $\delta v$, $\theta \rightarrow 0$. 

We now turn to the case in which the tunnel barrier is located in the bulk of the wire. The tunneling process corresponds to a particle transiting from the boundary of a {\TLL} into the boundary of the other, as illustrated in Fig.~\ref{Fig:illustration}(b). It gives rise to a current-voltage curve of the form of Eq.~\eqref{Eq:Balents} again, but with a different exponent,\footnote{Not to be confused with the exponent corresponding to the wire bulk density of states, relevant for tunneling into the bulk of the {\TLL}, considered in some references, such as Ref.~\cite{Bockrath:1999}.} namely 
\begin{eqnarray}
\abulk &=& \left( \frac{1}{g_{c}^{\prime}} + \frac{1}{g_{s}^{\prime}} \right) \left( \cos^2 \theta + g_0^2 \sin^2 \theta \right) -2.
\label{Eq:alpha_bulk}
\end{eqnarray}
Again, Eq.~\eqref{Eq:alpha_zero-so} follows for $\delta v$, $\theta \rightarrow 0$. Comparing to Eq.~\eqref{Eq:alpha_boundary}, we see that $\abulk$ is twice of $\aend$.
The exponent can be expanded in series of $\sin \theta$. In the leading order, the spin-orbit-induced change is
\begin{align}
& \abulk -\abulk (\delta v = 0) \approx - \frac{(g_c + g_s) \sin^2 \theta }{4 g_c^2 g_s^2 (g_c^2 + g_s^2) }\nonumber\\
& \hspace{0.5in} \times  \Big[  (g_c^2 - g_s^2)^2 + 4 g_c g_s \big(g_c^2 + g_s^2 - 2 g_c^2 g_s^2 \big) \Big]. 
\label{Eq:alpha_bulk_approx0}
\end{align}
For $g_c$ not close to $g_s$, we can further express it in terms of the intrinsic interaction and band distortion parameters. We get, up to second order in $\delta v/v_F$, 
\begin{eqnarray}
&& \abulk - \abulk (\delta v = 0) \nonumber\\
&\approx& - \frac{ \delta v^2 (g_c + g_s)}{8 v_F^2} \Big[ 1 + \frac{4g_c g_s}{ (g_c^2 - g_s^2)^2 } \big(g_c^2 + g_s^2 - 2 g_c^2 g_s^2 \big) \Big]. \nonumber\\
\label{Eq:alpha_bulk_approx}
\end{eqnarray}
For $g_c$ close to $g_s$ (including the noninteracting limit), on the other hand, the coefficient of the above expression diverges, so the approximation is inaccurate close to this limit. Nonetheless, Eq.~\eqref{Eq:alpha_bulk_approx} is a good approximation when the electron-electron interaction is sufficiently strong, say for $g_c \apprle 0.7$ and $g_s = 1$.

\begin{figure}[t]
 \includegraphics[width=\linewidth]{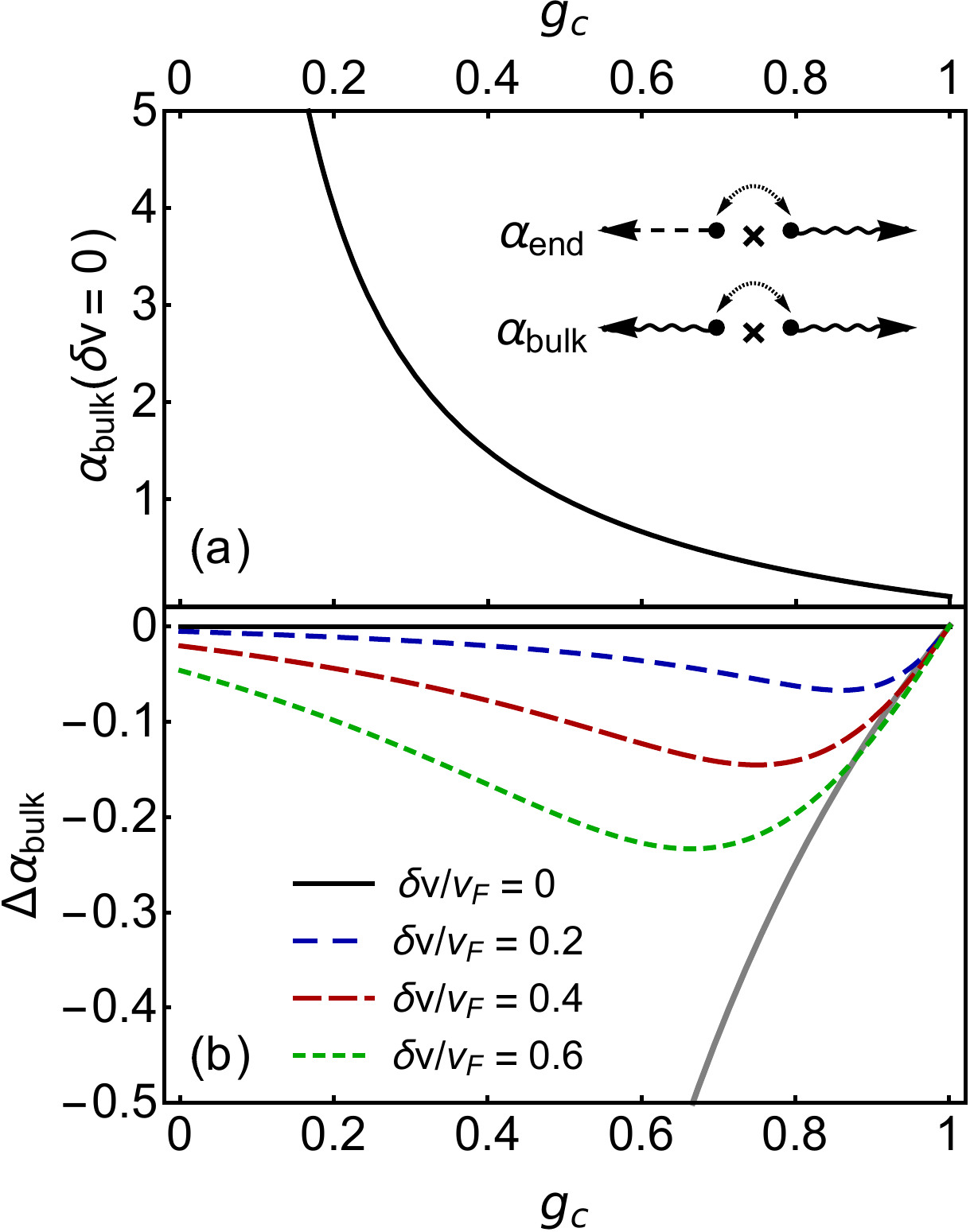}
\caption{(a) Interaction parameter ($g_c$) dependence of the bulk parameter ($\abulk$) for $g_{s}=1$ and $\delta v=0$.
The parameter $\abulk$ or $\aend$ characterizes the current-voltage curve Eq.~\eqref{Eq:Balents} and the power-law conductance given in Eq.~\eqref{Eq:dI/dV} for a bulk or end barrier, where the subscripts specify the barrier types illustrated in the inset, corresponding to Panels (a) and (b) of Fig.~\ref{Fig:illustration}. 
The associated parameter $\aend$ ($\abulk$) is given in Eq.~\eqref{Eq:alpha_boundary} [Eq.~\eqref{Eq:alpha_bulk}]. 
These two are related to each other by $\abulk = 2\aend$.
(b) Interaction parameter ($g_c$) dependence of the change ($\Delta \abulk$) due to the spin-orbit-induced band distortion $\delta v/v_{F}$ for $g_{s}=1$. 
The inverse of $\abulk$ from Panel (a) is plotted in gray. 
}
 \label{Fig:alpha_barrier}
 \end{figure}

To demonstrate how Eqs.~\eqref{Eq:alpha_boundary} and \eqref{Eq:alpha_bulk} depend on the band distortion, we plot in Fig.~\ref{Fig:alpha_barrier} the $g_c$ dependence of the $\delta v$-induced change $\Delta \abulk \equiv \abulk - \abulk (\delta v = 0)$ for several values of $\delta v/v_{F}$, as well as the zero-spin-orbit value of the exponent $\abulk$.\footnote{The discussion in this paragraph is valid for $\aend$ too, since $\abulk=2\aend$ for any value of $\delta v$ including zero.}
As shown in Fig.~\ref{Fig:alpha_barrier}(a), the parameter $\abulk (\delta v = 0)$ [see Eq.~\eqref{Eq:alpha_zero-so}] increases with a decreasing $g_c$. In other words, the suppression of the power-law conductance at low energies is stronger for systems with stronger interaction, a feature of the standard {\TLL}~\cite{Kane:1992,Furusaki:1993}.
For a nonzero $\delta v$, there is a non-monotonic change in $\abulk$, as shown in Fig.~\ref{Fig:alpha_barrier}(b).
For parameters of Ref.~\cite{Sato:2019}, the small value of $\delta v / v_F \apprle 0.1$ leads to negligible changes $\Delta \abulk$.
We include also larger values of $\delta v / v_F$, which are relevant for nanowires with weaker transverse confinement or stronger spin-orbit coupling. 
However, even with an exaggerated value of $\delta v / v_F=0.6$, one can see that the change $\Delta \abulk$ for small $g_c$ is small compared to its zero-spin-orbit value [the opposite value of $\abulk (\delta v = 0)$ is plotted in Fig.~\ref{Fig:alpha_barrier}(b) for comparison]. It means that {\it in the strong-interaction regime even very strong band distortion leads to negligible effects on the current-voltage curves.} 
On the other hand, in the moderate-interaction regime where the band distortion does modify the parameters, $\abulk$ decreases (rather than increases) upon increasing the degree of the band distortion. 
These features were the main argument for our conclusion in Ref.~\cite{Sato:2019} that the large extracted $\alpha$ values from the experimental data indeed reflect the strong electron-electron interaction in the system, instead of arising from strong spin-orbit coupling of InAs nanowires.

Here, we additionally point out that, in general, the change $\Delta \alpha$ of the parameter can be sizable compared to its zero-spin-orbit value $\alpha (\delta v = 0)$. 
As an example, for $\delta v / v_F=0.1$ and $g_c =0.9$ we obtain $\Delta \alpha / \alpha (\delta v = 0) \approx -27~\%$, which could be observable. 
As displayed in Fig.~\ref{Fig:alpha_barrier}, for $g_c$ even closer to unity, the magnitude of the correction can be comparable with the zero-spin-orbit value, resulting in a vanishing exponent. 
This feature implies that, for weakly interacting systems, the spin-orbit coupling can quench the transport signature for a {\TLL} subject to tunnel barriers.
Overall, we expect that the band-distortion effects found here become most significant in the moderate-interaction regime or even weak-(but finite-)interaction regime.

We now employ an alternative, renormalization-group (RG), approach~\cite{Giamarchi:2003,Kane:1992,Furusaki:1993} to compute the conductance.
To this end, we derive the RG flow equation for the tunnel amplitude, which is related to the scaling dimension of the equal-space correlation function at the origin. 
Following a similar procedure as presented in Appendices~\ref{App:Correlator} and \ref{App:FT}, we get
\begin{eqnarray}
\frac{d \tilde{t} (\ell)}{d \ell} &=& -\frac{1}{2} \Big( \aendone  + \aendtwo \Big)  \tilde{t} (\ell),
\label{Eq:RG_t}
\end{eqnarray}
with the dimensionless tunnel amplitude $\tilde{t} (\ell) \equiv t_{\rm tun} (\ell) / \Delta_a (\ell)$ and dimensionless length scale defined through $a(\ell) = a(0) e^{\ell}$.
Since for repulsive interaction the parameter $\aendone + \aendtwo$ is positive, the tunnel amplitude flows to zero, implying an insulating phase at low energies. 
However, to get the relevant solution, the RG flow equations should be stopped at a scale $\ell^*$ associated with the shorter of $\ln [\Delta_a/(k_B T)]$ and $\ln [\Delta_a/(eV)]$. The conductance through the tunnel barrier is obtained by integrating the RG flow up to the scale $\ell^*$, leading to
\begin{eqnarray}
G &\propto& \frac{2e^2}{h} \left[ \tilde{t} (\ell^*) \right]^2
\propto \big[{\rm Max}  (eV, k_B T) \big]^{\aendone + \aendtwo},
\end{eqnarray}
which holds for both types of barriers. 

We now specify the barrier type. For a boundary barrier, the conductance is given by
\begin{eqnarray}
\Gend (T,V) &\propto& \left\{
\begin{array}{l} 
T^{\aend}, \;\; {\rm for} \;\; eV\ll k_BT, \\
V^{\aend}, \;\; {\rm for} \;\; eV\gg k_BT,
 \end{array} \right. 
\label{Eq:G_boundary} 
\end{eqnarray}
which is consistent with both the high-temperature and the high-bias behavior of Eq.~\eqref{Eq:Balents}. 
Similarly, for a tunnel barrier in the bulk, we get
\begin{eqnarray}
\Gbulk (T,V) &\propto& \left\{
\begin{array}{l} 
T^{\abulk}, \;\; {\rm for} \;\; eV\ll k_BT, \\
V^{\abulk}, \;\; {\rm for} \;\; eV\gg k_BT,
 \end{array} \right. 
\label{Eq:G_bulk} 
\end{eqnarray}
again consistent with the tunnel current. 
The result that the RG approach gives the same power-law conductance as the tunnel current approach should not be surprising: both of them are essentially calculating the density of states at both sides of a barrier. 
Compared to the tunnel current, the RG approach gives only the asymptotic behavior of the conductance (in certain limits and without prefactors). On the other hand, it can be used to compute the conductance in the case of weak impurities, where the other method is not feasible.

\subsection{Weak impurities--potential disorder~\label{SubSec:Weak}}
We now consider the transport properties in the presence of weak impurities, each of which acts as a backscattering center. Accordingly, we retain Eq.~\eqref{Eq:diag} for the entire wire. Let us consider first one of such impurities at the origin, as illustrated in Fig.~\ref{Fig:illustration}(c). We model it as generating a delta-like potential 
\begin{eqnarray}
V_{\rm imp}(x) &=&  V_{0} \delta(x),
\label{Eq:Vimp}
\end{eqnarray}
with the strength $V_0$ and the Dirac delta function $\delta(x)$. 
By coupling to the charge density, it leads to the following term,
\begin{eqnarray}
H_{\rm imp} &=&  \sum_{r r' \sigma}   \int dx \; V_{\rm imp}(x) \left[  \psi_{r\sigma}^\dagger (x) \psi_{r'\sigma} (x)  \right] \nonumber \\
&\approx&   \frac{2V_{0}}{\pi a} \cos [\sqrt{2} \phi_{c}(0)] \cos [\sqrt{2} \phi_{s}(0)],
\end{eqnarray}
where in the second line we keep only the backscattering term, as the forward scattering does not affect the conductance. 
The above term contributes to the second-order terms of the effective action in $V_0$~\cite{Kane:1992,Furusaki:1993,Giamarchi:2003}, from which we derive the RG flow equation for the backscattering strength,
\begin{eqnarray}
\frac{d \tilde{V}_0 (\ell)}{d \ell} &=& \frac{ \aimp }{2}  \tilde{V}_0 (\ell).
\label{Eq:RG_V0} 
\end{eqnarray}
In the above, we introduce the dimensionless coupling constant $\tilde{V}_0 (\ell) \equiv V_{0} (\ell) / \Delta_{a}(\ell)$ and the parameter 
\begin{eqnarray}
\aimp &=& 2 - \left[ \cos^2 \theta (g_{c}^{\prime} + g_{s}^{\prime} ) + g_0^2 \sin^2 \theta \left( \frac{1}{g_{c}^{\prime}} + \frac{1}{g_{s}^{\prime}} \right) \right]. \nonumber \\
\label{Eq:alpha_imp}
\end{eqnarray}
In the absence of spin-orbit coupling, it becomes $\aimp (\delta v = 0) = 2 - g_{c}- g_{s}$, consistent with Refs.~\cite{Kane:1992,Furusaki:1993,Maslov:1995b}.
Up to the second order in $\sin \theta$, we get 
\begin{eqnarray}
&& \aimp - \aimp (\delta v=0) \nonumber \\ 
&\approx&  - \frac{\sin^2 \theta (g_c + g_s)(g_c - g_s)^4}{ 4 g_c^2 g_s^2 (g_c^2 + g_s^2)} \big( g_c^2 + g_s^2 + 3 g_c g_s \big). 
\label{Eq:alpha_imp_approx0}
\end{eqnarray}
In terms of the intrinsic parameters, we derive the following approximate formulas for the change of $\aimp$,
\begin{equation}
\aimp (\delta v) - \aimp (0) \approx  - \frac{\delta v^2 }{8 v_F^2}  \frac{(g_s - g_c)^2}{g_c + g_s} \big( g_c^2 + g_s^2 + 3 g_c g_s \big). 
\label{Eq:alpha_imp_approx}
\end{equation}
In contrast to Eq.~\eqref{Eq:alpha_bulk_approx}, the divergence upon expanding $\sin^2 \theta \propto (g_c - g_s)^{-2}$ is eliminated by the factor $(g_c - g_s)^4$ in Eq.~\eqref{Eq:alpha_imp_approx0}. As a result, the above approximation holds also for $g_c \approx g_s$, including the noninteracting limit. 

We now comment on the RG flow equation \eqref{Eq:RG_V0}.
For repulsive interaction we have $\aimp>0$, so that the backscattering strength grows under the RG flow.
Therefore, a single weak impurity gives rise to the conductance correction $\Gsingleimp$, with
\begin{eqnarray}
\frac{ \Gsingleimp }{G_0} &\propto&  - \tilde{V}^2_0 (\ell^*) 
\sim -\tilde{V}^2_0 (0) e^{\ell^* \aimp},
\label{Eq:Gsingle}
\end{eqnarray}
with the conductance quantum $G_0=2e^2/h$. The scale $\ell^*$, again, depends on other parameters, which we specify later.

Before moving on to the discussion of the many-impurity case, we have a few comments on the above result. 
First, with the RG approach we construct two flow equations--one for the tunnel amplitude $\tilde{t}$ derived for a tunnel barrier and the other for the backscattering strength $\tilde{V}_{0}$ derived for a weak impurity. 
For repulsive interaction, the former equation [Eq.~\eqref{Eq:RG_t}] indicates that the tunnel amplitude is RG irrelevant and flows toward zero, so the two semi-infinite {\TLL}s become isolated at low energies. The same conclusion follows from Eq.~\eqref{Eq:RG_V0}, where the backscattering strength increases under the RG flow, so that the conductance is suppressed by repulsive interaction.\footnote{A related calculation was done in Ref.~\cite{Moroz:2000b}, which studied the conductance correction due to a single weak impurity for a spin-orbit-coupled wire. There, it was found that the conductance correction can always be neglected because their corresponding $\aimp$ is negative. The discrepancy comes from a different form of the electron-electron interaction considered there (see the discussion in Ref.~\cite{Moroz:2000b} and also in Appendix~\ref{App:Correlator}). In contrast, here we find that $V_0$ is a relevant perturbation for repulsive interaction. We note that, in the limit of zero spin-orbit coupling, our results recover those in Refs.~\cite{Kane:1992,Furusaki:1993,Giamarchi:2003}. 
}
As a consequence, the RG approach provides consistent results for the two complementary limits of impurities, as in the absence of the spin-orbit coupling~\cite{Voit:1995,Giamarchi:2003}.

Second, in contrast to a standard {\TLL}, the presence of the spin-orbit coupling defies the duality mapping between a bulk barrier and a weak impurity. Namely, in the absence of the spin-orbit coupling, the corresponding RG flow equations 
[see Eqs.~\eqref{Eq:RG_t} and \eqref{Eq:RG_V0}] can be mapped into each other upon swapping the parameters $g_{\nu} \leftrightarrow 1/g_{\nu}$~\cite{Furusaki:1993}. In a spin-orbit-coupled wire, however, such a duality mapping is absent.

We now demonstrate how a power-law conductance can arise in the scenario of many weak impurities, relevant for a wire much longer than the average impurity separation. 
We assume that these impurities can be treated as independent (see Appendix~\ref{App:weak} for a discussion of this assumption) and each of them causes a conductance correction as computed above. 
Namely, the RG flow for $\tilde{V}_0$ is integrated up to the scale $\ell^*$, leading to the wire conductance in the presence of a single weak impurity
\begin{eqnarray} 
G_0 + \Gsingleimp & = &  G_0 - c_{\rm 1} G_0 \Big[ \frac{\Delta_a}{ {\rm Max}(k_B T, eV)} \Big]^{\aimp}, 
\label{Eq:G_single-imp}
\end{eqnarray}
with a dimensionless constant $c_{\rm 1}$ independent of temperature and voltage. 
From Eq.~\eqref{Eq:G_single-imp}, we obtain the resistance induced by a single impurity, 
\begin{align}
\Rsingleimp \approx - \frac{ \Gsingleimp }{G_0^2} \propto \big[{\rm Max}(k_B T, eV)\big]^{-\aimp}.
\end{align}
If there are $N_{\rm imp}$ impurities in the wire, adding their resistances in series gives the total resistance as $1/G_0 + N_{\rm imp} \Rsingleimp$ with the lead-wire contact resistance $1/G_0$.
For large $N_{\rm imp}$, the total resistance (defined as $1/\Gimp$) is dominated by the contribution from the impurities (so that the contact resistance is negligible), leading to
\begin{eqnarray}
\Gimp (T,V) &\propto& \left\{
\begin{array}{l} 
T^{\aimp}, \;\; {\rm for} \;\; eV\ll k_BT, \\
V^{\aimp}, \;\; {\rm for} \;\; eV\gg k_BT,
 \end{array} \right. 
\label{Eq:G_imp}
\end{eqnarray}
which is a power law with the same exponent in the high-temperature and high-bias regimes.\footnote{We note that the same power-law conductance can be obtained by starting with many impurities which are not independent, as discussed in Appendix~\ref{App:weak}. Such extended disorder generates random backscattering potential and causes resistance, which can be calculated upon applying the replica method. However, assuming that the renormalization of the interaction parameters due to the extended disorder is negligible, the power-law resistance will be the same as the isolated impurities considered here~\cite{Giamarchi:2003}.
}

Importantly, the two power laws in the opposite limits with the same exponent can be grasped by a single function using Eq.~\eqref{Eq:Balents} upon replacing the parameter $\alpha$ by $\aimp$. In other words, we take Eq.~\eqref{Eq:Balents} as an interpolation formula valid for arbitrary bias and temperature. It can be then used as the fitting curve of data displaying universal scaling. 
In this many-weak-impurity scenario, the variable $V$ denotes the bias voltage across the entire wire. In contrast to the tunnel barrier scenario, where the replacement $V \rightarrow V/N_{\rm b}$ in Eq.~\eqref{Eq:Balents} is necessary for multiple barriers, here Eq.~\eqref{Eq:Balents} remains unchanged regardless of the number of weak impurities.

\begin{figure}[t]
 \includegraphics[width=\linewidth]{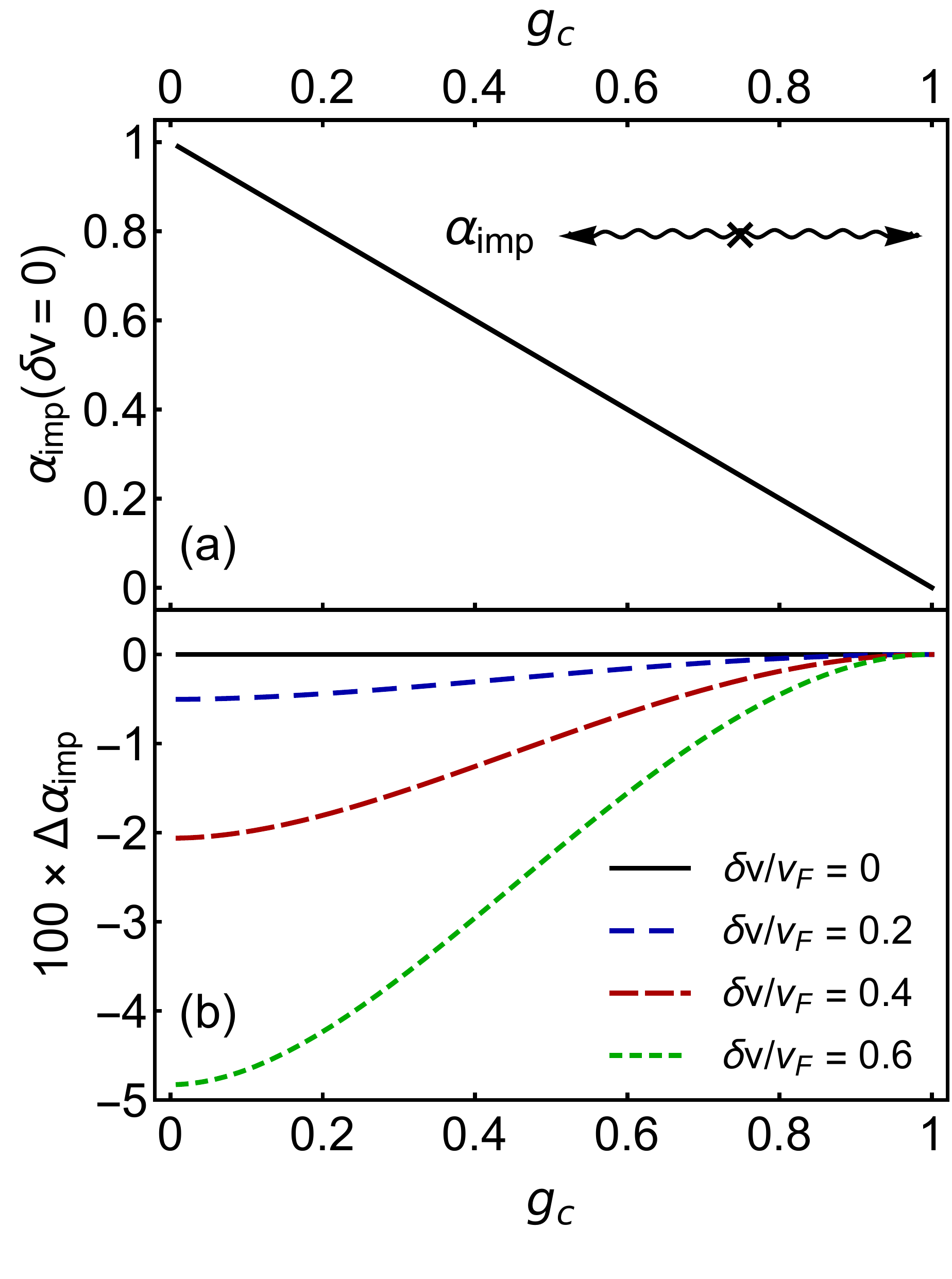}
 \caption{
(a) Interaction parameter ($g_c$) dependence of the parameter ($\aimp$) for $g_{s}=1$ and $\delta v=0$.
The inset illustrates a weak impurity corresponding to Fig.~\ref{Fig:illustration}(c). 
Many such weak impurities cause a power-law conductance [see Eq.~\eqref{Eq:G_imp}] with the exponent $\aimp$ defined in Eq.~\eqref{Eq:alpha_imp}.
(b) Interaction parameter ($g_c$) dependence of the change ($\Delta \aimp$, multiplied by 100 for clarity) of the exponent $\aimp$ with respect to its value for $\delta v =0$ for $g_s=1$ and several values of $\delta v /v_F$. 
}
 \label{Fig:alpha_imp}
\end{figure}

In Fig.~\ref{Fig:alpha_imp}, we plot the exponent $\aimp$ at zero spin-orbit coupling and its change $\Delta \aimp$ for several values of $\delta v /v_F$ as functions of $g_c$.
Similar to Fig.~\ref{Fig:alpha_barrier}, the exponent $\aimp$ for $\delta v =0$ [see Fig.~\ref{Fig:alpha_imp}(a)] increases with a larger strength of the interaction (that is, a smaller $g_{\rm c}$ value), and the band distortion can only reduce $\aimp$ [see Fig.~\ref{Fig:alpha_imp}(b)]. 
In comparison with Fig.~\ref{Fig:alpha_barrier}, on the other hand, the effect of the band distortion on the exponent $\aimp$ is quantitatively much weaker.
Moreover, in contrast to Fig.~\ref{Fig:alpha_barrier}, where $\aend$ and $\abulk$ are unbounded in the strong-interaction regime, the corresponding parameter for weak impurities is bounded in the range $\aimp \in [0,1]$.
Therefore, it is possible to rule out weak impurities as the dominant resistance contribution if the $\alpha$ value extracted from the current-voltage measurements exceeds unity.
Furthermore, if both types of impurities are present, the resistance due to tunnel barriers dominates weak impurities for strong interaction, while the relation is opposite for weak interaction. Therefore, we predict a transition of the power-law conductance by varying electron-electron interaction, as discussed below.

\begin{figure}[t]
\includegraphics[width=\linewidth]{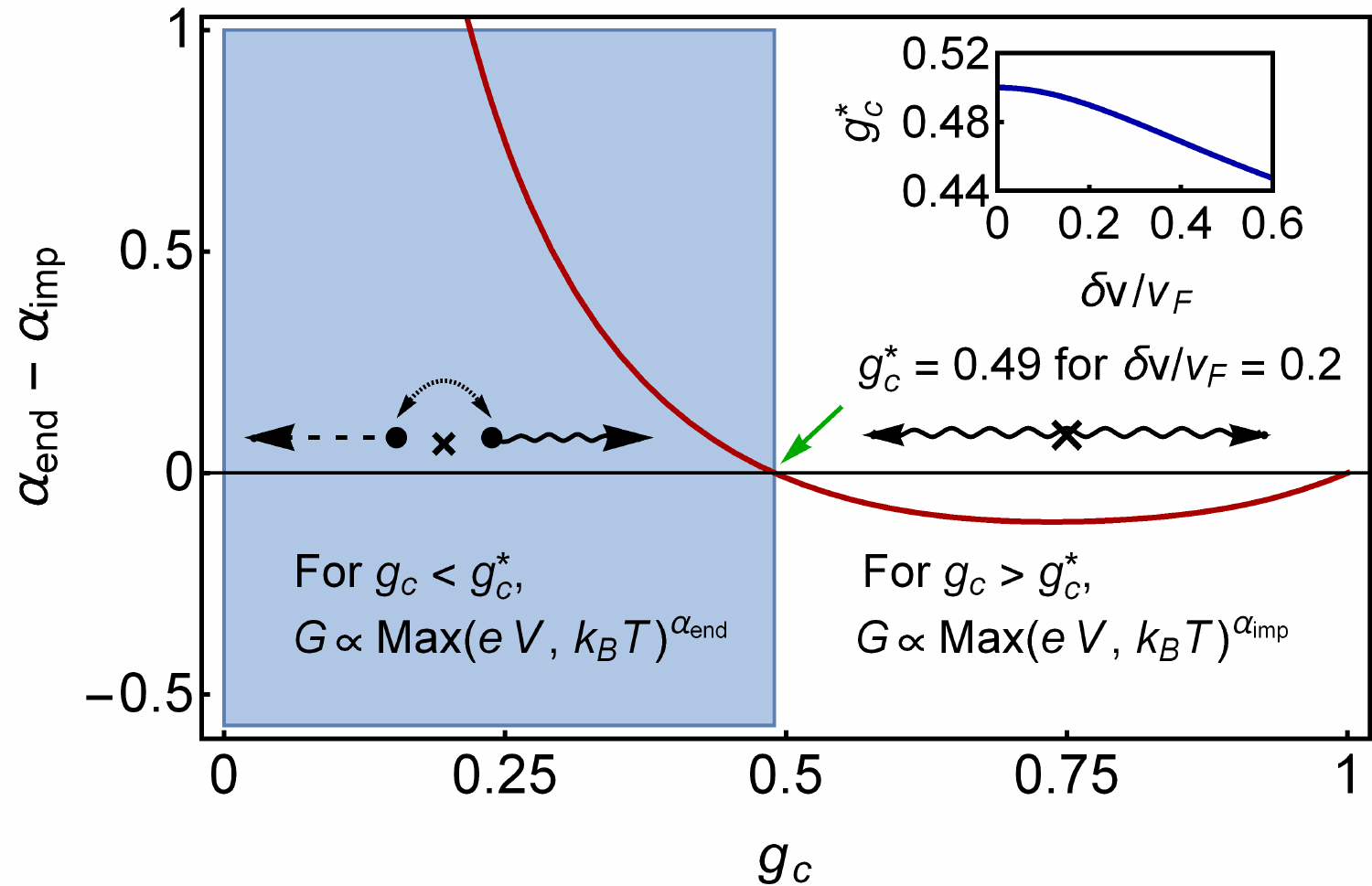}
\caption{Transition of the power-law conductance. 
The main panel shows the difference $(\aend - \aimp)$ as a function of the interaction parameter $(g_{c})$ for $g_s=1$ and $\delta v/v_{F}=0.2$. 
The exponents $\aend$ and $\aimp$ are given in Eqs.~\eqref{Eq:alpha_boundary} and \eqref{Eq:alpha_imp}, respectively.
The point at which $\aend = \aimp$ (denoted as $g_{c}^{*}$) indicates a transition between the regimes with different power-law conductance.
For $g_{c}<g_{c}^{*}$ (blue shaded region), the power-law conductance is characterized by $\aend$, whereas for $g_{c}>g_{c}^{*}$ it is characterized by $\aimp$.
The inset shows the dependence of $g_{c}^{*}$ value on $\delta v/v_F $.
}
 \label{Fig:alpha_transition}
 \end{figure}

\subsection{Coexisting strong and weak impurities~\label{SubSec:Coexist}}
Here we discuss the scenario in which impurities of all the types are present. Provided that the effects of the tunnel barriers and weak impurities on the resistance do not interfere with each other so that each resistance source can be treated separately as in Secs.~\ref{SubSec:Strong} and \ref{SubSec:Weak}, their contributions can be added into the total resistance of the entire wire. In general, the three resistance sources [corresponding to Eqs.~\eqref{Eq:G_boundary}, \eqref{Eq:G_bulk}, and \eqref{Eq:G_imp}] enter the total resistance as 
\begin{eqnarray}
R_{\rm tot} &\propto& \frac{h}{2e^2} \sum_{n} c_{n}\big[\frac{\Delta_{a}}{{\rm Max}(eV, k_B T)} \big]^{\alpha_{n}},
\end{eqnarray}
where $n \in \{{\rm end},~{\rm bulk},~{\rm imp} \}$ indicates the resistance source.
In the above, $c_{n}$'s are the corresponding prefactors. 
Since in typical experiments we have $\Delta_a \gg eV, k_B T$, the total resistance of a wire is dominated by the contribution with the largest exponent. 
Therefore, for any repulsive interaction, as long as there exists a tunnel barrier in the bulk of the wire, the charge transport of the wire is characterized by the current-voltage curve and the dc conductance with $\abulk$.

Interestingly, if there exist both boundary barriers and weak impurities, but no bulk barriers, the dominant exponent depends on the strength of the electron-electron interaction. 
In Fig.~\ref{Fig:alpha_transition}, we plot the difference between the exponents representing the boundary barrier and the weak impurity as a function of the interaction parameter $g_{c}$ for the value of $\delta v/v_F = 0.2$. 
The sign of the difference $(\aend - \aimp)$ then indicates whether the boundary barriers or the weak impurities dominate. 
The transition happens at a point denoted as $g_{c}^{*}$. 
In the strong-interaction ($g_{c}<g_{c}^{*}$) regime, the current-voltage curve and the dc conductance are characterized by $\aend$, whereas in the weak-interaction ($g_{c}>g_{c}^{*}$) regime, they are characterized by $\aimp$. The value of $g_{c}^{*}$ weakly depends on the strength of the spin-orbit coupling (see the figure inset). 
For $\delta v/v_F = 0.2$, we find $g_{c}^{*} \approx 0.49$, very close to $g_{c}^{*}=1/2$ for $\delta v/v_F = 0$.
Concluding, the transport properties of a {\TLL} strongly depend on the types and locations of the impurities.

\begin{table*}[t]
\caption{Parameters $\alpha$ characterizing the current-voltage curve and the power-law (differential) conductance of a spin-orbit-coupled {\TLL}\footnote{See Table~II in Ref.~\cite{Sato:2019} for a summary of the corresponding parameters of various {\TLL}s without spin-orbit coupling.}
 subject to various types of impurities. The first and second columns give the impurity type and their illustration, respectively. 
The third (fourth) column gives the notation (expression) of the corresponding parameter.
The fifth (sixth) column gives the corresponding equation (figure) number. 
The parameters $g_{c}^\prime$, $g_{s}^\prime$, $g_{0}$, and $\theta$ are given in Eqs.~\eqref{Eq:parameters} and \eqref{Eq:g_nu_prime}.
}
\begin{tabular}[c]{ l c c c c c}
\hline \hline
Impurity or defect type & Illustration & Notation & Expression & Eq. & Fig.\\
\hline
Strong impurity (tunnel barrier) near the wire end & Fig.~\ref{Fig:illustration}(a) & $\aend$ & $\frac{1}{2}\left( \frac{1}{g_{c}^{\prime}} + \frac{1}{g_{s}^{\prime}} \right) \left( \cos^2 \theta + g_0^2 \sin^2 \theta \right) -1$ & Eq.~\eqref{Eq:alpha_boundary} & Fig.~\ref{Fig:alpha_barrier} \\
\hline
Strong impurity (tunnel barrier) within the wire & Fig.~\ref{Fig:illustration}(b) & $\abulk$  &  $\left( \frac{1}{g_{c}^{\prime}} + \frac{1}{g_{s}^{\prime}} \right) \left( \cos^2 \theta + g_0^2 \sin^2 \theta \right) -2$ & Eq.~\eqref{Eq:alpha_bulk} & Fig.~\ref{Fig:alpha_barrier} \\
\hline
Many weak impurities (potential disorder) & Fig.~\ref{Fig:illustration}(c) & $\aimp$ & $2 - \left[ \cos^2 \theta (g_{c}^{\prime} + g_{s}^{\prime} ) + g_0^2 \sin^2 \theta \left( \frac{1}{g_{c}^{\prime}} + \frac{1}{g_{s}^{\prime}} \right) \right]$ & Eq.~\eqref{Eq:alpha_imp} & Fig.~\ref{Fig:alpha_imp} \\
\hline \hline
\end{tabular}
\label{Table:summary}
\end{table*}

\section{Discussion~\label{Sec:Discussion}}
\subsection{Effects of multiple subbands~\label{SubSec:Subband}}
Having analyzed wires with a single occupied subband, we now look at the case with the Fermi energy intersecting multiple transverse subbands. 
In the absence of spin-orbit coupling, the corresponding problems were solved for tunnel barriers~\cite{Matveev:1993} and weak impurities~\cite{Sandler:1997}. Instead of repeating similar calculations, here we discuss what we expect for a spin-orbit-coupled system.

In the tunneling regime, the current through a barrier depends on the density of states on the two sides of the barrier. For each subband, the power-law density of states is characterized by an effective exponent, which can be obtained by solving an eigenvalue problem as in Ref.~\cite{Matveev:1993}. 
The tunneling current through a multi-subband wire is determined by the sum of the currents through each subband.
As a result, the total current is dominated by the subband with the largest conductance, or, equivalently, the smallest effective exponent among the subbands. 
In the absence of the spin-orbit coupling, the smallest exponent corresponds to the lowest transverse subband~\cite{Matveev:1993}.
Since, based on our single-subband results, we expect that the spin-orbit coupling leads to small modifications of the exponents, we expect that the total tunnel current will show universal scaling with an exponent corresponding to the lowest subband.

Weak impurities, on the other hand, induce backscattering in the highest occupied subband~\cite{Sandler:1997}.
It leads to a conductance correction with an exponent, which can be computed as in Ref.~\cite{Sandler:1997}.
Again, based on our single-subband results, we expect little effects of the spin-orbit coupling on this exponent.
Provided that there are many weak impurities, their resistance contributions dominate the contact resistance with the leads, leading to a power-law conductance characterized by the same exponent. 

For both tunneling and disorder regimes, we expect that the effective exponents reduce to zero when the subband number becomes infinity, thereby recovering the Fermi-liquid behavior in higher dimensions. As pointed out in Ref.~\cite{Sandler:1997}, we expect that the {\TLL} behavior can be observable for wires in which not many subbands are populated.

\subsection{Transport signatures for a {\TLL}~\label{SubSec:TLL}}
Here, we comment on the robustness of the {\TLLadj} behavior displayed in the charge transport of quantum wires.
The power-law resistances induced by various types of impurities, which we obtain from the RG analysis, allows us to determine the dominant contribution from their corresponding exponents, assuming these resistance sources are independent.
As discussed above, when there are multiple resistances due to barriers or impurities in series, 
the current and power-law conductance of the wire are characterized by the largest exponent $\alpha$ among the constituent resistance sources. 
On the other hand, when there are multiple subbands or wires in parallel, the total current and power-law conductance are characterized by the smallest $\alpha$. 
In any case, the universal scaling behavior can be observed even for wires in few-transverse-mode regime~\cite{Sato:2019} and nanotube bundles~\cite{Bockrath:1999}.
Consequently, the universal scaling behavior persists in rather general situations [with or without spin-orbit coupling, with (single or multiple) barriers or disorder potential, in the single- or multi-mode regime, in a single or multiple wires], providing quite robust signatures for a {\TLL}.

Such signatures provide a useful tool to characterize the interacting one-dimensional electron systems through their transport properties. 
In addition to the effects of spin-orbit coupling, our work points out that extracting the interaction strength is complicated by the impurity character in the system. In order to make sensible extraction of the interaction strength, it requires assumptions on the impurity type and location.
A similar complication has been discussed in the context of the edge conductance of a two-dimensional topological insulator~\cite{Vayrynen:2016,Stuhler:2019}, where both an isolated strong magnetic impurity and many weak magnetic impurities can cause power-law conductance, though with distinct exponents~\cite{Maciejko:2009,Vayrynen:2016,Hsu:2017,Hsu:2018}.

Nonetheless, here we find two ways to overcome such complications. 
First, we point out that the exponent of the power-law conductance due to weak impurities is bounded, so an experimental value exceeding this bounded value can rule out weak impurities as the dominant resistance source. 
Second, by making use of the full current-voltage curve, one can extract also the barrier number, which can serve as an indicator of the dominant resistance source.
For concreteness, let us assume that there are two boundary barriers and many weak impurities coexisting in a wire with $\delta v / v_F = 0.2$.
Even though both resistance sources lead to Eq.~\eqref{Eq:Balents} upon replacing $V \rightarrow V/N$, the corresponding curves are quantitatively distinguishable, with $(\alpha, N) \rightarrow (\aend, 2)$ for boundary barriers and $(\alpha, N) \rightarrow (\aimp, 1)$ for weak impurities.
 In addition, as demonstrated in Fig.~\ref{Fig:alpha_transition}, the relative strength of their contribution to resistance changes with the interaction strength, which can be varied by applying gate voltage. 
It leads to a transition of the power-law conductance at the point $g_c^* \approx 1/2$ (corresponding to $\alpha^* \approx 1/2$).
Across the transition point, the exponent changes from $\aend$ (for $\alpha \apprge \alpha^*$) to $\aimp$ (for $\alpha \apprle \alpha^*$), whereas the $N$ value changes from two to one {\it at the same point $\alpha^*$}.
Remarkably, such behavior was indeed observed in Ref.~\cite{Sato:2019}.
In conclusion, by fitting the full current-voltage curve, the extracted barrier number can indicate the dominant resistance source and can be used for an independent check.

\section{Summary~\label{Sec:Summary}}
In summary, we theoretically investigate the transport properties of a spin-orbit-coupled {\TLL} in a quasi-one-dimensional confinement. 
We calculate the temperature and bias-voltage dependence of the tunnel current and conductance subject to various types of impurities.
Our main conclusion is that, for realistic strengths of the spin-orbit coupling, the current-voltage curves follow the universal scaling relation of a non-spin-orbit-coupled {\TLL} with a modified parameter $\alpha$. 
For convenience, we summarize these results in Table~\ref{Table:summary}.
Importantly, the spin-orbit coupling leads to mostly negligible modifications if the electron-electron interaction is strong.
Our findings can be applied to characterize spin-orbit-coupled quantum wires such as InAs and InSb in order to design devices for spintronics and topological matter.

\begin{acknowledgments}
We thank T. Aono, T.-M. Chen, and Y. Tokura for interesting discussions. This work was supported financially by the JSPS Kakenhi Grant No. 16H02204 and No. 19H05610, the Swiss National Science Foundation (Switzerland), and the NCCR QSIT. 
YS acknowledges support from JSPS Research Fellowship for Young Scientists (Grant No. JP18J14172).
SM acknowledges supports from JSPS Grant-in-Aid for Scientific Research (B) (Grant No. JP18H01813), JST, PRESTO (Grant No. JPMJPR18L8), and JSPS Grant-in-Aid for Scientific Research (S) (Grant No. JP26220710).
\end{acknowledgments}

\appendix

\section{Single-particle correlation function~\label{App:Correlator}}
In this appendix we calculate the single-particle equal-space correlation function at the origin $x=0$,  
\begin{align}
G_{r\sigma}(0,t) \equiv \Big\langle \psi_{r\sigma}^\dagger (0,t) \psi_{r\sigma} (0,0) \Big\rangle,
\end{align}
where the subscripts $r$ and $\sigma$ label the right-/left-movers and up-/down-spins, respectively, and the argument $(x,t)$ is given by the spatial and real-time coordinates. 
In terms of the boson fields [see Eg.~\eqref{Eq:bosonization}], the correlator reads
\begin{align}
G_{r\sigma}(0,t) = \frac{1}{2\pi a} \Big\langle e^{\frac{i}{\sqrt{2}} [ r \delta\phi_{c} - \delta\theta_{c} + r\sigma\delta\phi_{s} -\sigma \delta\theta_{s} ] }  \Big\rangle,
\label{Eq:G_sp}
\end{align}
where, since we are interested in the equal-space correlator, we define the following notations for simplicity,
\begin{align}
\label{Eq:abbreviation_delta}
\begin{split}
\delta \phi_{\nu} \equiv & \; \phi_{\nu}(0,t)-\phi_{\nu}(0,0),\\
\delta\theta_{\nu} \equiv & \; \theta_{\nu}(0,t)-\theta_{\nu}(0,0).
\end{split}
\end{align}
Transforming into the diagonalized basis [see Eq.~\eqref{Eq:NewFields}], the bracket in Eq.~\eqref{Eq:G_sp} becomes
\begin{widetext}
\begin{equation}
\begin{split}
\Big\langle {\rm Exp}\Big\{\frac{i}{\sqrt{2}} \Big[ (r\cos \theta -\frac{\sigma}{g_0} \sin \theta) \delta\phi_{c}^\prime 
 + ( r \sigma g_0 \sin \theta - \cos \theta ) \delta\theta_{c}^\prime 
 - ( r  g_0 \sin \theta + \sigma\cos \theta ) \delta\theta_{s}^\prime 
+ ( r \sigma \cos \theta + \frac{1}{g_0} \sin \theta) \delta\phi_{s}^\prime \Big] \Big\}  \Big\rangle.
\end{split}
\end{equation}
\end{widetext}
The correlator depends on whether we are looking at the boundary or at the bulk of the wire. 
The correlator at the wire boundary behaves differently from the one in the bulk and needs more caution~\cite{Kane:1992,Giamarchi:2003}. On the other hand, it is straightforward to compute the correlator in the bulk using Eq.~\eqref{Eq:diag}, which allows us to obtain the density of states $\rho_{r\sigma}^{\rm bulk}(\epsilon) \propto \epsilon^{\dosbulk}$ in the bulk. For a given set of $(r,\sigma)$, the exponent is
\begin{align}
\dosbulk = 
\frac{g_{c}^{\prime}}{4}  \left( \cos \theta - \frac{r \sigma}{ g_0} \sin \theta \right) ^2 
+  \frac{\left( \cos \theta - r \sigma g_0 \sin \theta \right) ^2}{4g_{c}^{\prime}} \nonumber\\
 + \frac{g_{s}^{\prime}}{4}  \left( \cos \theta + \frac{r \sigma}{ g_0} \sin \theta \right) ^2 
+  \frac{\left( \cos \theta + r \sigma g_0 \sin \theta \right) ^2}{4g_{s}^{\prime}}  -1.
\label{Eq:DOS}
\end{align}
It becomes $(g_c + g_s + 1/g_c+1/g_s)/4-1$ in the absence of spin-orbit coupling. The bulk density of states $\rho_{r\sigma}^{\rm bulk}(\epsilon) $ can be probed by scanning tunneling spectroscopy~\cite{Kane:1992,Balents:1999,Moroz:2000,Iucci:2003}. 
Before continuing, let us comment on the difference of the exponent for the single-particle correlation function obtained here and those in Refs.~\cite{Moroz:2000,Moroz:2000b}. The discrepancy arises from the different form of the electron-electron interaction.
Namely, here we follow Refs.~\cite{Kane:1992,Giamarchi:2003}, and keep $g_s=1$ in the limit of zero spin-orbit coupling. 
On the other hand, in Refs.~\cite{Moroz:2000,Moroz:2000b} the interaction parameters in the charge and spin sectors $g_c$ and $g_s$ are dependent (see the discussion in Sec.~IV there for details~\cite{Moroz:2000b}),  
such that they have $g_s>1$ even in the absence of spin-orbit coupling. In consequence, these different choices result in distinct exponents of the correlation functions.

Next, we consider the correlator at the boundary of the wire by assuming that the {\TLL} given by Eq.~\eqref{Eq:diag} extends over semi-infinite space ($x>0$) and terminates at the origin $x=0$. To proceed, we use the trick from Ref.~\cite{Giamarchi:2003}, which makes use of chiral boson fields to map the semi-infinite system onto an infinite system. Specifically, we express the boson fields in the sector $\nu$ as 
\begin{subequations}
\label{Eq:xfmn1}
\begin{eqnarray}
\phi_{\nu}^\prime(x,t) = \frac{\sqrt{g_{\nu}^\prime}}{2} \Big[ \phi_{\nu}^{L}(x,t) -  \phi_{\nu}^{R}(x,t) \Big],\\
\theta_{\nu}^\prime(x,t) = \frac{1}{2\sqrt{g_{\nu}^\prime}} \Big[ \phi_{\nu}^{L}(x,t) +  \phi_{\nu}^{R}(x,t) \Big],
\end{eqnarray}
\end{subequations}
where $\phi_{\nu}^{R/L}$ are right-/left-moving chiral boson fields. In the above, we rescaled the fields by the interaction parameters $g_{\nu}^\prime$ such that $\phi_{\nu}^{R/L}$ represent free chiral bosons. These chiral fields allow us to define
\begin{subequations}
\label{Eq:xfmn2}
\begin{eqnarray}
\phi_{\nu}^{R}(x,t) &\rightarrow& \tilde{\phi}_{\nu}^{\infty} (x,t), \\ 
\phi_{\nu}^{L}(x,t) &\rightarrow& \tilde{\phi}_{\nu}^{\infty} (-x,t),
\end{eqnarray}
\end{subequations}
where $\tilde{\phi}_{\nu}^{\infty}$ is a free chiral boson field defined in a system extending over the entire one-dimensional space. Finally, we can reexpress the chiral fields as
\begin{eqnarray}
\label{Eq:xfmn3}
\tilde{\phi}_{\nu}^{\infty} (x,t) &\rightarrow& \theta_{\nu}^{\infty} (x,t) - \phi_{\nu}^{\infty} (x,t), 
\end{eqnarray}
where the fields $\phi_{\nu}^{\infty}$ and $\theta_{\nu}^{\infty}$ are analogous to $\phi_{\nu}^{\prime}$ and $\theta_{\nu}^{\prime}$ except that they are free and defined in an infinite space. 

Performing the transformations \eqref{Eq:xfmn1}--\eqref{Eq:xfmn3}, we rewrite the correlation function as 
\begin{widetext}
\begin{equation}
G_{r\sigma}(0,t) = \frac{1}{2\pi a} \Big\langle {\rm Exp}\Big\{ \frac{i}{\sqrt{2}} \Big[
 \frac{( r \sigma g_0 \sin \theta - \cos \theta )}{\sqrt{g_{c}^\prime}} (\delta\theta_{c}^{\infty} - \delta\phi_{c}^{\infty} )
 - \frac{( r  g_0 \sin \theta + \sigma\cos \theta )}{\sqrt{g_{s}^{\prime}}} (\delta\theta_{s}^{\infty} - \delta\phi_{s}^{\infty} ) \Big] \Big\}  \Big\rangle,
\end{equation}
where we have introduced the notations $\delta\phi_{\nu}^{\infty}$ and $\delta\theta_{\nu}^{\infty}$ analogous to Eq.~\eqref{Eq:abbreviation_delta}. 
Since in the above formula the fields $\phi_{\nu}^{\infty}$ and $\theta_{\nu}^{\infty}$ are free and defined in an infinite space, their correlation functions can be computed directly, leading to the finite-temperature correlation function
\begin{equation}
G_{r\sigma}(0,t) = \frac{1}{2\pi a} 
\left[ \frac{ \pi a k_B T/ (\hbar v_F) }{ i \sinh \Big( \pi k_B T t/\hbar \Big)} \right]^{ (\cos \theta - r \sigma g_0 \sin \theta) ^2 / (2 g_{c}^\prime) }  
\left[ \frac{ \pi a k_B T/ (\hbar v_F) }{ i \sinh \Big( \pi k_B T t/\hbar \Big)} \right]^{ (\cos \theta + r \sigma g_0 \sin \theta) ^2 / (2 g_{s}^\prime) }. 
\end{equation}
\end{widetext} 
The expression simplifies to 
\begin{equation}
G_{r\sigma}(0,t) = \frac{1}{2\pi a} \left[ \frac{ \pi k_B T/\Delta_a }{ i \sinh \Big( \pi k_B T t/\hbar \Big)} \right]^{\dosend + 1 },
\label{Eq:DOS_end}
\end{equation}
with $\Delta_{a} \equiv \hbar v_F/a$ denoting the bandwidth associated with the short-distance cutoff. The parameter in the exponent is 
\begin{align}
\dosend =& 
 \frac{\left( \cos \theta - r \sigma g_0 \sin \theta \right) ^2}{2g_{c}^{\prime}} 
+  \frac{\left( \cos \theta + r \sigma g_0 \sin \theta \right) ^2}{2g_{s}^{\prime}}  -1,
\label{Eq:alpha_DOS_end}
\end{align}
which becomes $(1/g_c+1/g_s)/2-1$ in the absence of spin-orbit coupling. 

The correlation function at the boundary of the wire, Eq.~\eqref{Eq:DOS_end}, is directly related to the current through a tunnel barrier, as presented in Sec.~\ref{SubSec:Strong}.
In addition, Eq.~\eqref{Eq:DOS_end} allows us to get the density of states at the boundary of the wire $\rho_{r\sigma}^{\rm end}(\epsilon) \propto \epsilon^{\dosend}$ for a given set of $(r,\sigma)$, which can be probed using scanning tunneling spectroscopy~\cite{Kane:1992,Balents:1999,Moroz:2000,Iucci:2003}. In Appendix~\ref{App:GZ}, the zero-spin-orbit values of the bulk and boundary exponents [see Eqs.~\eqref{Eq:DOS} and \eqref{Eq:alpha_DOS_end}] are plotted in Fig.~\ref{Fig:beta_GZ}. 
The exponent Eq.~\eqref{Eq:alpha_DOS_end} is used to derive the RG flow equation for the tunnel amplitude, as discussed in Sec.~\ref{SubSec:Strong}.

\section{Correlation function and current-voltage characteristics~\label{App:FT}}
In this appendix we present the calculation of the correlation function $\chi_{\sigma}^{\rm ret}(\omega)$ given in Eq.~\eqref{Eq:chi_ret}, which is used to compute the current-voltage characteristics for a wire with a tunnel barrier.
We follow the procedure presented in Appendix~\ref{App:Correlator} to obtain two single-particle correlation functions, each of which corresponds to one of the subsystems in Eq.~\eqref{Eq:H_12}. The result is given by Eq.~\eqref{Eq:DOS_end}, with the exponent Eq.~\eqref{Eq:alpha_DOS_end} generalized in order to incorporate the two subsystems on the two sides of the barrier. 
Namely, let us define the exponent corresponding to the subsystem $j$,
\begin{subequations} 
\label{Eq:alpha_DOS_end_j}
\begin{align}
\dosendj \equiv & \aendj + r \sigma \deltaj, \\
\aendj \equiv & 
\left( \frac{1}{2g_{jc}^{\prime}} + \frac{1}{2g_{js}^{\prime}}  \right) \left( \cos^2 \theta_j + g_{j0}^2 \sin^2 \theta_j \right)-1,
\\
\deltaj \equiv &
\left( -\frac{1}{2g_{jc}^{\prime}} + \frac{1}{2g_{js}^{\prime}}  \right) g_{j0} \sin (2\theta_j), 
\end{align}
\end{subequations}
with the parameters $g_{jc}^{\prime}$, $g_{js}^{\prime}$, $g_{j0}$, and $\theta_j$ corresponding to the subsystem $j$ defined in Eq.~\eqref{Eq:H_12}. 
Introducing the above notations in Eqs.~\eqref{Eq:DOS_end}--\eqref{Eq:alpha_DOS_end} and plugging the latter into Eq.~\eqref{Eq:chi_all},
we get the sum of four terms (for a given $\sigma$). Each of the four terms can be written as
\begin{eqnarray}
\chi_{\sigma}^{\rm ret} (\omega ) &=& \frac{\sin(\alpha \pi/2)}{2 \pi^2 a^2} \Big(\frac{\pi k_B T}{\Delta_a}\Big)^{\alpha+2} \nonumber \\
&& \times  \int_{0}^{\infty} dt \; e^{i \omega t} \Big| \sinh \Big( \frac{\pi k_B T t}{\hbar} \Big) \Big|^{-\alpha-2}, 
\label{Eq:time-integral}
\end{eqnarray}
with the parameter $\alpha$ given by one of the following,
\begin{eqnarray}
\label{Eq:alpha_diff}
\alpha &\in& \Big \{ \aendone + \aendtwo + \sigma (\deltaone + \deltatwo),\; \aendone + \aendtwo - \sigma (\deltaone + \deltatwo) , \nonumber \\
 && \hspace{5pt} \aendone + \aendtwo + \sigma (\deltaone - \deltatwo),\; \aendone + \aendtwo - \sigma (\deltaone - \deltatwo) \Big\}. \nonumber \\
\end{eqnarray}
These four terms with different $\alpha$'s can be computed separately and then summed up. The integral over time in the second line of Eq.~\eqref{Eq:time-integral} gives the beta function, which can be converted into the gamma function with the relation $B(x,y)=\Gamma(x)\Gamma(y)/\Gamma(x+y)$. It gives
\begin{eqnarray}
\frac{ \Gamma (-\alpha -1) \Gamma \Big( 1 + \frac{\alpha}{2} -\frac{i \hbar \omega }{ 2\pi k_B T} \Big)    }{  2 \Gamma \Big( - \frac{\alpha}{2} -\frac{i \hbar \omega }{ 2\pi k_B T} \Big)  }. 
\end{eqnarray}
Applying Euler's reflection formula $ \Gamma (1-z) \Gamma (z) = \pi / \sin(\pi z)$ for non-integer $z$ and taking the imaginary part, we get, for a given $\alpha$ in Eq.~\eqref{Eq:alpha_diff},
\begin{eqnarray}
{\rm Im}\left[ \chi_{\sigma}^{\rm ret} (\omega) \right] &=& -\frac{1}{8 \pi^3 a^2} \frac{\hbar}{k_B T} \Big(\frac{2\pi k_B T}{\Delta_a}\Big)^{\alpha+2} \sinh \Big(\frac{\hbar \omega }{ 2 k_B T} \Big) \nonumber \\
&& \times \frac{1}{\Gamma(\alpha+2)} \left| \Gamma \Big( 1 + \frac{\alpha}{2} + \frac{i \hbar \omega }{ 2\pi k_B T} \Big) \right|^2.
\label{Eq:chi_Im}
\end{eqnarray}
While the sum of the contributions with distinct $\alpha$'s does not produce a single curve, we note that, for realistic values of $\delta v /v_F \apprle 0.1$, the deviations $\delta \alpha$'s in Eq.~\eqref{Eq:alpha_diff} are negligible. In addition, since both the terms with $\alpha + \delta \alpha$ and $\alpha - \delta \alpha$ contribute to the sum, the leading-order correction in current caused by the small parameter $\delta \alpha$ here will be $\delta I ({\delta \alpha}) \propto \delta \alpha^2$. In contrast, the band-distortion-induced change $\Delta \alpha$ in the main text results in the first-order correction $\delta I ({\Delta \alpha}) \propto \Delta \alpha$. 
Since for typical parameters we have $\delta \alpha^2 \ll |\Delta \alpha|$, it allows us to neglect $\delta \alpha$ and to approximate $\alpha $ as $\aendone + \aendtwo$. As a result, we can write the sum as Eq.~\eqref{Eq:chi_Im} multiplied by a factor of 4, with $\alpha$ given in Eq.~\eqref{Eq:alpha_general}.
Finally, inserting Eq.~\eqref{Eq:chi_Im} into Eq.~\eqref{Eq:I_tun2} gives Eq.~\eqref{Eq:Balents} in the main text. 
We remark that the approximation on negligible $\delta \alpha$ is justified for the experiment in Ref.~\cite{Sato:2019}, which clearly observed the universal scaling behavior of the current-voltage characteristics in InAs nanowires in spite of presumably strong spin-orbit coupling of the material. 

Finally, we demonstrate that the asymptotic behavior of Eq.~\eqref{Eq:Balents} in the high-temperature and high-bias regimes are indeed consistent with the power-law conductance obtained from the RG approach. In the high-temperature $(k_B T \gg eV)$ regime, we expand Eq.~\eqref{Eq:Balents} in powers of $V$ and retain the leading-order term, resulting in the linear response $I \propto V T^{\alpha}$. In the high-bias $(eV \gg k_B T)$ regime, on the other hand, the following asymptotic form of the gamma function can be used~\cite{Gradshteyn:1980} 
\begin{align}
\lim_{|y| \rightarrow \infty} \Big| \Gamma (x+iy) \Big| = \sqrt{2\pi} |y|^{x-\frac{1}{2}} e^{-\pi|y|/2},
\end{align}
which leads to $I \propto V^{\alpha +1}$. 
We note that there is a factor of $\pi$ in the exponential on the right-hand side, which is crucial for the cancellation of the gamma function and hyperbolic sine function with different arguments. 
In summary, the asymptotic behavior of Eq.~\eqref{Eq:Balents} gives the conductance Eq.~\eqref{Eq:dI/dV}, so the current-voltage characteristics obtained by computing the tunnel current is consistent with the conductance derived from the RG approach.

\section{Density of states of the Governale-Z{\"u}licke model~\label{App:GZ}}
In this appendix we discuss the density of states of the bosonized model proposed by Governale and Z{\"u}licke. 
We assume that the system parameters are (fine-)tuned to the regime in which the electrons have the spin orientation described in Refs.~\cite{Governale:2002,Governale:2004}: In our notation, the spins of $R_A$ and $L_B$ are antiparallel and the overlap $P_{AB}$ is zero. In this configuration the wire is helical and immune against backscattering on charge impurities. Within our model, the Luttinger liquid has no resistance. Nevertheless, one can inspect the effects of the spin-orbit coupling on other physical quantities. 
Here we look at the density of states.

Upon bosonization using Eq.~\eqref{Eq:bosonization}, we get $H^{\prime} \equiv H_{0} + H_{\textrm{so}}^{\prime}$, where $H_{0}$ retains the same form as Eq.~\eqref{Eq:H_TLL}, with the index $\nu \in \{c,s\}$ now meaning the symmetric and anti-symmetric combination of the fields involving branch $A$ and $B$, respectively. In other words, instead of real spin, the pseudospin index $\sigma$ now indicates the branch $\sigma \in \{ A,B\}$.
In contrast to Refs.~\cite{Moroz:2000,Moroz:2000b}, the charge-spin mixing term now takes the following form,
\begin{align}
H_{\textrm{so}}^{\prime} =&
 \delta v \int \frac{\hbar dx}{2\pi}  \left[ (\partial_x \phi_{c}) (\partial_x \phi_{s}) + (\partial_x \theta_{c}) (\partial_x \theta_{s}) \right] ,
\label{Eq:H_SOI_GZ}
\end{align}
where the coordinates of the fields are suppressed for simplicity.

\begin{figure}[t]
 \includegraphics[width=\linewidth]{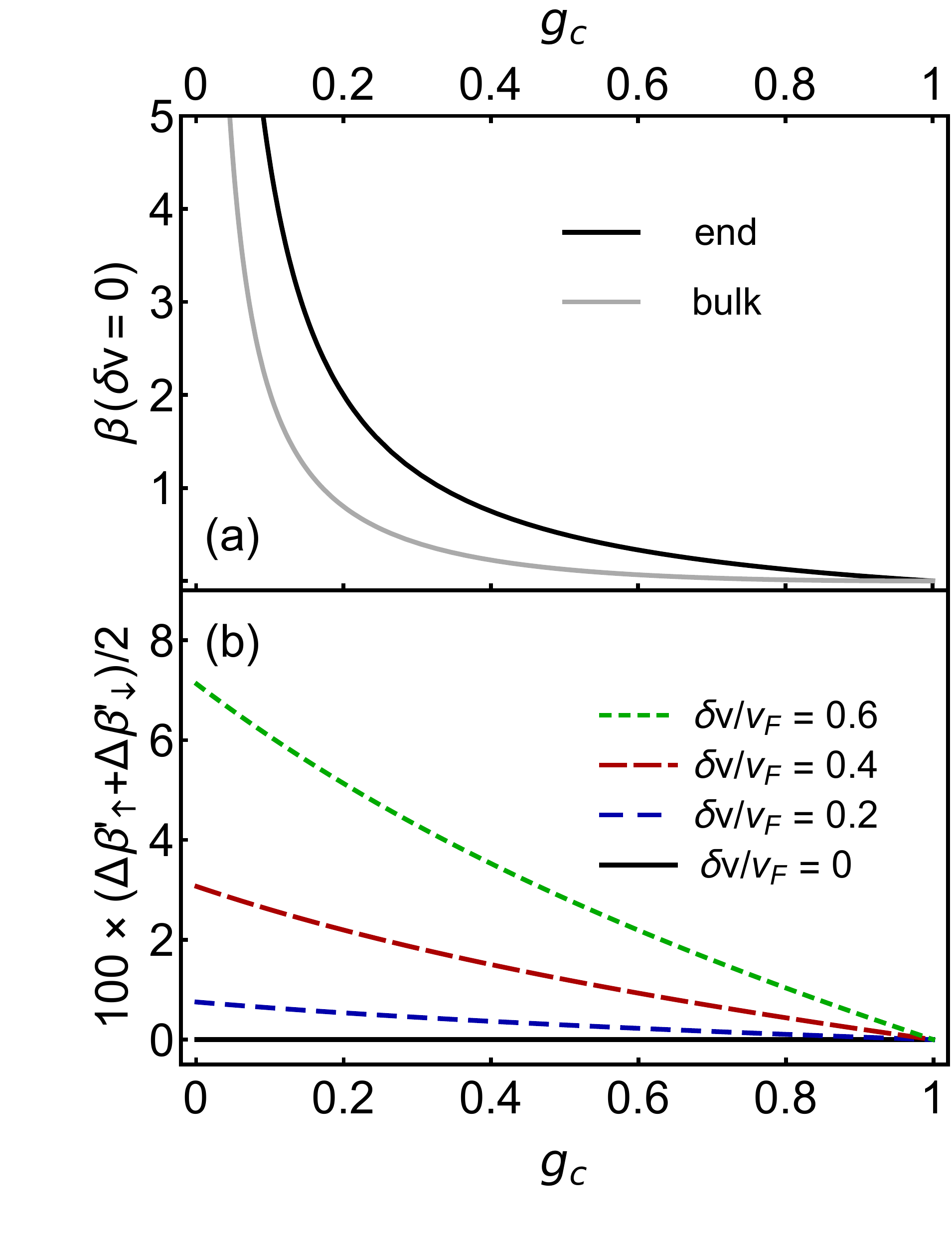}
 \caption{(a) Exponents of the density of states as a function of the interaction parameter ($g_c$) for $g_{s}=1$ and $\delta v=0$.
The black (gray) curve corresponds to the boundary (bulk) exponent $\beta_{\sigma}^{\prime}$ ($\beta_{\sigma}^{\prime~{\rm bulk}} $) given in Eq.~\eqref{Eq:alpha_DOS_end_GZ} [Eq.~\eqref{Eq:DOS_GZ}]. Since we are plotting the zero-spin-orbit value, these curves are identical to those given in Eqs.~\eqref{Eq:DOS} and \eqref{Eq:alpha_DOS_end} and do not depend on the species $r$ or $\sigma$. We therefore neglect the superscript and the subscript. 
(b) Interaction parameter ($g_c$) dependence of the change of the boundary exponent (multiplied by 100 for clarity) with respect to its zero-spin-orbit value for $g_s=1$ and several values of $\delta v /v_F$. 
The curves correspond to the boundary exponent $\beta_{\sigma}^{\prime}$ given in Eq.~\eqref{Eq:alpha_DOS_end_GZ}, upon taking average over $\sigma = 1$ and $\sigma = -1$.
}
 \label{Fig:beta_GZ}
\end{figure}

Diagonalizing the Hamiltonian $H^{\prime}$, we get
\begin{eqnarray}
H^{\prime} &=& \sum_{\nu} \int \frac{\hbar dx}{2\pi} \left[ u_{\nu}^{\prime\prime} g_{\nu}^{\prime\prime} \left( \partial_x \theta_{\nu}^{\prime\prime} \right)^2 + \frac{u_{\nu}^{\prime\prime}}{g_{\nu}^{\prime\prime}} \left( \partial_x \phi_{\nu}^{\prime\prime} \right)^2
\right] , 
\label{Eq:diag_GZ}
\end{eqnarray}
where the new fields are 
\begin{subequations}
\label{Eq:NewFields_GZ}
\begin{eqnarray}
\left( \begin{array}{c} \phi_c^{\prime\prime} \\ \phi_s^{\prime\prime}  \end{array} \right) &=& 
\left( \begin{array}{cc} \cos \theta^{\prime} & g_0^{\prime} \sin \theta^{\prime} \\ -\frac{1}{g_0^{\prime}}\sin \theta^{\prime} & \cos \theta^{\prime}  \end{array} \right) 
\left( \begin{array}{c} \phi_c \\ \phi_s  \end{array} \right), \\
\left( \begin{array}{c} \theta_c^{\prime\prime} \\ \theta_s^{\prime\prime} \end{array} \right) &=& 
\left( \begin{array}{cc} \cos \theta^{\prime} & \frac{1}{g_0^{\prime}} \sin \theta^{\prime} \\ -g_0^{\prime} \sin \theta^{\prime} & \cos \theta^{\prime}  \end{array} \right) 
\left( \begin{array}{c} \theta_c \\ \theta_s \end{array} \right),
\end{eqnarray}
\end{subequations}
with the parameters
\begin{subequations}
\label{Eq:parameters_GZ}
\begin{eqnarray}
g_0^{\prime} &=& \frac{ g_c }{ g_s } \sqrt{ \frac{ 1 + g_s ^2 }{ 1 + g_c^2 } }, \\
\theta^{\prime} &=& \frac{1}{2} \arctan \left( \frac{\delta v }{v_F}  \frac{  g_c g_s \sqrt{(1 + g_s^2) (1 + g_c^2)} }{ g_s^2 - g_c^2} \right).
\end{eqnarray}
\end{subequations}
The modified interaction parameters and velocities are related to the original parameters through
\begin{subequations}
\label{Eq:g_nu_prime_GZ}
\begin{align}
&\frac{u_{c}^{\prime\prime}}{g_{c}^{\prime\prime}} = 
\frac{u_c }{g_c} \cos^2 \theta^{\prime} + \frac{u_s }{g_s (g_0^{\prime})^2}  \sin^2 \theta^{\prime}
+ \frac{\delta v }{2 g_0^{\prime} }  \sin (2 \theta^{\prime}), \\
&\frac{u_{s}^{\prime\prime}}{g_{s}^{\prime\prime}} = 
\frac{u_s }{g_s} \cos^2 \theta^{\prime} + \frac{u_c (g_0^{\prime})^2}{g_c}  \sin^2 \theta^{\prime}
- \frac{\delta v g_0^{\prime}}{2 }  \sin (2 \theta^{\prime}),\\
&u_{c}^{\prime\prime} g_{c}^{\prime\prime} = u_c g_c \cos^2 \theta^{\prime} + u_s g_s (g_0^{\prime})^2 \sin^2 \theta^{\prime}
+ \frac{\delta v g_0^{\prime} }{2 }  \sin (2 \theta^{\prime}), \\
&u_{s}^{\prime\prime} g_{s}^{\prime\prime} = u_s g_s \cos^2 \theta^{\prime} + \frac{ u_c g_c }{ (g_0^{\prime})^2 } \sin^2 \theta^{\prime}
- \frac{\delta v }{2 g_0^{\prime} }  \sin (2 \theta^{\prime}).
\end{align}
\end{subequations}
Following the same procedure as in Appendix~\ref{App:Correlator}, we compute the density of states in the bulk and at the end of the wire. For a given $\sigma$, the former is given by $\rho_{\sigma}^{\rm bulk}(\epsilon) \propto \epsilon^{\beta_{\sigma}^{\prime{\rm bulk}}}$ with the exponent,
\begin{align}
\beta_{\sigma}^{\prime{\rm bulk}} =
\frac{g_{c}^{\prime\prime}}{4}  \left( \cos \theta^\prime + \frac{\sigma}{g_0^\prime} \sin \theta^\prime \right) ^2 
+  \frac{\left( \cos \theta^\prime + \sigma g_0^\prime \sin \theta^\prime \right) ^2}{4g_{c}^{\prime\prime}} \nonumber\\
+ \frac{g_{s}^{\prime\prime}}{4}  \left( \cos \theta^\prime - \sigma g_0^\prime \sin \theta^\prime \right) ^2 
+  \frac{\left( \cos \theta^\prime - \frac{ \sigma }{ g_0^\prime }\sin \theta^\prime \right) ^2}{4g_{s}^{\prime\prime}}  -1.
\label{Eq:DOS_GZ}
\end{align}
On the other hand, at the end of the wire, the density of states becomes $\rho_{\sigma}^{\rm end}(\epsilon) \propto \epsilon^{\beta_{\sigma}^{\prime}}$ with 
\begin{align}
\beta_{\sigma}^{\prime} =
\frac{\left( \cos \theta^\prime + \sigma g_0^\prime \sin \theta^\prime \right) ^2}{2g_{c}^{\prime\prime}}  
+  \frac{\left( \cos \theta^\prime - \frac{ \sigma }{ g_0^\prime }\sin \theta^\prime \right) ^2}{2g_{s}^{\prime\prime}}  -1.
\label{Eq:alpha_DOS_end_GZ}
\end{align}
The exponents in Eqs.~\eqref{Eq:DOS_GZ}--\eqref{Eq:alpha_DOS_end_GZ} are analogous to Eqs.~\eqref{Eq:DOS} and \eqref{Eq:alpha_DOS_end} of the bosonized model introduced in Refs.~\cite{Moroz:2000,Moroz:2000b}. The exponents can be extracted through the density of states measurement using scanning tunneling spectroscopy. Their behavior is displayed in Fig.~\ref{Fig:beta_GZ}. 
Similar to the standard {\TLL}~\cite{Kane:1992,Balents:1999}, the suppression of the density of states is stronger at the end than in the bulk of the wire. In contrast to the strong dependence on the position of the probe [see Fig.~\ref{Fig:beta_GZ}(a)], the spin-orbit-induced change is negligible [see Fig.~\ref{Fig:beta_GZ}(b)].

\section{An alternative approach for the weak-impurity analysis~\label{App:weak}}
Here we discuss an alternative approach for the analysis in the weak-impurity regime.
In Sec.~\ref{SubSec:Weak} we start our analysis by treating a weak impurity as an isolated object (referred to as weak barrier in Ref.~\cite{Kane:1992}), such that it creates a potential which is nonzero only near $x=0$ with the strength $V_{0}$. 
Following Ref.~\cite{Kane:1992} to construct the RG flow equation for $\tilde{V}_0$, we obtain the exponent $\aimp$ of the conductance correction due to a single impurity.
Then, as discussed in Ref.~\cite{Giamarchi:2003}, assuming that the contributions from multiple impurities are additive, many weak impurities lead to a power-law conductance characterized by the parameter $\aimp$.

Alternatively, one can start with random backscattering potential generated by impurities which are not isolated, as in Ref.~\cite{Giamarchi:2003}. In that reference, such disorder is named ``extended disorder'' or ``uniform disorder'' and assumed to be of Gaussian type. Then, one can apply the replica method to average over the disorder and then perform the RG analysis. In this case, there would be additional RG flow equations for the interaction parameters $g_c$ and $g_s$ and velocities, depending on the disorder strength. The additional RG flow equations arise because in this case impurities can affect bulk quantities, in contrast to isolated impurities, which cannot. Importantly for us, the renormalization of $g_c$, $g_s$ due to weak disorder is typically negligible, which would lead to the same power-law conductance in the high-$T$ or high-$V$ regimes as in the isolated-impurities scenario. 
As a result, there would be no significant difference for the power-law conductance, which is the main focus of this work.
We note that, if other phenomena such as localization are concerned, the ``extended disorder'' scenario would better describe the physical picture, as discussed in Ref.~\cite{Giamarchi:2003}.
Nevertheless, to give a better connection between the strong- and weak-impurity cases, we adopt the isolated-impurity picture for our discussion throughout the article.


\begin{thebibliography}{76}%
\makeatletter
\providecommand \@ifxundefined [1]{%
 \@ifx{#1\undefined}
}%
\providecommand \@ifnum [1]{%
 \ifnum #1\expandafter \@firstoftwo
 \else \expandafter \@secondoftwo
 \fi
}%
\providecommand \@ifx [1]{%
 \ifx #1\expandafter \@firstoftwo
 \else \expandafter \@secondoftwo
 \fi
}%
\providecommand \natexlab [1]{#1}%
\providecommand \enquote  [1]{``#1''}%
\providecommand \bibnamefont  [1]{#1}%
\providecommand \bibfnamefont [1]{#1}%
\providecommand \citenamefont [1]{#1}%
\providecommand \href@noop [0]{\@secondoftwo}%
\providecommand \href [0]{\begingroup \@sanitize@url \@href}%
\providecommand \@href[1]{\@@startlink{#1}\@@href}%
\providecommand \@@href[1]{\endgroup#1\@@endlink}%
\providecommand \@sanitize@url [0]{\catcode `\\12\catcode `\$12\catcode
  `\&12\catcode `\#12\catcode `\^12\catcode `\_12\catcode `\%12\relax}%
\providecommand \@@startlink[1]{}%
\providecommand \@@endlink[0]{}%
\providecommand \url  [0]{\begingroup\@sanitize@url \@url }%
\providecommand \@url [1]{\endgroup\@href {#1}{\urlprefix }}%
\providecommand \urlprefix  [0]{URL }%
\providecommand \Eprint [0]{\href }%
\providecommand \doibase [0]{http://dx.doi.org/}%
\providecommand \selectlanguage [0]{\@gobble}%
\providecommand \bibinfo  [0]{\@secondoftwo}%
\providecommand \bibfield  [0]{\@secondoftwo}%
\providecommand \translation [1]{[#1]}%
\providecommand \BibitemOpen [0]{}%
\providecommand \bibitemStop [0]{}%
\providecommand \bibitemNoStop [0]{.\EOS\space}%
\providecommand \EOS [0]{\spacefactor3000\relax}%
\providecommand \BibitemShut  [1]{\csname bibitem#1\endcsname}%
\let\auto@bib@innerbib\@empty
\bibitem [{\citenamefont {Tomonaga}(1950)}]{Tomonaga:1950}%
  \BibitemOpen
  \bibfield  {author} {\bibinfo {author} {\bibfnamefont {S.}~\bibnamefont
  {Tomonaga}},\ }\bibfield  {title} {\bibinfo {title} {{Remarks on Bloch's
  Method of Sound Waves applied to Many-Fermion Problems}},\ }\href {\doibase
  10.1143/ptp/5.4.544} {\bibfield  {journal} {\bibinfo  {journal} {Prog. Theor.
  Phys.}\ }\textbf {\bibinfo {volume} {5}},\ \bibinfo {pages} {544} (\bibinfo
  {year} {1950})}\BibitemShut {NoStop}%
\bibitem [{\citenamefont {Luttinger}(1963)}]{Luttinger:1963}%
  \BibitemOpen
  \bibfield  {author} {\bibinfo {author} {\bibfnamefont {J.~M.}\ \bibnamefont
  {Luttinger}},\ }\bibfield  {title} {\bibinfo {title} {{An Exactly Soluble
  Model of a Many-Fermion System}},\ }\href {\doibase 10.1063/1.1704046}
  {\bibfield  {journal} {\bibinfo  {journal} {J. Math. Phys.}\ }\textbf
  {\bibinfo {volume} {4}},\ \bibinfo {pages} {1154} (\bibinfo {year}
  {1963})}\BibitemShut {NoStop}%
\bibitem [{\citenamefont {Haldane}(1981)}]{Haldane:1981}%
  \BibitemOpen
  \bibfield  {author} {\bibinfo {author} {\bibfnamefont {F.~D.~M.}\
  \bibnamefont {Haldane}},\ }\bibfield  {title} {\bibinfo {title} {{Luttinger
  liquid theory of one-dimensional quantum fluids. I. Properties of the
  Luttinger model and their extension to the general 1D interacting spinless
  Fermi gas}},\ }\href {http://stacks.iop.org/0022-3719/14/i=19/a=010}
  {\bibfield  {journal} {\bibinfo  {journal} {J. Phys. C: Solid State Physics}\
  }\textbf {\bibinfo {volume} {14}},\ \bibinfo {pages} {2585} (\bibinfo {year}
  {1981})}\BibitemShut {NoStop}%
\bibitem [{\citenamefont {Voit}(1995)}]{Voit:1995}%
  \BibitemOpen
  \bibfield  {author} {\bibinfo {author} {\bibfnamefont {J.}~\bibnamefont
  {Voit}},\ }\bibfield  {title} {\bibinfo {title} {{One-dimensional Fermi
  liquids}},\ }\href {http://stacks.iop.org/0034-4885/58/i=9/a=002} {\bibfield
  {journal} {\bibinfo  {journal} {Rep. Prog. Phys.}\ }\textbf {\bibinfo
  {volume} {58}},\ \bibinfo {pages} {977} (\bibinfo {year} {1995})}\BibitemShut
  {NoStop}%
\bibitem [{\citenamefont {von Delft}\ and\ \citenamefont
  {Schoeller}(1998)}]{vonDelft:1998}%
  \BibitemOpen
  \bibfield  {author} {\bibinfo {author} {\bibfnamefont {J.}~\bibnamefont {von
  Delft}}\ and\ \bibinfo {author} {\bibfnamefont {H.}~\bibnamefont
  {Schoeller}},\ }\bibfield  {title} {\bibinfo {title} {{Bosonization for
  beginners--refermionization for experts}},\ }\href {\doibase
  10.1002/(SICI)1521-3889(199811)7:4<225::AID-ANDP225>3.0.CO;2-L} {\bibfield
  {journal} {\bibinfo  {journal} {Ann. Phys.}\ }\textbf {\bibinfo {volume}
  {7}},\ \bibinfo {pages} {225} (\bibinfo {year} {1998})}\BibitemShut {NoStop}%
\bibitem [{\citenamefont {Giamarchi}(2003)}]{Giamarchi:2003}%
  \BibitemOpen
  \bibfield  {author} {\bibinfo {author} {\bibfnamefont {T.}~\bibnamefont
  {Giamarchi}},\ }\href@noop {} {\emph {\bibinfo {title} {{Quantum Physics in
  One Dimension}}}}\ (\bibinfo  {publisher} {Oxford University Press, New
  York},\ \bibinfo {year} {2003})\BibitemShut {NoStop}%
\bibitem [{\citenamefont {Maslov}\ and\ \citenamefont
  {Stone}(1995)}]{Maslov:1995}%
  \BibitemOpen
  \bibfield  {author} {\bibinfo {author} {\bibfnamefont {D.~L.}\ \bibnamefont
  {Maslov}}\ and\ \bibinfo {author} {\bibfnamefont {M.}~\bibnamefont {Stone}},\
  }\bibfield  {title} {\bibinfo {title} {{Landauer conductance of Luttinger
  liquids with leads}},\ }\href {\doibase 10.1103/PhysRevB.52.R5539} {\bibfield
   {journal} {\bibinfo  {journal} {Phys. Rev. B}\ }\textbf {\bibinfo {volume}
  {52}},\ \bibinfo {pages} {R5539} (\bibinfo {year} {1995})}\BibitemShut
  {NoStop}%
\bibitem [{\citenamefont {Ponomarenko}(1995)}]{Ponomarenko:1995}%
  \BibitemOpen
  \bibfield  {author} {\bibinfo {author} {\bibfnamefont {V.~V.}\ \bibnamefont
  {Ponomarenko}},\ }\bibfield  {title} {\bibinfo {title} {{Renormalization of
  the one-dimensional conductance in the Luttinger-liquid model}},\ }\href
  {\doibase 10.1103/PhysRevB.52.R8666} {\bibfield  {journal} {\bibinfo
  {journal} {Phys. Rev. B}\ }\textbf {\bibinfo {volume} {52}},\ \bibinfo
  {pages} {R8666} (\bibinfo {year} {1995})}\BibitemShut {NoStop}%
\bibitem [{\citenamefont {Safi}\ and\ \citenamefont
  {Schulz}(1995)}]{Safi:1995}%
  \BibitemOpen
  \bibfield  {author} {\bibinfo {author} {\bibfnamefont {I.}~\bibnamefont
  {Safi}}\ and\ \bibinfo {author} {\bibfnamefont {H.~J.}\ \bibnamefont
  {Schulz}},\ }\bibfield  {title} {\bibinfo {title} {{Transport in an
  inhomogeneous interacting one-dimensional system}},\ }\href {\doibase
  10.1103/PhysRevB.52.R17040} {\bibfield  {journal} {\bibinfo  {journal} {Phys.
  Rev. B}\ }\textbf {\bibinfo {volume} {52}},\ \bibinfo {pages} {R17040}
  (\bibinfo {year} {1995})}\BibitemShut {NoStop}%
\bibitem [{\citenamefont {Kane}\ and\ \citenamefont
  {Fisher}(1992{\natexlab{a}})}]{Kane:1992a}%
  \BibitemOpen
  \bibfield  {author} {\bibinfo {author} {\bibfnamefont {C.~L.}\ \bibnamefont
  {Kane}}\ and\ \bibinfo {author} {\bibfnamefont {M.~P.~A.}\ \bibnamefont
  {Fisher}},\ }\bibfield  {title} {\bibinfo {title} {{Transport in a
  one-channel Luttinger liquid}},\ }\href {\doibase
  10.1103/PhysRevLett.68.1220} {\bibfield  {journal} {\bibinfo  {journal}
  {Phys. Rev. Lett.}\ }\textbf {\bibinfo {volume} {68}},\ \bibinfo {pages}
  {1220} (\bibinfo {year} {1992}{\natexlab{a}})}\BibitemShut {NoStop}%
\bibitem [{\citenamefont {Kane}\ and\ \citenamefont
  {Fisher}(1992{\natexlab{b}})}]{Kane:1992}%
  \BibitemOpen
  \bibfield  {author} {\bibinfo {author} {\bibfnamefont {C.~L.}\ \bibnamefont
  {Kane}}\ and\ \bibinfo {author} {\bibfnamefont {M.~P.~A.}\ \bibnamefont
  {Fisher}},\ }\bibfield  {title} {\bibinfo {title} {{Transmission through
  barriers and resonant tunneling in an interacting one-dimensional electron
  gas}},\ }\href {\doibase 10.1103/PhysRevB.46.15233} {\bibfield  {journal}
  {\bibinfo  {journal} {Phys. Rev. B}\ }\textbf {\bibinfo {volume} {46}},\
  \bibinfo {pages} {15233} (\bibinfo {year} {1992}{\natexlab{b}})}\BibitemShut
  {NoStop}%
\bibitem [{\citenamefont {Furusaki}\ and\ \citenamefont
  {Nagaosa}(1993)}]{Furusaki:1993}%
  \BibitemOpen
  \bibfield  {author} {\bibinfo {author} {\bibfnamefont {A.}~\bibnamefont
  {Furusaki}}\ and\ \bibinfo {author} {\bibfnamefont {N.}~\bibnamefont
  {Nagaosa}},\ }\bibfield  {title} {\bibinfo {title} {{Single-barrier problem
  and Anderson localization in a one-dimensional interacting electron
  system}},\ }\href {\doibase 10.1103/PhysRevB.47.4631} {\bibfield  {journal}
  {\bibinfo  {journal} {Phys. Rev. B}\ }\textbf {\bibinfo {volume} {47}},\
  \bibinfo {pages} {4631} (\bibinfo {year} {1993})}\BibitemShut {NoStop}%
\bibitem [{\citenamefont {Matveev}\ and\ \citenamefont
  {Glazman}(1993)}]{Matveev:1993}%
  \BibitemOpen
  \bibfield  {author} {\bibinfo {author} {\bibfnamefont {K.~A.}\ \bibnamefont
  {Matveev}}\ and\ \bibinfo {author} {\bibfnamefont {L.~I.}\ \bibnamefont
  {Glazman}},\ }\bibfield  {title} {\bibinfo {title} {{Coulomb blockade of
  tunneling into a quasi-one-dimensional wire}},\ }\href {\doibase
  10.1103/PhysRevLett.70.990} {\bibfield  {journal} {\bibinfo  {journal} {Phys.
  Rev. Lett.}\ }\textbf {\bibinfo {volume} {70}},\ \bibinfo {pages} {990}
  (\bibinfo {year} {1993})}\BibitemShut {NoStop}%
\bibitem [{\citenamefont {Maslov}(1995)}]{Maslov:1995b}%
  \BibitemOpen
  \bibfield  {author} {\bibinfo {author} {\bibfnamefont {D.~L.}\ \bibnamefont
  {Maslov}},\ }\bibfield  {title} {\bibinfo {title} {{Transport through dirty
  Luttinger liquids connected to reservoirs}},\ }\href {\doibase
  10.1103/PhysRevB.52.R14368} {\bibfield  {journal} {\bibinfo  {journal} {Phys.
  Rev. B}\ }\textbf {\bibinfo {volume} {52}},\ \bibinfo {pages} {R14368}
  (\bibinfo {year} {1995})}\BibitemShut {NoStop}%
\bibitem [{\citenamefont {Sandler}\ and\ \citenamefont
  {Maslov}(1997)}]{Sandler:1997}%
  \BibitemOpen
  \bibfield  {author} {\bibinfo {author} {\bibfnamefont {N.~P.}\ \bibnamefont
  {Sandler}}\ and\ \bibinfo {author} {\bibfnamefont {D.~L.}\ \bibnamefont
  {Maslov}},\ }\bibfield  {title} {\bibinfo {title} {{Interactions and disorder
  in multichannel quantum wires}},\ }\href {\doibase 10.1103/PhysRevB.55.13808}
  {\bibfield  {journal} {\bibinfo  {journal} {Phys. Rev. B}\ }\textbf {\bibinfo
  {volume} {55}},\ \bibinfo {pages} {13808} (\bibinfo {year}
  {1997})}\BibitemShut {NoStop}%
\bibitem [{\citenamefont {{Balents}}(1999)}]{Balents:1999}%
  \BibitemOpen
  \bibfield  {author} {\bibinfo {author} {\bibfnamefont {L.}~\bibnamefont
  {{Balents}}},\ }\bibfield  {title} {\bibinfo {title} {{Orthogonality
  Catastrophes in Carbon Nanotubes}},\ }\href@noop {} {\bibfield  {journal}
  {\bibinfo  {journal} {eprint arXiv:cond-mat/9906032}\ } (\bibinfo {year}
  {1999})},\ \Eprint {http://arxiv.org/abs/cond-mat/9906032} {cond-mat/9906032}
  \BibitemShut {NoStop}%
\bibitem [{\citenamefont {Tarucha}\ \emph {et~al.}(1995)\citenamefont
  {Tarucha}, \citenamefont {Honda},\ and\ \citenamefont {Saku}}]{Tarucha:1995}%
  \BibitemOpen
  \bibfield  {author} {\bibinfo {author} {\bibfnamefont {S.}~\bibnamefont
  {Tarucha}}, \bibinfo {author} {\bibfnamefont {T.}~\bibnamefont {Honda}}, \
  and\ \bibinfo {author} {\bibfnamefont {T.}~\bibnamefont {Saku}},\ }\bibfield
  {title} {\bibinfo {title} {{Reduction of quantized conductance at low
  temperatures observed in 2 to 10 $\mu$m-long quantum wires}},\ }\href
  {\doibase https://doi.org/10.1016/0038-1098(95)00102-6} {\bibfield  {journal}
  {\bibinfo  {journal} {Solid State Commun.}\ }\textbf {\bibinfo {volume}
  {94}},\ \bibinfo {pages} {413} (\bibinfo {year} {1995})}\BibitemShut
  {NoStop}%
\bibitem [{\citenamefont {Datta}\ and\ \citenamefont {Das}(1990)}]{Datta:1990}%
  \BibitemOpen
  \bibfield  {author} {\bibinfo {author} {\bibfnamefont {S.}~\bibnamefont
  {Datta}}\ and\ \bibinfo {author} {\bibfnamefont {B.}~\bibnamefont {Das}},\
  }\bibfield  {title} {\bibinfo {title} {{Electronic analog of the
  electro-optic modulator}},\ }\href {\doibase 10.1063/1.102730} {\bibfield
  {journal} {\bibinfo  {journal} {Appl. Phys. Lett.}\ }\textbf {\bibinfo
  {volume} {56}},\ \bibinfo {pages} {665} (\bibinfo {year} {1990})}\BibitemShut
  {NoStop}%
\bibitem [{\citenamefont {Nitta}\ \emph {et~al.}(1997)\citenamefont {Nitta},
  \citenamefont {Akazaki}, \citenamefont {Takayanagi},\ and\ \citenamefont
  {Enoki}}]{Nitta:1997}%
  \BibitemOpen
  \bibfield  {author} {\bibinfo {author} {\bibfnamefont {J.}~\bibnamefont
  {Nitta}}, \bibinfo {author} {\bibfnamefont {T.}~\bibnamefont {Akazaki}},
  \bibinfo {author} {\bibfnamefont {H.}~\bibnamefont {Takayanagi}}, \ and\
  \bibinfo {author} {\bibfnamefont {T.}~\bibnamefont {Enoki}},\ }\bibfield
  {title} {\bibinfo {title} {{Gate Control of Spin-Orbit Interaction in an
  Inverted
  I${\mathrm{n}}_{0.53}$G${\mathrm{a}}_{0.47}$As/I${\mathrm{n}}_{0.52}$A${\mathrm{l}}_{0.48}$As
  Heterostructure}},\ }\href {\doibase 10.1103/PhysRevLett.78.1335} {\bibfield
  {journal} {\bibinfo  {journal} {Phys. Rev. Lett.}\ }\textbf {\bibinfo
  {volume} {78}},\ \bibinfo {pages} {1335} (\bibinfo {year}
  {1997})}\BibitemShut {NoStop}%
\bibitem [{\citenamefont {Fasth}\ \emph {et~al.}(2007)\citenamefont {Fasth},
  \citenamefont {Fuhrer}, \citenamefont {Samuelson}, \citenamefont {Golovach},\
  and\ \citenamefont {Loss}}]{Fasth:2007}%
  \BibitemOpen
  \bibfield  {author} {\bibinfo {author} {\bibfnamefont {C.}~\bibnamefont
  {Fasth}}, \bibinfo {author} {\bibfnamefont {A.}~\bibnamefont {Fuhrer}},
  \bibinfo {author} {\bibfnamefont {L.}~\bibnamefont {Samuelson}}, \bibinfo
  {author} {\bibfnamefont {V.~N.}\ \bibnamefont {Golovach}}, \ and\ \bibinfo
  {author} {\bibfnamefont {D.}~\bibnamefont {Loss}},\ }\bibfield  {title}
  {\bibinfo {title} {{Direct Measurement of the Spin-Orbit Interaction in a
  Two-Electron InAs Nanowire Quantum Dot}},\ }\href {\doibase
  10.1103/PhysRevLett.98.266801} {\bibfield  {journal} {\bibinfo  {journal}
  {Phys. Rev. Lett.}\ }\textbf {\bibinfo {volume} {98}},\ \bibinfo {pages}
  {266801} (\bibinfo {year} {2007})}\BibitemShut {NoStop}%
\bibitem [{\citenamefont {Fabian}\ \emph {et~al.}(2007)\citenamefont {Fabian},
  \citenamefont {Matos-Abiague}, \citenamefont {Ertler}, \citenamefont
  {Stano},\ and\ \citenamefont {{\v{Z}}uti{\'{c}}}}]{Fabian:2007}%
  \BibitemOpen
  \bibfield  {author} {\bibinfo {author} {\bibfnamefont {J.}~\bibnamefont
  {Fabian}}, \bibinfo {author} {\bibfnamefont {A.}~\bibnamefont
  {Matos-Abiague}}, \bibinfo {author} {\bibfnamefont {C.}~\bibnamefont
  {Ertler}}, \bibinfo {author} {\bibfnamefont {P.}~\bibnamefont {Stano}}, \
  and\ \bibinfo {author} {\bibfnamefont {I.}~\bibnamefont
  {{\v{Z}}uti{\'{c}}}},\ }\bibfield  {title} {\bibinfo {title} {{Semiconductor
  spintronics}},\ }\href {\doibase 10.2478/v10155-010-0086-8} {\bibfield
  {journal} {\bibinfo  {journal} {Acta Phys. Slovaca}\ }\textbf {\bibinfo
  {volume} {57}},\ \bibinfo {pages} {565} (\bibinfo {year} {2007})}\BibitemShut
  {NoStop}%
\bibitem [{\citenamefont {Nadj-Perge}\ \emph {et~al.}(2010)\citenamefont
  {Nadj-Perge}, \citenamefont {Frolov}, \citenamefont {Bakkers},\ and\
  \citenamefont {Kouwenhoven}}]{Nadj-Perge:2010}%
  \BibitemOpen
  \bibfield  {author} {\bibinfo {author} {\bibfnamefont {S.}~\bibnamefont
  {Nadj-Perge}}, \bibinfo {author} {\bibfnamefont {S.~M.}\ \bibnamefont
  {Frolov}}, \bibinfo {author} {\bibfnamefont {E.~P. A.~M.}\ \bibnamefont
  {Bakkers}}, \ and\ \bibinfo {author} {\bibfnamefont {L.~P.}\ \bibnamefont
  {Kouwenhoven}},\ }\bibfield  {title} {\bibinfo {title} {{Spin-orbit qubit in
  a semiconductor nanowire}},\ }\href@noop {} {\bibfield  {journal} {\bibinfo
  {journal} {Nature}\ }\textbf {\bibinfo {volume} {468}},\ \bibinfo {pages}
  {1084} (\bibinfo {year} {2010})}\BibitemShut {NoStop}%
\bibitem [{\citenamefont {Lutchyn}\ \emph {et~al.}(2010)\citenamefont
  {Lutchyn}, \citenamefont {Sau},\ and\ \citenamefont
  {Das~Sarma}}]{Lutchyn:2010}%
  \BibitemOpen
  \bibfield  {author} {\bibinfo {author} {\bibfnamefont {R.~M.}\ \bibnamefont
  {Lutchyn}}, \bibinfo {author} {\bibfnamefont {J.~D.}\ \bibnamefont {Sau}}, \
  and\ \bibinfo {author} {\bibfnamefont {S.}~\bibnamefont {Das~Sarma}},\
  }\bibfield  {title} {\bibinfo {title} {{Majorana Fermions and a Topological
  Phase Transition in Semiconductor-Superconductor Heterostructures}},\ }\href
  {\doibase 10.1103/PhysRevLett.105.077001} {\bibfield  {journal} {\bibinfo
  {journal} {Phys. Rev. Lett.}\ }\textbf {\bibinfo {volume} {105}},\ \bibinfo
  {pages} {077001} (\bibinfo {year} {2010})}\BibitemShut {NoStop}%
\bibitem [{\citenamefont {Oreg}\ \emph {et~al.}(2010)\citenamefont {Oreg},
  \citenamefont {Refael},\ and\ \citenamefont {von Oppen}}]{Oreg:2010}%
  \BibitemOpen
  \bibfield  {author} {\bibinfo {author} {\bibfnamefont {Y.}~\bibnamefont
  {Oreg}}, \bibinfo {author} {\bibfnamefont {G.}~\bibnamefont {Refael}}, \ and\
  \bibinfo {author} {\bibfnamefont {F.}~\bibnamefont {von Oppen}},\ }\bibfield
  {title} {\bibinfo {title} {{Helical Liquids and Majorana Bound States in
  Quantum Wires}},\ }\href {\doibase 10.1103/PhysRevLett.105.177002} {\bibfield
   {journal} {\bibinfo  {journal} {Phys. Rev. Lett.}\ }\textbf {\bibinfo
  {volume} {105}},\ \bibinfo {pages} {177002} (\bibinfo {year}
  {2010})}\BibitemShut {NoStop}%
\bibitem [{\citenamefont {Mourik}\ \emph {et~al.}(2012)\citenamefont {Mourik},
  \citenamefont {Zuo}, \citenamefont {Frolov}, \citenamefont {Plissard},
  \citenamefont {Bakkers},\ and\ \citenamefont {Kouwenhoven}}]{Mourik:2012}%
  \BibitemOpen
  \bibfield  {author} {\bibinfo {author} {\bibfnamefont {V.}~\bibnamefont
  {Mourik}}, \bibinfo {author} {\bibfnamefont {K.}~\bibnamefont {Zuo}},
  \bibinfo {author} {\bibfnamefont {S.~M.}\ \bibnamefont {Frolov}}, \bibinfo
  {author} {\bibfnamefont {S.~R.}\ \bibnamefont {Plissard}}, \bibinfo {author}
  {\bibfnamefont {E.~P. A.~M.}\ \bibnamefont {Bakkers}}, \ and\ \bibinfo
  {author} {\bibfnamefont {L.~P.}\ \bibnamefont {Kouwenhoven}},\ }\bibfield
  {title} {\bibinfo {title} {{Signatures of Majorana Fermions in Hybrid
  Superconductor-Semiconductor Nanowire Devices}},\ }\href {\doibase
  10.1126/science.1222360} {\bibfield  {journal} {\bibinfo  {journal}
  {Science}\ }\textbf {\bibinfo {volume} {336}},\ \bibinfo {pages} {1003}
  (\bibinfo {year} {2012})}\BibitemShut {NoStop}%
\bibitem [{\citenamefont {Das}\ \emph {et~al.}(2012)\citenamefont {Das},
  \citenamefont {Ronen}, \citenamefont {Most}, \citenamefont {Oreg},
  \citenamefont {Heiblum},\ and\ \citenamefont {Shtrikman}}]{Das:2012}%
  \BibitemOpen
  \bibfield  {author} {\bibinfo {author} {\bibfnamefont {A.}~\bibnamefont
  {Das}}, \bibinfo {author} {\bibfnamefont {Y.}~\bibnamefont {Ronen}}, \bibinfo
  {author} {\bibfnamefont {Y.}~\bibnamefont {Most}}, \bibinfo {author}
  {\bibfnamefont {Y.}~\bibnamefont {Oreg}}, \bibinfo {author} {\bibfnamefont
  {M.}~\bibnamefont {Heiblum}}, \ and\ \bibinfo {author} {\bibfnamefont
  {H.}~\bibnamefont {Shtrikman}},\ }\bibfield  {title} {\bibinfo {title}
  {{Zero-bias peaks and splitting in an {I}n{A}s nanowire topological
  superconductor as a signature of {M}ajorana fermions}},\ }\href {\doibase
  10.1038/nphys2479} {\bibfield  {journal} {\bibinfo  {journal} {Nat. Phys.}\
  }\textbf {\bibinfo {volume} {8}},\ \bibinfo {pages} {887} (\bibinfo {year}
  {2012})}\BibitemShut {NoStop}%
\bibitem [{\citenamefont {Deng}\ \emph {et~al.}(2012)\citenamefont {Deng},
  \citenamefont {Yu}, \citenamefont {Huang}, \citenamefont {Larsson},
  \citenamefont {Caroff},\ and\ \citenamefont {Xu}}]{Deng:2012}%
  \BibitemOpen
  \bibfield  {author} {\bibinfo {author} {\bibfnamefont {M.~T.}\ \bibnamefont
  {Deng}}, \bibinfo {author} {\bibfnamefont {C.~L.}\ \bibnamefont {Yu}},
  \bibinfo {author} {\bibfnamefont {G.~Y.}\ \bibnamefont {Huang}}, \bibinfo
  {author} {\bibfnamefont {M.}~\bibnamefont {Larsson}}, \bibinfo {author}
  {\bibfnamefont {P.}~\bibnamefont {Caroff}}, \ and\ \bibinfo {author}
  {\bibfnamefont {H.~Q.}\ \bibnamefont {Xu}},\ }\bibfield  {title} {\bibinfo
  {title} {{Anomalous Zero-Bias Conductance Peak in a Nb--InSb Nanowire--Nb
  Hybrid Device}},\ }\href {\doibase 10.1021/nl303758w} {\bibfield  {journal}
  {\bibinfo  {journal} {Nano Lett.}\ }\textbf {\bibinfo {volume} {12}},\
  \bibinfo {pages} {6414} (\bibinfo {year} {2012})}\BibitemShut {NoStop}%
\bibitem [{\citenamefont {Rokhinson}\ \emph {et~al.}(2012)\citenamefont
  {Rokhinson}, \citenamefont {Liu},\ and\ \citenamefont
  {Furdyna}}]{Rokhinson:2012}%
  \BibitemOpen
  \bibfield  {author} {\bibinfo {author} {\bibfnamefont {L.~P.}\ \bibnamefont
  {Rokhinson}}, \bibinfo {author} {\bibfnamefont {X.}~\bibnamefont {Liu}}, \
  and\ \bibinfo {author} {\bibfnamefont {J.~K.}\ \bibnamefont {Furdyna}},\
  }\bibfield  {title} {\bibinfo {title} {{The fractional a.c. Josephson effect
  in a semiconductor--superconductor nanowire as a signature of Majorana
  particles}},\ }\href {\doibase 10.1038/nphys2429} {\bibfield  {journal}
  {\bibinfo  {journal} {Nat. Phys.}\ }\textbf {\bibinfo {volume} {8}},\
  \bibinfo {pages} {795} (\bibinfo {year} {2012})}\BibitemShut {NoStop}%
\bibitem [{\citenamefont {Churchill}\ \emph {et~al.}(2013)\citenamefont
  {Churchill}, \citenamefont {Fatemi}, \citenamefont {Grove-Rasmussen},
  \citenamefont {Deng}, \citenamefont {Caroff}, \citenamefont {Xu},\ and\
  \citenamefont {Marcus}}]{Churchill:2013}%
  \BibitemOpen
  \bibfield  {author} {\bibinfo {author} {\bibfnamefont {H.~O.~H.}\
  \bibnamefont {Churchill}}, \bibinfo {author} {\bibfnamefont {V.}~\bibnamefont
  {Fatemi}}, \bibinfo {author} {\bibfnamefont {K.}~\bibnamefont
  {Grove-Rasmussen}}, \bibinfo {author} {\bibfnamefont {M.~T.}\ \bibnamefont
  {Deng}}, \bibinfo {author} {\bibfnamefont {P.}~\bibnamefont {Caroff}},
  \bibinfo {author} {\bibfnamefont {H.~Q.}\ \bibnamefont {Xu}}, \ and\ \bibinfo
  {author} {\bibfnamefont {C.~M.}\ \bibnamefont {Marcus}},\ }\bibfield  {title}
  {\bibinfo {title} {{Superconductor-nanowire devices from tunneling to the
  multichannel regime: Zero-bias oscillations and magnetoconductance
  crossover}},\ }\href {\doibase 10.1103/PhysRevB.87.241401} {\bibfield
  {journal} {\bibinfo  {journal} {Phys. Rev. B}\ }\textbf {\bibinfo {volume}
  {87}},\ \bibinfo {pages} {241401(R)} (\bibinfo {year} {2013})}\BibitemShut
  {NoStop}%
\bibitem [{\citenamefont {Gaidamauskas}\ \emph {et~al.}(2014)\citenamefont
  {Gaidamauskas}, \citenamefont {Paaske},\ and\ \citenamefont
  {Flensberg}}]{Gaidamauskas:2014}%
  \BibitemOpen
  \bibfield  {author} {\bibinfo {author} {\bibfnamefont {E.}~\bibnamefont
  {Gaidamauskas}}, \bibinfo {author} {\bibfnamefont {J.}~\bibnamefont
  {Paaske}}, \ and\ \bibinfo {author} {\bibfnamefont {K.}~\bibnamefont
  {Flensberg}},\ }\bibfield  {title} {\bibinfo {title} {{Majorana Bound States
  in Two-Channel Time-Reversal-Symmetric Nanowire Systems}},\ }\href {\doibase
  10.1103/PhysRevLett.112.126402} {\bibfield  {journal} {\bibinfo  {journal}
  {Phys. Rev. Lett.}\ }\textbf {\bibinfo {volume} {112}},\ \bibinfo {pages}
  {126402} (\bibinfo {year} {2014})}\BibitemShut {NoStop}%
\bibitem [{\citenamefont {Ebisu}\ \emph {et~al.}(2016)\citenamefont {Ebisu},
  \citenamefont {Lu}, \citenamefont {Klinovaja},\ and\ \citenamefont
  {Tanaka}}]{Ebisu:2016}%
  \BibitemOpen
  \bibfield  {author} {\bibinfo {author} {\bibfnamefont {H.}~\bibnamefont
  {Ebisu}}, \bibinfo {author} {\bibfnamefont {B.}~\bibnamefont {Lu}}, \bibinfo
  {author} {\bibfnamefont {J.}~\bibnamefont {Klinovaja}}, \ and\ \bibinfo
  {author} {\bibfnamefont {Y.}~\bibnamefont {Tanaka}},\ }\bibfield  {title}
  {\bibinfo {title} {{Theory of time-reversal topological superconductivity in
  double Rashba wires: symmetries of Cooper pairs and Andreev bound states}},\
  }\href {\doibase 10.1093/ptep/ptw094} {\bibfield  {journal} {\bibinfo
  {journal} {Prog. Theo. Exp. Phys.}\ }\textbf {\bibinfo {volume} {2016}},\
  \bibinfo {pages} {083I01} (\bibinfo {year} {2016})}\BibitemShut {NoStop}%
\bibitem [{\citenamefont {Reeg}\ \emph {et~al.}(2017)\citenamefont {Reeg},
  \citenamefont {Klinovaja},\ and\ \citenamefont {Loss}}]{Reeg:2017}%
  \BibitemOpen
  \bibfield  {author} {\bibinfo {author} {\bibfnamefont {C.}~\bibnamefont
  {Reeg}}, \bibinfo {author} {\bibfnamefont {J.}~\bibnamefont {Klinovaja}}, \
  and\ \bibinfo {author} {\bibfnamefont {D.}~\bibnamefont {Loss}},\ }\bibfield
  {title} {\bibinfo {title} {{Destructive interference of direct and crossed
  {A}ndreev pairing in a system of two nanowires coupled via an $s$-wave
  superconductor}},\ }\href {\doibase 10.1103/PhysRevB.96.081301} {\bibfield
  {journal} {\bibinfo  {journal} {Phys. Rev. B}\ }\textbf {\bibinfo {volume}
  {96}},\ \bibinfo {pages} {081301(R)} (\bibinfo {year} {2017})}\BibitemShut
  {NoStop}%
\bibitem [{\citenamefont {Schrade}\ \emph {et~al.}(2017)\citenamefont
  {Schrade}, \citenamefont {Thakurathi}, \citenamefont {Reeg}, \citenamefont
  {Hoffman}, \citenamefont {Klinovaja},\ and\ \citenamefont
  {Loss}}]{Schrade:2017}%
  \BibitemOpen
  \bibfield  {author} {\bibinfo {author} {\bibfnamefont {C.}~\bibnamefont
  {Schrade}}, \bibinfo {author} {\bibfnamefont {M.}~\bibnamefont {Thakurathi}},
  \bibinfo {author} {\bibfnamefont {C.}~\bibnamefont {Reeg}}, \bibinfo {author}
  {\bibfnamefont {S.}~\bibnamefont {Hoffman}}, \bibinfo {author} {\bibfnamefont
  {J.}~\bibnamefont {Klinovaja}}, \ and\ \bibinfo {author} {\bibfnamefont
  {D.}~\bibnamefont {Loss}},\ }\bibfield  {title} {\bibinfo {title} {{Low-field
  topological threshold in Majorana double nanowires}},\ }\href {\doibase
  10.1103/PhysRevB.96.035306} {\bibfield  {journal} {\bibinfo  {journal} {Phys.
  Rev. B}\ }\textbf {\bibinfo {volume} {96}},\ \bibinfo {pages} {035306}
  (\bibinfo {year} {2017})}\BibitemShut {NoStop}%
\bibitem [{\citenamefont {Thakurathi}\ \emph {et~al.}(2018)\citenamefont
  {Thakurathi}, \citenamefont {Simon}, \citenamefont {Mandal}, \citenamefont
  {Klinovaja},\ and\ \citenamefont {Loss}}]{Thakurathi:2018}%
  \BibitemOpen
  \bibfield  {author} {\bibinfo {author} {\bibfnamefont {M.}~\bibnamefont
  {Thakurathi}}, \bibinfo {author} {\bibfnamefont {P.}~\bibnamefont {Simon}},
  \bibinfo {author} {\bibfnamefont {I.}~\bibnamefont {Mandal}}, \bibinfo
  {author} {\bibfnamefont {J.}~\bibnamefont {Klinovaja}}, \ and\ \bibinfo
  {author} {\bibfnamefont {D.}~\bibnamefont {Loss}},\ }\bibfield  {title}
  {\bibinfo {title} {{{M}ajorana {K}ramers pairs in {R}ashba double nanowires
  with interactions and disorder}},\ }\href {\doibase
  10.1103/PhysRevB.97.045415} {\bibfield  {journal} {\bibinfo  {journal} {Phys.
  Rev. B}\ }\textbf {\bibinfo {volume} {97}},\ \bibinfo {pages} {045415}
  (\bibinfo {year} {2018})}\BibitemShut {NoStop}%
\bibitem [{\citenamefont {Klinovaja}\ and\ \citenamefont
  {Loss}(2014{\natexlab{a}})}]{Klinovaja:2014a}%
  \BibitemOpen
  \bibfield  {author} {\bibinfo {author} {\bibfnamefont {J.}~\bibnamefont
  {Klinovaja}}\ and\ \bibinfo {author} {\bibfnamefont {D.}~\bibnamefont
  {Loss}},\ }\bibfield  {title} {\bibinfo {title} {{Parafermions in an
  Interacting Nanowire Bundle}},\ }\href {\doibase
  10.1103/PhysRevLett.112.246403} {\bibfield  {journal} {\bibinfo  {journal}
  {Phys. Rev. Lett.}\ }\textbf {\bibinfo {volume} {112}},\ \bibinfo {pages}
  {246403} (\bibinfo {year} {2014}{\natexlab{a}})}\BibitemShut {NoStop}%
\bibitem [{\citenamefont {Klinovaja}\ and\ \citenamefont
  {Loss}(2014{\natexlab{b}})}]{Klinovaja:2014}%
  \BibitemOpen
  \bibfield  {author} {\bibinfo {author} {\bibfnamefont {J.}~\bibnamefont
  {Klinovaja}}\ and\ \bibinfo {author} {\bibfnamefont {D.}~\bibnamefont
  {Loss}},\ }\bibfield  {title} {\bibinfo {title} {{Time-reversal invariant
  parafermions in interacting Rashba nanowires}},\ }\href {\doibase
  10.1103/PhysRevB.90.045118} {\bibfield  {journal} {\bibinfo  {journal} {Phys.
  Rev. B}\ }\textbf {\bibinfo {volume} {90}},\ \bibinfo {pages} {045118}
  (\bibinfo {year} {2014}{\natexlab{b}})}\BibitemShut {NoStop}%
\bibitem [{\citenamefont {Oreg}\ \emph {et~al.}(2014)\citenamefont {Oreg},
  \citenamefont {Sela},\ and\ \citenamefont {Stern}}]{Oreg:2014}%
  \BibitemOpen
  \bibfield  {author} {\bibinfo {author} {\bibfnamefont {Y.}~\bibnamefont
  {Oreg}}, \bibinfo {author} {\bibfnamefont {E.}~\bibnamefont {Sela}}, \ and\
  \bibinfo {author} {\bibfnamefont {A.}~\bibnamefont {Stern}},\ }\bibfield
  {title} {\bibinfo {title} {{Fractional helical liquids in quantum wires}},\
  }\href {\doibase 10.1103/PhysRevB.89.115402} {\bibfield  {journal} {\bibinfo
  {journal} {Phys. Rev. B}\ }\textbf {\bibinfo {volume} {89}},\ \bibinfo
  {pages} {115402} (\bibinfo {year} {2014})}\BibitemShut {NoStop}%
\bibitem [{\citenamefont {Alicea}\ and\ \citenamefont
  {Fendley}(2016)}]{Alicea:2016}%
  \BibitemOpen
  \bibfield  {author} {\bibinfo {author} {\bibfnamefont {J.}~\bibnamefont
  {Alicea}}\ and\ \bibinfo {author} {\bibfnamefont {P.}~\bibnamefont
  {Fendley}},\ }\bibfield  {title} {\bibinfo {title} {{Topological Phases with
  Parafermions: Theory and Blueprints}},\ }\href {\doibase
  10.1146/annurev-conmatphys-031115-011336} {\bibfield  {journal} {\bibinfo
  {journal} {Ann. Rev. Condens. Matter Phys.}\ }\textbf {\bibinfo {volume}
  {7}},\ \bibinfo {pages} {119} (\bibinfo {year} {2016})}\BibitemShut {NoStop}%
\bibitem [{\citenamefont {{Ueda}}\ \emph {et~al.}(2019)\citenamefont {{Ueda}},
  \citenamefont {{Matsuo}}, \citenamefont {{Kamata}}, \citenamefont {{Baba}},
  \citenamefont {{Sato}}, \citenamefont {{Takeshige}}, \citenamefont {{Li}},
  \citenamefont {{Jeppesen}}, \citenamefont {{Samuelson}}, \citenamefont
  {{Xu}},\ and\ \citenamefont {{Tarucha}}}]{Ueda:2018}%
  \BibitemOpen
  \bibfield  {author} {\bibinfo {author} {\bibfnamefont {K.}~\bibnamefont
  {{Ueda}}}, \bibinfo {author} {\bibfnamefont {S.}~\bibnamefont {{Matsuo}}},
  \bibinfo {author} {\bibfnamefont {H.}~\bibnamefont {{Kamata}}}, \bibinfo
  {author} {\bibfnamefont {S.}~\bibnamefont {{Baba}}}, \bibinfo {author}
  {\bibfnamefont {Y.}~\bibnamefont {{Sato}}}, \bibinfo {author} {\bibfnamefont
  {Y.}~\bibnamefont {{Takeshige}}}, \bibinfo {author} {\bibfnamefont
  {K.}~\bibnamefont {{Li}}}, \bibinfo {author} {\bibfnamefont {S.}~\bibnamefont
  {{Jeppesen}}}, \bibinfo {author} {\bibfnamefont {L.}~\bibnamefont
  {{Samuelson}}}, \bibinfo {author} {\bibfnamefont {H.}~\bibnamefont {{Xu}}}, \
  and\ \bibinfo {author} {\bibfnamefont {S.}~\bibnamefont {{Tarucha}}},\
  }\bibfield  {title} {\bibinfo {title} {{{Dominant nonlocal superconducting
  proximity effect due to electron-electron interaction in a ballistic double
  nanowire}}},\ }\href {\doibase 10.1126/sciadv.aaw2194} {\bibfield  {journal}
  {\bibinfo  {journal} {Sci. Adv.}\ }\textbf {\bibinfo {volume} {5}},\ \bibinfo
  {pages} {eaaw2194} (\bibinfo {year} {2019})}\BibitemShut {NoStop}%
\bibitem [{\citenamefont {Sato}\ \emph {et~al.}(2019)\citenamefont {Sato},
  \citenamefont {Matsuo}, \citenamefont {Hsu}, \citenamefont {Stano},
  \citenamefont {Ueda}, \citenamefont {Takeshige}, \citenamefont {Kamata},
  \citenamefont {Lee}, \citenamefont {Shojaei}, \citenamefont {Wickramasinghe},
  \citenamefont {Shabani}, \citenamefont {{Palmstr{\o}m}}, \citenamefont
  {Tokura}, \citenamefont {Loss},\ and\ \citenamefont {Tarucha}}]{Sato:2019}%
  \BibitemOpen
  \bibfield  {author} {\bibinfo {author} {\bibfnamefont {Y.}~\bibnamefont
  {Sato}}, \bibinfo {author} {\bibfnamefont {S.}~\bibnamefont {Matsuo}},
  \bibinfo {author} {\bibfnamefont {C.-H.}\ \bibnamefont {Hsu}}, \bibinfo
  {author} {\bibfnamefont {P.}~\bibnamefont {Stano}}, \bibinfo {author}
  {\bibfnamefont {K.}~\bibnamefont {Ueda}}, \bibinfo {author} {\bibfnamefont
  {Y.}~\bibnamefont {Takeshige}}, \bibinfo {author} {\bibfnamefont
  {H.}~\bibnamefont {Kamata}}, \bibinfo {author} {\bibfnamefont {J.~S.}\
  \bibnamefont {Lee}}, \bibinfo {author} {\bibfnamefont {B.}~\bibnamefont
  {Shojaei}}, \bibinfo {author} {\bibfnamefont {K.}~\bibnamefont
  {Wickramasinghe}}, \bibinfo {author} {\bibfnamefont {J.}~\bibnamefont
  {Shabani}}, \bibinfo {author} {\bibfnamefont {C.}~\bibnamefont
  {{Palmstr{\o}m}}}, \bibinfo {author} {\bibfnamefont {Y.}~\bibnamefont
  {Tokura}}, \bibinfo {author} {\bibfnamefont {D.}~\bibnamefont {Loss}}, \ and\
  \bibinfo {author} {\bibfnamefont {S.}~\bibnamefont {Tarucha}},\ }\bibfield
  {title} {\bibinfo {title} {{Strong electron-electron interactions of a
  Tomonaga-Luttinger liquid observed in InAs quantum wires}},\ }\href {\doibase
  10.1103/PhysRevB.99.155304} {\bibfield  {journal} {\bibinfo  {journal} {Phys.
  Rev. B}\ }\textbf {\bibinfo {volume} {99}},\ \bibinfo {pages} {155304}
  (\bibinfo {year} {2019})}\BibitemShut {NoStop}%
\bibitem [{\citenamefont {Bockrath}\ \emph {et~al.}(1999)\citenamefont
  {Bockrath}, \citenamefont {Cobden}, \citenamefont {Lu}, \citenamefont
  {Rinzler}, \citenamefont {Smalley}, \citenamefont {Balents},\ and\
  \citenamefont {McEuen}}]{Bockrath:1999}%
  \BibitemOpen
  \bibfield  {author} {\bibinfo {author} {\bibfnamefont {M.}~\bibnamefont
  {Bockrath}}, \bibinfo {author} {\bibfnamefont {D.~H.}\ \bibnamefont
  {Cobden}}, \bibinfo {author} {\bibfnamefont {J.}~\bibnamefont {Lu}}, \bibinfo
  {author} {\bibfnamefont {A.~G.}\ \bibnamefont {Rinzler}}, \bibinfo {author}
  {\bibfnamefont {R.~E.}\ \bibnamefont {Smalley}}, \bibinfo {author}
  {\bibfnamefont {L.}~\bibnamefont {Balents}}, \ and\ \bibinfo {author}
  {\bibfnamefont {P.~L.}\ \bibnamefont {McEuen}},\ }\bibfield  {title}
  {\bibinfo {title} {{Luttinger-liquid behaviour in carbon nanotubes}},\
  }\href@noop {} {\bibfield  {journal} {\bibinfo  {journal} {Nature}\ }\textbf
  {\bibinfo {volume} {397}},\ \bibinfo {pages} {598} (\bibinfo {year}
  {1999})}\BibitemShut {NoStop}%
\bibitem [{\citenamefont {Luo}\ \emph {et~al.}(1988)\citenamefont {Luo},
  \citenamefont {Munekata}, \citenamefont {Fang},\ and\ \citenamefont
  {Stiles}}]{Luo:1988}%
  \BibitemOpen
  \bibfield  {author} {\bibinfo {author} {\bibfnamefont {J.}~\bibnamefont
  {Luo}}, \bibinfo {author} {\bibfnamefont {H.}~\bibnamefont {Munekata}},
  \bibinfo {author} {\bibfnamefont {F.~F.}\ \bibnamefont {Fang}}, \ and\
  \bibinfo {author} {\bibfnamefont {P.~J.}\ \bibnamefont {Stiles}},\ }\bibfield
   {title} {\bibinfo {title} {{Observation of the zero-field spin splitting of
  the ground electron subband in GaSb/InAs/GaSb quantum wells}},\ }\href
  {\doibase 10.1103/PhysRevB.38.10142} {\bibfield  {journal} {\bibinfo
  {journal} {Phys. Rev. B}\ }\textbf {\bibinfo {volume} {38}},\ \bibinfo
  {pages} {10142} (\bibinfo {year} {1988})}\BibitemShut {NoStop}%
\bibitem [{\citenamefont {Luo}\ \emph {et~al.}(1990)\citenamefont {Luo},
  \citenamefont {Munekata}, \citenamefont {Fang},\ and\ \citenamefont
  {Stiles}}]{Luo:1990}%
  \BibitemOpen
  \bibfield  {author} {\bibinfo {author} {\bibfnamefont {J.}~\bibnamefont
  {Luo}}, \bibinfo {author} {\bibfnamefont {H.}~\bibnamefont {Munekata}},
  \bibinfo {author} {\bibfnamefont {F.~F.}\ \bibnamefont {Fang}}, \ and\
  \bibinfo {author} {\bibfnamefont {P.~J.}\ \bibnamefont {Stiles}},\ }\bibfield
   {title} {\bibinfo {title} {{Effects of inversion asymmetry on electron
  energy band structures in GaSb/InAs/GaSb quantum wells}},\ }\href {\doibase
  10.1103/PhysRevB.41.7685} {\bibfield  {journal} {\bibinfo  {journal} {Phys.
  Rev. B}\ }\textbf {\bibinfo {volume} {41}},\ \bibinfo {pages} {7685}
  (\bibinfo {year} {1990})}\BibitemShut {NoStop}%
\bibitem [{\citenamefont {Grundler}(2000)}]{Grundler:2000}%
  \BibitemOpen
  \bibfield  {author} {\bibinfo {author} {\bibfnamefont {D.}~\bibnamefont
  {Grundler}},\ }\bibfield  {title} {\bibinfo {title} {{Large Rashba Splitting
  in InAs Quantum Wells due to Electron Wave Function Penetration into the
  Barrier Layers}},\ }\href {\doibase 10.1103/PhysRevLett.84.6074} {\bibfield
  {journal} {\bibinfo  {journal} {Phys. Rev. Lett.}\ }\textbf {\bibinfo
  {volume} {84}},\ \bibinfo {pages} {6074} (\bibinfo {year}
  {2000})}\BibitemShut {NoStop}%
\bibitem [{\citenamefont {{Heedt}}\ \emph {et~al.}(2017)\citenamefont
  {{Heedt}}, \citenamefont {{Traverso Ziani}}, \citenamefont {{Cr{\'e}pin}},
  \citenamefont {{Prost}}, \citenamefont {{Trellenkamp}}, \citenamefont
  {{Schubert}}, \citenamefont {{Gr{\"u}tzmacher}}, \citenamefont
  {{Trauzettel}},\ and\ \citenamefont {{Sch{\"a}pers}}}]{Heedt:2017}%
  \BibitemOpen
  \bibfield  {author} {\bibinfo {author} {\bibfnamefont {S.}~\bibnamefont
  {{Heedt}}}, \bibinfo {author} {\bibfnamefont {N.}~\bibnamefont {{Traverso
  Ziani}}}, \bibinfo {author} {\bibfnamefont {F.}~\bibnamefont {{Cr{\'e}pin}}},
  \bibinfo {author} {\bibfnamefont {W.}~\bibnamefont {{Prost}}}, \bibinfo
  {author} {\bibfnamefont {{\relax St}.}~\bibnamefont {{Trellenkamp}}},
  \bibinfo {author} {\bibfnamefont {J.}~\bibnamefont {{Schubert}}}, \bibinfo
  {author} {\bibfnamefont {D.}~\bibnamefont {{Gr{\"u}tzmacher}}}, \bibinfo
  {author} {\bibfnamefont {B.}~\bibnamefont {{Trauzettel}}}, \ and\ \bibinfo
  {author} {\bibfnamefont {{\relax Th}.}~\bibnamefont {{Sch{\"a}pers}}},\
  }\bibfield  {title} {\bibinfo {title} {{Signatures of interaction-induced
  helical gaps in nanowire quantum point contacts}},\ }\href {\doibase
  10.1038/nphys4070} {\bibfield  {journal} {\bibinfo  {journal} {Nat. Phys.}\
  }\textbf {\bibinfo {volume} {13}},\ \bibinfo {pages} {563} (\bibinfo {year}
  {2017})}\BibitemShut {NoStop}%
\bibitem [{\citenamefont {Braunecker}\ \emph {et~al.}(2010)\citenamefont
  {Braunecker}, \citenamefont {Japaridze}, \citenamefont {Klinovaja},\ and\
  \citenamefont {Loss}}]{Braunecker:2010}%
  \BibitemOpen
  \bibfield  {author} {\bibinfo {author} {\bibfnamefont {B.}~\bibnamefont
  {Braunecker}}, \bibinfo {author} {\bibfnamefont {G.~I.}\ \bibnamefont
  {Japaridze}}, \bibinfo {author} {\bibfnamefont {J.}~\bibnamefont
  {Klinovaja}}, \ and\ \bibinfo {author} {\bibfnamefont {D.}~\bibnamefont
  {Loss}},\ }\bibfield  {title} {\bibinfo {title} {{Spin-selective Peierls
  transition in interacting one-dimensional conductors with spin-orbit
  interaction}},\ }\href {\doibase 10.1103/PhysRevB.82.045127} {\bibfield
  {journal} {\bibinfo  {journal} {Phys. Rev. B}\ }\textbf {\bibinfo {volume}
  {82}},\ \bibinfo {pages} {045127} (\bibinfo {year} {2010})}\BibitemShut
  {NoStop}%
\bibitem [{\citenamefont {Meng}\ \emph {et~al.}(2014)\citenamefont {Meng},
  \citenamefont {Klinovaja},\ and\ \citenamefont {Loss}}]{Meng:2014}%
  \BibitemOpen
  \bibfield  {author} {\bibinfo {author} {\bibfnamefont {T.}~\bibnamefont
  {Meng}}, \bibinfo {author} {\bibfnamefont {J.}~\bibnamefont {Klinovaja}}, \
  and\ \bibinfo {author} {\bibfnamefont {D.}~\bibnamefont {Loss}},\ }\bibfield
  {title} {\bibinfo {title} {{Renormalization of anticrossings in interacting
  quantum wires with Rashba and Dresselhaus spin-orbit couplings}},\ }\href
  {\doibase 10.1103/PhysRevB.89.205133} {\bibfield  {journal} {\bibinfo
  {journal} {Phys. Rev. B}\ }\textbf {\bibinfo {volume} {89}},\ \bibinfo
  {pages} {205133} (\bibinfo {year} {2014})}\BibitemShut {NoStop}%
\bibitem [{\citenamefont {Kainaris}\ and\ \citenamefont
  {Carr}(2015)}]{Kainaris:2015}%
  \BibitemOpen
  \bibfield  {author} {\bibinfo {author} {\bibfnamefont {N.}~\bibnamefont
  {Kainaris}}\ and\ \bibinfo {author} {\bibfnamefont {S.~T.}\ \bibnamefont
  {Carr}},\ }\bibfield  {title} {\bibinfo {title} {{Emergent topological
  properties in interacting one-dimensional systems with spin-orbit
  coupling}},\ }\href {\doibase 10.1103/PhysRevB.92.035139} {\bibfield
  {journal} {\bibinfo  {journal} {Phys. Rev. B}\ }\textbf {\bibinfo {volume}
  {92}},\ \bibinfo {pages} {035139} (\bibinfo {year} {2015})}\BibitemShut
  {NoStop}%
\bibitem [{\citenamefont {Moroz}\ and\ \citenamefont
  {Barnes}(1999)}]{Moroz:1999}%
  \BibitemOpen
  \bibfield  {author} {\bibinfo {author} {\bibfnamefont {A.~V.}\ \bibnamefont
  {Moroz}}\ and\ \bibinfo {author} {\bibfnamefont {C.~H.~W.}\ \bibnamefont
  {Barnes}},\ }\bibfield  {title} {\bibinfo {title} {{Effect of the spin-orbit
  interaction on the band structure and conductance of quasi-one-dimensional
  systems}},\ }\href {\doibase 10.1103/PhysRevB.60.14272} {\bibfield  {journal}
  {\bibinfo  {journal} {Phys. Rev. B}\ }\textbf {\bibinfo {volume} {60}},\
  \bibinfo {pages} {14272} (\bibinfo {year} {1999})}\BibitemShut {NoStop}%
\bibitem [{\citenamefont {Governale}\ and\ \citenamefont
  {Z\"ulicke}(2002)}]{Governale:2002}%
  \BibitemOpen
  \bibfield  {author} {\bibinfo {author} {\bibfnamefont {M.}~\bibnamefont
  {Governale}}\ and\ \bibinfo {author} {\bibfnamefont {U.}~\bibnamefont
  {Z\"ulicke}},\ }\bibfield  {title} {\bibinfo {title} {{Spin accumulation in
  quantum wires with strong Rashba spin-orbit coupling}},\ }\href {\doibase
  10.1103/PhysRevB.66.073311} {\bibfield  {journal} {\bibinfo  {journal} {Phys.
  Rev. B}\ }\textbf {\bibinfo {volume} {66}},\ \bibinfo {pages} {073311}
  (\bibinfo {year} {2002})}\BibitemShut {NoStop}%
\bibitem [{\citenamefont {Moroz}\ \emph
  {et~al.}(2000{\natexlab{a}})\citenamefont {Moroz}, \citenamefont {Samokhin},\
  and\ \citenamefont {Barnes}}]{Moroz:2000}%
  \BibitemOpen
  \bibfield  {author} {\bibinfo {author} {\bibfnamefont {A.~V.}\ \bibnamefont
  {Moroz}}, \bibinfo {author} {\bibfnamefont {K.~V.}\ \bibnamefont {Samokhin}},
  \ and\ \bibinfo {author} {\bibfnamefont {C.~H.~W.}\ \bibnamefont {Barnes}},\
  }\bibfield  {title} {\bibinfo {title} {{Spin-Orbit Coupling in Interacting
  Quasi-One-Dimensional Electron Systems}},\ }\href {\doibase
  10.1103/PhysRevLett.84.4164} {\bibfield  {journal} {\bibinfo  {journal}
  {Phys. Rev. Lett.}\ }\textbf {\bibinfo {volume} {84}},\ \bibinfo {pages}
  {4164} (\bibinfo {year} {2000}{\natexlab{a}})}\BibitemShut {NoStop}%
\bibitem [{\citenamefont {Moroz}\ \emph
  {et~al.}(2000{\natexlab{b}})\citenamefont {Moroz}, \citenamefont {Samokhin},\
  and\ \citenamefont {Barnes}}]{Moroz:2000b}%
  \BibitemOpen
  \bibfield  {author} {\bibinfo {author} {\bibfnamefont {A.~V.}\ \bibnamefont
  {Moroz}}, \bibinfo {author} {\bibfnamefont {K.~V.}\ \bibnamefont {Samokhin}},
  \ and\ \bibinfo {author} {\bibfnamefont {C.~H.~W.}\ \bibnamefont {Barnes}},\
  }\bibfield  {title} {\bibinfo {title} {{Theory of quasi-one-dimensional
  electron liquids with spin-orbit coupling}},\ }\href {\doibase
  10.1103/PhysRevB.62.16900} {\bibfield  {journal} {\bibinfo  {journal} {Phys.
  Rev. B}\ }\textbf {\bibinfo {volume} {62}},\ \bibinfo {pages} {16900}
  (\bibinfo {year} {2000}{\natexlab{b}})}\BibitemShut {NoStop}%
\bibitem [{\citenamefont {Gritsev}\ \emph {et~al.}(2005)\citenamefont
  {Gritsev}, \citenamefont {Japaridze}, \citenamefont {Pletyukhov},\ and\
  \citenamefont {Baeriswyl}}]{Gritsev:2005}%
  \BibitemOpen
  \bibfield  {author} {\bibinfo {author} {\bibfnamefont {V.}~\bibnamefont
  {Gritsev}}, \bibinfo {author} {\bibfnamefont {G.~I.}\ \bibnamefont
  {Japaridze}}, \bibinfo {author} {\bibfnamefont {M.}~\bibnamefont
  {Pletyukhov}}, \ and\ \bibinfo {author} {\bibfnamefont {D.}~\bibnamefont
  {Baeriswyl}},\ }\bibfield  {title} {\bibinfo {title} {{Competing Effects of
  Interactions and Spin-Orbit Coupling in a Quantum Wire}},\ }\href {\doibase
  10.1103/PhysRevLett.94.137207} {\bibfield  {journal} {\bibinfo  {journal}
  {Phys. Rev. Lett.}\ }\textbf {\bibinfo {volume} {94}},\ \bibinfo {pages}
  {137207} (\bibinfo {year} {2005})}\BibitemShut {NoStop}%
\bibitem [{\citenamefont {Schulz}\ \emph {et~al.}(2009)\citenamefont {Schulz},
  \citenamefont {De~Martino}, \citenamefont {Ingenhoven},\ and\ \citenamefont
  {Egger}}]{Schulz:2009}%
  \BibitemOpen
  \bibfield  {author} {\bibinfo {author} {\bibfnamefont {A.}~\bibnamefont
  {Schulz}}, \bibinfo {author} {\bibfnamefont {A.}~\bibnamefont {De~Martino}},
  \bibinfo {author} {\bibfnamefont {P.}~\bibnamefont {Ingenhoven}}, \ and\
  \bibinfo {author} {\bibfnamefont {R.}~\bibnamefont {Egger}},\ }\bibfield
  {title} {\bibinfo {title} {{Low-energy theory and RKKY interaction for
  interacting quantum wires with Rashba spin-orbit coupling}},\ }\href
  {\doibase 10.1103/PhysRevB.79.205432} {\bibfield  {journal} {\bibinfo
  {journal} {Phys. Rev. B}\ }\textbf {\bibinfo {volume} {79}},\ \bibinfo
  {pages} {205432} (\bibinfo {year} {2009})}\BibitemShut {NoStop}%
\bibitem [{\citenamefont {Sun}\ \emph {et~al.}(2007)\citenamefont {Sun},
  \citenamefont {Gangadharaiah},\ and\ \citenamefont {Starykh}}]{Sun:2007}%
  \BibitemOpen
  \bibfield  {author} {\bibinfo {author} {\bibfnamefont {J.}~\bibnamefont
  {Sun}}, \bibinfo {author} {\bibfnamefont {S.}~\bibnamefont {Gangadharaiah}},
  \ and\ \bibinfo {author} {\bibfnamefont {O.~A.}\ \bibnamefont {Starykh}},\
  }\bibfield  {title} {\bibinfo {title} {{Spin-Orbit-Induced Spin-Density Wave
  in a Quantum Wire}},\ }\href {\doibase 10.1103/PhysRevLett.98.126408}
  {\bibfield  {journal} {\bibinfo  {journal} {Phys. Rev. Lett.}\ }\textbf
  {\bibinfo {volume} {98}},\ \bibinfo {pages} {126408} (\bibinfo {year}
  {2007})}\BibitemShut {NoStop}%
\bibitem [{\citenamefont {Gangadharaiah}\ \emph {et~al.}(2008)\citenamefont
  {Gangadharaiah}, \citenamefont {Sun},\ and\ \citenamefont
  {Starykh}}]{Gangadharaiah:2008}%
  \BibitemOpen
  \bibfield  {author} {\bibinfo {author} {\bibfnamefont {S.}~\bibnamefont
  {Gangadharaiah}}, \bibinfo {author} {\bibfnamefont {J.}~\bibnamefont {Sun}},
  \ and\ \bibinfo {author} {\bibfnamefont {O.~A.}\ \bibnamefont {Starykh}},\
  }\bibfield  {title} {\bibinfo {title} {{Spin-orbital effects in magnetized
  quantum wires and spin chains}},\ }\href {\doibase
  10.1103/PhysRevB.78.054436} {\bibfield  {journal} {\bibinfo  {journal} {Phys.
  Rev. B}\ }\textbf {\bibinfo {volume} {78}},\ \bibinfo {pages} {054436}
  (\bibinfo {year} {2008})}\BibitemShut {NoStop}%
\bibitem [{\citenamefont {Schmidt}(2013)}]{Schmidt:2013}%
  \BibitemOpen
  \bibfield  {author} {\bibinfo {author} {\bibfnamefont {T.~L.}\ \bibnamefont
  {Schmidt}},\ }\bibfield  {title} {\bibinfo {title} {{Finite-temperature
  conductance of interacting quantum wires with Rashba spin-orbit coupling}},\
  }\href {\doibase 10.1103/PhysRevB.88.235429} {\bibfield  {journal} {\bibinfo
  {journal} {Phys. Rev. B}\ }\textbf {\bibinfo {volume} {88}},\ \bibinfo
  {pages} {235429} (\bibinfo {year} {2013})}\BibitemShut {NoStop}%
\bibitem [{\citenamefont {Tretiakov}\ \emph {et~al.}(2013)\citenamefont
  {Tretiakov}, \citenamefont {Tikhonov},\ and\ \citenamefont
  {Pokrovsky}}]{Tretiakov:2013}%
  \BibitemOpen
  \bibfield  {author} {\bibinfo {author} {\bibfnamefont {O.~A.}\ \bibnamefont
  {Tretiakov}}, \bibinfo {author} {\bibfnamefont {K.~S.}\ \bibnamefont
  {Tikhonov}}, \ and\ \bibinfo {author} {\bibfnamefont {V.~L.}\ \bibnamefont
  {Pokrovsky}},\ }\bibfield  {title} {\bibinfo {title} {{Spin resonance in a
  Luttinger liquid with spin-orbit interaction}},\ }\href {\doibase
  10.1103/PhysRevB.88.125143} {\bibfield  {journal} {\bibinfo  {journal} {Phys.
  Rev. B}\ }\textbf {\bibinfo {volume} {88}},\ \bibinfo {pages} {125143}
  (\bibinfo {year} {2013})}\BibitemShut {NoStop}%
\bibitem [{\citenamefont {Pedder}\ \emph {et~al.}(2016)\citenamefont {Pedder},
  \citenamefont {Meng}, \citenamefont {Tiwari},\ and\ \citenamefont
  {Schmidt}}]{Pedder:2016}%
  \BibitemOpen
  \bibfield  {author} {\bibinfo {author} {\bibfnamefont {C.~J.}\ \bibnamefont
  {Pedder}}, \bibinfo {author} {\bibfnamefont {T.}~\bibnamefont {Meng}},
  \bibinfo {author} {\bibfnamefont {R.~P.}\ \bibnamefont {Tiwari}}, \ and\
  \bibinfo {author} {\bibfnamefont {T.~L.}\ \bibnamefont {Schmidt}},\
  }\bibfield  {title} {\bibinfo {title} {{Dynamic response functions and
  helical gaps in interacting Rashba nanowires with and without magnetic
  fields}},\ }\href {\doibase 10.1103/PhysRevB.94.245414} {\bibfield  {journal}
  {\bibinfo  {journal} {Phys. Rev. B}\ }\textbf {\bibinfo {volume} {94}},\
  \bibinfo {pages} {245414} (\bibinfo {year} {2016})}\BibitemShut {NoStop}%
\bibitem [{\citenamefont {Pedder}\ \emph {et~al.}(2017)\citenamefont {Pedder},
  \citenamefont {Meng}, \citenamefont {Tiwari},\ and\ \citenamefont
  {Schmidt}}]{Pedder:2017}%
  \BibitemOpen
  \bibfield  {author} {\bibinfo {author} {\bibfnamefont {C.~J.}\ \bibnamefont
  {Pedder}}, \bibinfo {author} {\bibfnamefont {T.}~\bibnamefont {Meng}},
  \bibinfo {author} {\bibfnamefont {R.~P.}\ \bibnamefont {Tiwari}}, \ and\
  \bibinfo {author} {\bibfnamefont {T.~L.}\ \bibnamefont {Schmidt}},\
  }\bibfield  {title} {\bibinfo {title} {{Missing Shapiro steps and the
  $8\ensuremath{\pi}$-periodic Josephson effect in interacting helical electron
  systems}},\ }\href {\doibase 10.1103/PhysRevB.96.165429} {\bibfield
  {journal} {\bibinfo  {journal} {Phys. Rev. B}\ }\textbf {\bibinfo {volume}
  {96}},\ \bibinfo {pages} {165429} (\bibinfo {year} {2017})}\BibitemShut
  {NoStop}%
\bibitem [{\citenamefont {Iucci}(2003)}]{Iucci:2003}%
  \BibitemOpen
  \bibfield  {author} {\bibinfo {author} {\bibfnamefont {A.}~\bibnamefont
  {Iucci}},\ }\bibfield  {title} {\bibinfo {title} {{Correlation functions for
  one-dimensional interacting fermions with spin-orbit coupling}},\ }\href
  {\doibase 10.1103/PhysRevB.68.075107} {\bibfield  {journal} {\bibinfo
  {journal} {Phys. Rev. B}\ }\textbf {\bibinfo {volume} {68}},\ \bibinfo
  {pages} {075107} (\bibinfo {year} {2003})}\BibitemShut {NoStop}%
\bibitem [{\citenamefont {Governale}\ and\ \citenamefont
  {Z\"ulicke}(2004)}]{Governale:2004}%
  \BibitemOpen
  \bibfield  {author} {\bibinfo {author} {\bibfnamefont {M.}~\bibnamefont
  {Governale}}\ and\ \bibinfo {author} {\bibfnamefont {U.}~\bibnamefont
  {Z\"ulicke}},\ }\bibfield  {title} {\bibinfo {title} {{Rashba spin splitting
  in quantum wires}},\ }\href {\doibase
  https://doi.org/10.1016/j.ssc.2004.05.047} {\bibfield  {journal} {\bibinfo
  {journal} {Solid State Commun.}\ }\textbf {\bibinfo {volume} {131}},\
  \bibinfo {pages} {581 } (\bibinfo {year} {2004})}\BibitemShut {NoStop}%
\bibitem [{\citenamefont {Rainis}\ and\ \citenamefont
  {Loss}(2014)}]{Rainis:2014}%
  \BibitemOpen
  \bibfield  {author} {\bibinfo {author} {\bibfnamefont {D.}~\bibnamefont
  {Rainis}}\ and\ \bibinfo {author} {\bibfnamefont {D.}~\bibnamefont {Loss}},\
  }\bibfield  {title} {\bibinfo {title} {{Conductance behavior in nanowires
  with spin-orbit interaction: A numerical study}},\ }\href {\doibase
  10.1103/PhysRevB.90.235415} {\bibfield  {journal} {\bibinfo  {journal} {Phys.
  Rev. B}\ }\textbf {\bibinfo {volume} {90}},\ \bibinfo {pages} {235415}
  (\bibinfo {year} {2014})}\BibitemShut {NoStop}%
\bibitem [{\citenamefont {Hsu}\ \emph {et~al.}(2017)\citenamefont {Hsu},
  \citenamefont {Stano}, \citenamefont {Klinovaja},\ and\ \citenamefont
  {Loss}}]{Hsu:2017}%
  \BibitemOpen
  \bibfield  {author} {\bibinfo {author} {\bibfnamefont {C.-H.}\ \bibnamefont
  {Hsu}}, \bibinfo {author} {\bibfnamefont {P.}~\bibnamefont {Stano}}, \bibinfo
  {author} {\bibfnamefont {J.}~\bibnamefont {Klinovaja}}, \ and\ \bibinfo
  {author} {\bibfnamefont {D.}~\bibnamefont {Loss}},\ }\bibfield  {title}
  {\bibinfo {title} {{Nuclear-spin-induced localization of edge states in
  two-dimensional topological insulators}},\ }\href {\doibase
  10.1103/PhysRevB.96.081405} {\bibfield  {journal} {\bibinfo  {journal} {Phys.
  Rev. B}\ }\textbf {\bibinfo {volume} {96}},\ \bibinfo {pages} {081405(R)}
  (\bibinfo {year} {2017})}\BibitemShut {NoStop}%
\bibitem [{\citenamefont {Hsu}\ \emph {et~al.}(2018)\citenamefont {Hsu},
  \citenamefont {Stano}, \citenamefont {Klinovaja},\ and\ \citenamefont
  {Loss}}]{Hsu:2018}%
  \BibitemOpen
  \bibfield  {author} {\bibinfo {author} {\bibfnamefont {C.-H.}\ \bibnamefont
  {Hsu}}, \bibinfo {author} {\bibfnamefont {P.}~\bibnamefont {Stano}}, \bibinfo
  {author} {\bibfnamefont {J.}~\bibnamefont {Klinovaja}}, \ and\ \bibinfo
  {author} {\bibfnamefont {D.}~\bibnamefont {Loss}},\ }\bibfield  {title}
  {\bibinfo {title} {{Effects of nuclear spins on the transport properties of
  the edge of two-dimensional topological insulators}},\ }\href {\doibase
  10.1103/PhysRevB.97.125432} {\bibfield  {journal} {\bibinfo  {journal} {Phys.
  Rev. B}\ }\textbf {\bibinfo {volume} {97}},\ \bibinfo {pages} {125432}
  (\bibinfo {year} {2018})}\BibitemShut {NoStop}%
\bibitem [{\citenamefont {Egger}\ and\ \citenamefont
  {Grabert}(1998)}]{Egger:1998b}%
  \BibitemOpen
  \bibfield  {author} {\bibinfo {author} {\bibfnamefont {R.}~\bibnamefont
  {Egger}}\ and\ \bibinfo {author} {\bibfnamefont {H.}~\bibnamefont
  {Grabert}},\ }\bibfield  {title} {\bibinfo {title} {{Applying voltage sources
  to a Luttinger liquid with arbitrary transmission}},\ }\href {\doibase
  10.1103/PhysRevB.58.10761} {\bibfield  {journal} {\bibinfo  {journal} {Phys.
  Rev. B}\ }\textbf {\bibinfo {volume} {58}},\ \bibinfo {pages} {10761}
  (\bibinfo {year} {1998})}\BibitemShut {NoStop}%
\bibitem [{\citenamefont {Aseev}\ \emph {et~al.}(2018)\citenamefont {Aseev},
  \citenamefont {Loss},\ and\ \citenamefont {Klinovaja}}]{Aseev:2018}%
  \BibitemOpen
  \bibfield  {author} {\bibinfo {author} {\bibfnamefont {P.~P.}\ \bibnamefont
  {Aseev}}, \bibinfo {author} {\bibfnamefont {D.}~\bibnamefont {Loss}}, \ and\
  \bibinfo {author} {\bibfnamefont {J.}~\bibnamefont {Klinovaja}},\ }\bibfield
  {title} {\bibinfo {title} {{Conductance of fractional Luttinger liquids at
  finite temperatures}},\ }\href {\doibase 10.1103/PhysRevB.98.045416}
  {\bibfield  {journal} {\bibinfo  {journal} {Phys. Rev. B}\ }\textbf {\bibinfo
  {volume} {98}},\ \bibinfo {pages} {045416} (\bibinfo {year}
  {2018})}\BibitemShut {NoStop}%
\bibitem [{\citenamefont {Meden}\ \emph {et~al.}(2003)\citenamefont {Meden},
  \citenamefont {Andergassen}, \citenamefont {Metzner}, \citenamefont
  {Schollw{\"o}ck},\ and\ \citenamefont {Sch{\"o}nhammer}}]{Meden:2003}%
  \BibitemOpen
  \bibfield  {author} {\bibinfo {author} {\bibfnamefont {V.}~\bibnamefont
  {Meden}}, \bibinfo {author} {\bibfnamefont {S.}~\bibnamefont {Andergassen}},
  \bibinfo {author} {\bibfnamefont {W.}~\bibnamefont {Metzner}}, \bibinfo
  {author} {\bibfnamefont {U.}~\bibnamefont {Schollw{\"o}ck}}, \ and\ \bibinfo
  {author} {\bibfnamefont {K.}~\bibnamefont {Sch{\"o}nhammer}},\ }\bibfield
  {title} {\bibinfo {title} {{Scaling of the conductance in a quantum wire}},\
  }\href {\doibase 10.1209/epl/i2003-00624-x} {\bibfield  {journal} {\bibinfo
  {journal} {{Europhys. Lett.}}\ }\textbf {\bibinfo {volume} {64}},\ \bibinfo
  {pages} {769} (\bibinfo {year} {2003})}\BibitemShut {NoStop}%
\bibitem [{\citenamefont {Janzen}\ \emph {et~al.}(2006)\citenamefont {Janzen},
  \citenamefont {Meden},\ and\ \citenamefont {Sch\"onhammer}}]{Janzen:2006}%
  \BibitemOpen
  \bibfield  {author} {\bibinfo {author} {\bibfnamefont {K.}~\bibnamefont
  {Janzen}}, \bibinfo {author} {\bibfnamefont {V.}~\bibnamefont {Meden}}, \
  and\ \bibinfo {author} {\bibfnamefont {K.}~\bibnamefont {Sch\"onhammer}},\
  }\bibfield  {title} {\bibinfo {title} {{Influence of the contacts on the
  conductance of interacting quantum wires}},\ }\href {\doibase
  10.1103/PhysRevB.74.085301} {\bibfield  {journal} {\bibinfo  {journal} {Phys.
  Rev. B}\ }\textbf {\bibinfo {volume} {74}},\ \bibinfo {pages} {085301}
  (\bibinfo {year} {2006})}\BibitemShut {NoStop}%
\bibitem [{\citenamefont {Schrieffer}\ \emph {et~al.}(1963)\citenamefont
  {Schrieffer}, \citenamefont {Scalapino},\ and\ \citenamefont
  {Wilkins}}]{Schrieffer:1963}%
  \BibitemOpen
  \bibfield  {author} {\bibinfo {author} {\bibfnamefont {J.~R.}\ \bibnamefont
  {Schrieffer}}, \bibinfo {author} {\bibfnamefont {D.~J.}\ \bibnamefont
  {Scalapino}}, \ and\ \bibinfo {author} {\bibfnamefont {J.~W.}\ \bibnamefont
  {Wilkins}},\ }\bibfield  {title} {\bibinfo {title} {{Effective Tunneling
  Density of States in Superconductors}},\ }\href {\doibase
  10.1103/PhysRevLett.10.336} {\bibfield  {journal} {\bibinfo  {journal} {Phys.
  Rev. Lett.}\ }\textbf {\bibinfo {volume} {10}},\ \bibinfo {pages} {336}
  (\bibinfo {year} {1963})}\BibitemShut {NoStop}%
\bibitem [{\citenamefont {Wen}(1991)}]{Wen:1991}%
  \BibitemOpen
  \bibfield  {author} {\bibinfo {author} {\bibfnamefont {X.-G.}\ \bibnamefont
  {Wen}},\ }\bibfield  {title} {\bibinfo {title} {{Edge transport properties of
  the fractional quantum Hall states and weak-impurity scattering of a
  one-dimensional charge-density wave}},\ }\href {\doibase
  10.1103/PhysRevB.44.5708} {\bibfield  {journal} {\bibinfo  {journal} {Phys.
  Rev. B}\ }\textbf {\bibinfo {volume} {44}},\ \bibinfo {pages} {5708}
  (\bibinfo {year} {1991})}\BibitemShut {NoStop}%
\bibitem [{\citenamefont {Jakobs}\ \emph {et~al.}(2007)\citenamefont {Jakobs},
  \citenamefont {Meden}, \citenamefont {Schoeller},\ and\ \citenamefont
  {Enss}}]{Jakobs:2007}%
  \BibitemOpen
  \bibfield  {author} {\bibinfo {author} {\bibfnamefont {S.~G.}\ \bibnamefont
  {Jakobs}}, \bibinfo {author} {\bibfnamefont {V.}~\bibnamefont {Meden}},
  \bibinfo {author} {\bibfnamefont {H.}~\bibnamefont {Schoeller}}, \ and\
  \bibinfo {author} {\bibfnamefont {T.}~\bibnamefont {Enss}},\ }\bibfield
  {title} {\bibinfo {title} {{Temperature-induced phase averaging in
  one-dimensional mesoscopic systems}},\ }\href {\doibase
  10.1103/PhysRevB.75.035126} {\bibfield  {journal} {\bibinfo  {journal} {Phys.
  Rev. B}\ }\textbf {\bibinfo {volume} {75}},\ \bibinfo {pages} {035126}
  (\bibinfo {year} {2007})}\BibitemShut {NoStop}%
\bibitem [{\citenamefont {V\"ayrynen}\ \emph {et~al.}(2016)\citenamefont
  {V\"ayrynen}, \citenamefont {Geissler},\ and\ \citenamefont
  {Glazman}}]{Vayrynen:2016}%
  \BibitemOpen
  \bibfield  {author} {\bibinfo {author} {\bibfnamefont {J.~I.}\ \bibnamefont
  {V\"ayrynen}}, \bibinfo {author} {\bibfnamefont {F.}~\bibnamefont
  {Geissler}}, \ and\ \bibinfo {author} {\bibfnamefont {L.~I.}\ \bibnamefont
  {Glazman}},\ }\bibfield  {title} {\bibinfo {title} {{Magnetic moments in a
  helical edge can make weak correlations seem strong}},\ }\href {\doibase
  10.1103/PhysRevB.93.241301} {\bibfield  {journal} {\bibinfo  {journal} {Phys.
  Rev. B}\ }\textbf {\bibinfo {volume} {93}},\ \bibinfo {pages} {241301(R)}
  (\bibinfo {year} {2016})}\BibitemShut {NoStop}%
\bibitem [{\citenamefont {{St{\"u}hler}}\ \emph {et~al.}(2019)\citenamefont
  {{St{\"u}hler}}, \citenamefont {{Reis}}, \citenamefont {{M{\"u}ller}},
  \citenamefont {{Helbig}}, \citenamefont {{Schwemmer}}, \citenamefont
  {{Thomale}}, \citenamefont {{Sch{\"a}fer}},\ and\ \citenamefont
  {{Claessen}}}]{Stuhler:2019}%
  \BibitemOpen
  \bibfield  {author} {\bibinfo {author} {\bibfnamefont {R.}~\bibnamefont
  {{St{\"u}hler}}}, \bibinfo {author} {\bibfnamefont {F.}~\bibnamefont
  {{Reis}}}, \bibinfo {author} {\bibfnamefont {T.}~\bibnamefont
  {{M{\"u}ller}}}, \bibinfo {author} {\bibfnamefont {T.}~\bibnamefont
  {{Helbig}}}, \bibinfo {author} {\bibfnamefont {T.}~\bibnamefont
  {{Schwemmer}}}, \bibinfo {author} {\bibfnamefont {R.}~\bibnamefont
  {{Thomale}}}, \bibinfo {author} {\bibfnamefont {J.}~\bibnamefont
  {{Sch{\"a}fer}}}, \ and\ \bibinfo {author} {\bibfnamefont {R.}~\bibnamefont
  {{Claessen}}},\ }\bibfield  {title} {\bibinfo {title} {{Tomonaga-Luttinger
  liquid in the edge channels of a quantum spin Hall insulator}},\ }\href
  {\doibase 10.1038/s41567-019-0697-z} {\bibfield  {journal} {\bibinfo
  {journal} {Nat. Phys.}\ } (\bibinfo {year} {2019}),\
  10.1038/s41567-019-0697-z}\BibitemShut {NoStop}%
\bibitem [{\citenamefont {Maciejko}\ \emph {et~al.}(2009)\citenamefont
  {Maciejko}, \citenamefont {Liu}, \citenamefont {Oreg}, \citenamefont {Qi},
  \citenamefont {Wu},\ and\ \citenamefont {Zhang}}]{Maciejko:2009}%
  \BibitemOpen
  \bibfield  {author} {\bibinfo {author} {\bibfnamefont {J.}~\bibnamefont
  {Maciejko}}, \bibinfo {author} {\bibfnamefont {C.}~\bibnamefont {Liu}},
  \bibinfo {author} {\bibfnamefont {Y.}~\bibnamefont {Oreg}}, \bibinfo {author}
  {\bibfnamefont {X.-L.}\ \bibnamefont {Qi}}, \bibinfo {author} {\bibfnamefont
  {C.}~\bibnamefont {Wu}}, \ and\ \bibinfo {author} {\bibfnamefont {S.-C.}\
  \bibnamefont {Zhang}},\ }\bibfield  {title} {\bibinfo {title} {{Kondo Effect
  in the Helical Edge Liquid of the Quantum Spin Hall State}},\ }\href
  {\doibase 10.1103/PhysRevLett.102.256803} {\bibfield  {journal} {\bibinfo
  {journal} {Phys. Rev. Lett.}\ }\textbf {\bibinfo {volume} {102}},\ \bibinfo
  {pages} {256803} (\bibinfo {year} {2009})}\BibitemShut {NoStop}%
\bibitem [{\citenamefont {Gradshteyn}\ and\ \citenamefont
  {Ryzhik}(1980)}]{Gradshteyn:1980}%
  \BibitemOpen
  \bibfield  {author} {\bibinfo {author} {\bibfnamefont {I.}~\bibnamefont
  {Gradshteyn}}\ and\ \bibinfo {author} {\bibfnamefont {I.}~\bibnamefont
  {Ryzhik}},\ }\href@noop {} {\emph {\bibinfo {title} {{Table of Integrals,
  Series, and Products}}}}\ (\bibinfo  {publisher} {{Academic Press, New
  York}},\ \bibinfo {year} {1980})\BibitemShut {NoStop}%
\end{thebibliography}

%

\end{document}